\documentclass[12pt]{article}
\usepackage{amsfonts}
\newcommand {\ov} {\overline}
\newcommand{\ba}{\begin{eqnarray}}
\newcommand{\ea}{\end{eqnarray}}

\def\k{\kappa}

\newsavebox{\uuunit}
\sbox{\uuunit}
    {\setlength{\unitlength}{0.825em}
     \begin{picture}(0.6,0.7)
        \thinlines
        \put(0,0){\line(1,0){0.5}}
        \put(0.15,0){\line(0,1){0.7}}
        \put(0.35,0){\line(0,1){0.8}}
       \multiput(0.3,0.8)(-0.04,-0.02){12}{\rule{0.5pt}{0.5pt}}
     \end {picture}}

\begin{document}
\begin{titlepage}
\title
 {\bf \large Aspects of D-brane Dynamics in Supergravity Backgrounds with Fluxes,
 Kappa--symmetry and Equations of Motion. Part IIB
}

\author{Igor Bandos$^{1,2}$ and Dmitri Sorokin$^{2,3}$
\\\\
{\it\small $^1$ Akhiezer Institute for Theoretical Physics}\\
{\it\small NSC KIPT, National Academy of Sciences of Ukraine}\\
{\it\small Akademicheskaya Street 1, 61108  Kharkov, Ukraine}
\\
\\
{\it \small $^2$ Departamento de Fisica Teorica, Facultad de Fisica,
Universidad de
Valencia} \\
{\it\small C/Dr. Moliner, 50, 46100 Burjassot (Valencia), Spain}
\\
\\
$^3$ {\it \small INFN, Sezione di Padova}\\{\it \small  ${\&}$
Dipartimento di Fisica ``Galileo Galilei", Universit\`{a} degli
Studi
di Padova} \\
{\it\small via F. Marzolo 8, 35131 Padova, Italia}}
\date{}
\maketitle \thispagestyle{empty}
\begin{center}
{\bf Abstract}
\end{center}
We derive and carry out a detailed analysis of the equations of
motion of the type IIB \hbox{D branes} in generic supergravity
backgrounds with fluxes making account of the worldvolume
Born--Infeld gauge field and putting a special emphasis on the
structure of the Dirac equation for Dp brane fermionic modes. We
present an explicit form of the worldvolume field equations for each
of the Dp branes $(p=1,3,5,7,9)$ in the cases in which the
Neveu--Schwarz flux and the Ramond--Ramond p--form flux along the
Dp--brane worldvolume are zero and the supergravity backgrounds do
not necessarily induce the worldvolume Born--Infeld flux. We then
give several examples of D3, D5 and D7 brane configurations in which
the worldvolume Born--Infeld flux is intrinsically non--zero and
therefore must be taken into account in studying problems where such
branes are involved. The examples include D3 and D5 brane instantons
carrying (self--dual) worldvolume gauge fields in warped
compactification backgrounds.

\end{titlepage}

\newpage
\tableofcontents

\newpage
\section{Introduction}

The actions for the M2 \cite{M2} and M5 branes \cite{M5} in the
$D=11$ supergravity background  and those for the Dirichlet branes
in type II $D=10$ supergravity backgrounds \cite{Dp}--\cite{bt}, as
well as their equations of motion
\cite{M2,M5e,Howe:1997fb,Bandos:1997rq} have been used for studying
numerous problems of string theory and M theory. The superbrane
actions and equations of motion are constructed in curved target
superspaces in terms of the pullbacks to the brane worldvolume of
bulk supervielbeins and tensor gauge field superforms subject to
superfield constraints. The constraints (which actually put the
$D=10,11$ supergravity theories on the mass shell) ensure that the
superbranes possess local fermionic kappa--symmetry. In addition to
the physical fields of the $D=10,11$ supergravity supermultiplets
the supervielbeins and the tensor gauge field superforms contain an
enormous amount of other fields which upon imposing the supergravity
constraints can either be gauged away by local superdiffeomorphisms
and gauge symmetries or can be expressed
 in terms of the physical fields via algebraic equations. Hence, the
superbrane actions and equations of motion only implicitly describe
the interaction of the worldvolume modes of the superbranes with the
physical supergravity fields, especially as far as the worldvolume
fermions $\theta(\xi)$ are concerned. These fermionic modes are
Goldstone fields associated with brane fluctuations in the 32
Grassmann--odd directions $\theta$ of the target superspace.

For various applications of the M--branes and the D--branes it is
necessary to know the explicit dependence of their actions and
equations of motion on the physical fields in the bulk. To obtain
the explicit and entire expressions one should eliminate the
auxiliary fields by imposing on the supergravity superfields a
Wess--Zumino gauge \footnote{This conventional procedure deals with
the background field approximation to supergravity--superbrane
interactions. For further discussion and an approach to account the
back reaction of branes see \cite{sbs,sbs'}.}, expand these
superfields in power series of the Grassmann--odd coordinates
$\theta$ (up to the 32-nd order in $D=10,11$ superspaces) and insert
these polynomials into the superbrane actions and/or field
equations. One can perform such an expansion using the gauge
completion method \cite{deWit:1998tk} or a much more systematic
method of superspace normal coordinates
\cite{McArthur:1983fm,bz,Grisaru:2000ij,sbs',Tsimpis:2004gq}.

Though it might be technically possible to arrive at the final
expressions, in the general case they will be unpracticable even if
by gauge fixing the $\kappa$--symmetry one eliminates half of the 32
$\theta(\xi)$ components and reduces the polynomial order down to
16. In some cases of particular branes and backgrounds, such as flat
space--time (see {\it e.g.}
\cite{M2,Dp,DeCastro:2002vd}) and  $AdS_d\times S^{D-d}$
\cite{ads}, gauge fixed M2, M5 and Dp brane actions can be
constructed of the supervielbeins which are polynomials of up to the
second or fourth order in the physical modes of $\theta(\xi)$.

In some applications, for instance, for studying brane--world
scenarios and \linebreak non--perturbative effects due to brane
instantons on the generation of superpotentials in string/M--theory
compactifications (\cite{Becker:1995kb}--\cite{Buchbinder:2002pr}
and references therein) it is sufficient, at a first stage, to know
the explicit form of the brane actions up to the second order in
worldvolume fermions, and the equations of motion linear in
fermions. This approximation is admissible, {\it e.g.} in situations
where the Dirac operator on the brane has exactly two fermion zero
modes which one integrates over to calculate instanton contributions
to the effective action.

Quadratic actions for fermions on branes coupled to a generic
supergravity background with fluxes were derived in
\cite{Grisaru:2000ij} for the M2 brane, in \cite{Grana:2002tu} for a
D3 brane, in \cite{Marolf:2003vf,Martucci:2005rb} for the
Dp--branes, and in \cite{Kallosh:2005yu} for the M5--brane. The back
reaction of branes on the compactification setup has been discussed
in \cite{Lust:2005cu} and in a general context of interacting
superbrane--supergravity systems in \cite{Bandos:2005ww}.

Using these results, the zero modes of the Dirac operators of brane
fermions interacting with  bulk fluxes have been analyzed in
\cite{Saulina:2005ve,Kallosh:2005gs} for  M5--brane instantons in
M--theory on $M_{3}\times X_8$, and in
\cite{Tripathy:2005hv,Bergshoeff:2005yp,Park:2005hj} for D3 brane
instantons in type IIB String Theory on $M_{4}\times X_6$ (where $M$
is the effective physical space--time and $X$ are compactified
subspaces of String/M--theory). It has been shown that in some cases
including the $K3 \times T^2/Z^2$ orientifold compactification of
type IIB String Theory, brane instantons coupled to {\it bulk}
fluxes can produce non--perturbative corrections to the
superpotential and hence should be taken into account when carrying
out the analysis in search for phenomenologically relevant models of
particle interactions and cosmology.

A role of the {\it worldvolume} flux of D--branes in producing D-
and F- terms in effective $D=4$ theory and thus giving the
possibility of stabilizing moduli in type IIB String Theory
compactified on Calabi--Yau orientifolds has been discussed {\it
e.g.} in \cite{luis,garcia,Martucci:2006ij} and references therein.
The analysis of the latter paper has been based on the relation with
supersymmetry conditions for D branes with general worldvolume
fluxes on general $N=1$ flux backgrounds studied in detail in
\cite{Martucci:2005ht}.

Recently, effects of a worldvolume flux on solutions of the Dirac
equation for fermions on an M5 brane instanton wrapping a
Calabi--Yau space have been considered in \cite{Lilia}.

It seems of interest to look for more examples of brane
configurations generating scalar field potentials and to analyze
whether also the {\it worldvolume} fluxes on D branes and the M5
brane may produce or suppress some non--trivial effects. To this end
one should analyze the brane actions and equations of motion in the
presence of worldvolume Born--Infeld gauge fields. This can also be
useful for other brane applications such as brane worlds and AdS/CFT
correspondence. A main goal of this paper is to carry out such an
analysis for the Dp--branes of type IIB string theory, so
$p=1,3,5,7$ and 9. A reason why we restrict the consideration to the
type IIB D branes is the desire to present the results (at least
where it is possible) in an explicit form. The actions and equations
of motion for the type IIA branes can be analyzed in a similar way
elsewhere.

Though the explicit and geometrically suggestive quadratic actions
for the type IIB (and IIA) Dp--brane fermions in an arbitrary
bosonic supergravity background with the bulk and worldvolume fluxes
have already been given in \cite{Martucci:2005rb}, one should still
extract from these actions the Dirac equations and present them in a
hopefully tractable form. For instance, though the kappa--symmetry
gauge fixed actions for the Dp--branes in the string frame do not
contain terms with dilaton derivatives (\cite{Martucci:2005rb}),
these terms reappear in the fermion equations of motion, and they
should be taken into account in the analysis in the situations of a
non--trivial dilaton--axion. Also the consequences of different
gauge fixings of kappa--symmetry should be understood in more
detail. For this we shall elucidate the role of the kappa--symmetry
projector and present the fermionic equations in a gauge independent
covariant form. As a valuable byproduct of this analysis we observe
an interesting feature, which for some reason has not been
emphasized in the literature, that a kind of the Dirac equation for
the Dp--brane kappa--symmetry projector reproduces in a concise form
the full set of the D--brane bosonic field equations (for a concise
form of the $D=11$ supergravity equations see \cite{Bandos:2005mm}
and references therein  and  for a concise form of the (massive)
type IIA supergravity equations see \cite{Bovy:2005qq}).

Let us stress that when analyzing the fermionic equation in a
certain setup, one should take into consideration the consistency of
the setup with the Dp--brane bosonic equations. For instance, if one
ignores the contribution of the worldvolume flux this should be in
agreement with the BI equation for the worldvolume gauge field and
with the field equation for the worldvolume scalars. This also
concerns static brane configurations which do not fluctuate in the
transverse directions.

We also find it instructive and useful for further applications to
collect in one place and review  the complete set of the Dp--brane
equations of motion in an arbitrary superspace supergravity
background before truncating them to the linear approximation in
fermions. With this instructive purpose we present the equations
both in the Einstein and string frame.

The paper is organized as follows. In Sections 2, 3 and 4 we review
the generic structure of the actions and of the worldvolume field
equations for the Dp--branes coupled to the Neveu--Schwarz and
Ramond--Ramond fields in curved type IIB superspace using techniques
similar to those in the superembedding approach (see
\cite{se,ser,Howe:1997wf,Sorokin:1999jx} for a review). This simplifies the
derivation and the analysis of the Dp--brane equations. In Section 4
we study the properties of the D--brane kappa--symmetry and of its
projector and show that the projector matrix takes values in the
group $Spin(1,p)\subset Spin(1,9)$
 and that it is related via the spinor--vector
representation correspondence to the $SO(1,p)$--valued matrix
$k_a{}^b=(\eta+F)_{ac}(\eta-F)^{-1\,cb}$ \cite{Akulov:1998bq}, the
so called {\it Cayley image} of the worldvolume gauge field strength
$F_{ab}$ . Using this property we then show that a first--order
differential equation for the $\kappa$--symmetry projection matrix
amounts to the full set of equations and Bianchi identities for the
bosonic worldvolume fields on the brane. In Subsection \ref{linear}
we linearize the D--brane equations for the fermion field
$\theta(\xi)$ in purely bosonic supergravity backgrounds and in
Section \ref{analysis} we present them in a $\kappa$--symmetry
covariant form making use of the concise form of the bosonic
equations derived in Section
\ref{conciseboson} in terms of the kappa--symmetry projector. The
covariant (gauge independent) form of the fermionic equations allows
one to see how different kappa--symmetry gauge choices are related
to each other.

In Subsection \ref{withoutflux} we give the explicit form of the
fermionic equations for each of the type IIB Dp--branes
$(p=1,3,5,7,9)$ in the supergravity backgrounds which do not induce
a worldvolume flux. In Subsection \ref{D3i} we consider a $D3$ brane
instanton wrapping $K3$ and carrying an (anti)--self--dual
Born--Infeld instanton field. We show that if the compactified
space--time is warped by the $R_5$ flux, the consistency of the
equations of motion  of the D3 brane instanton wrapping a
four--cycle of the Calabi--Yau manifold can require the presence of
the worldvolume flux.

In Subsection \ref{warped} we give more examples of situations in
which the worldvolume BI field must be taken into account in the
study of the dynamics of D--branes in certain setups. For instance,
a D5 brane wrapping the compact manifold $X_5$ in an $AdS_5 \times
X_5$ background (where $X_5$ is an $S^5$ sphere or a
Sasaki--Einstein space) carries an effective electric charge
proportional to the inverse $X_5$ radius which induces an electric
BI field in its worldvolume. The D5 brane wrapping $S^5$ is
associated via AdS/CFT correspondence with the baryon vertex of the
effective $D=4$ gauge theory \cite{baryon,Gross:1998gk} and has been
extensively studied from various perspectives
\cite{Imamura:1998gk}--\cite{Yolanda}. We add to the known results
the explicit form of the linearized Dirac equation for the fermionic
modes on the  baryonic  D5 brane.

Other examples of branes with worldvolume fluxes are D5 and D7
branes in warped compactification backgrounds and, in particular, a
D5 brane instanton wrapping a Calabi--Yau manifold with non--zero
5--form and/or 3--form background fluxes. For all examples we give
an explicit form of the Dirac equation for the Dp--brane fermionic
modes.

We have tried to present the examples in a self--contained form so
that the reader interested in applications could directly use them
without the necessity of ploughing through the generic part of this
paper. Basic notation and conventions are given in the Appendix.

\setcounter{equation}0
\section{The action for the super-Dp-branes in $D=10$ type $IIB$ superspace}

The $p+1$ dimensional worldvolume ${\cal{W}}^{p+1}$ of a D--brane is
parametrized by the coordinates $\xi^m$ ($m=0,1\cdots,\, p$). On its
worldvolume the D--brane carries a vector gauge field $A_m(\xi)$.
The dynamics of the D--brane is described by how its worldvolume is
embedded into $D=10$ type $IIB$ superspace whose supercoordinates
$Z^{\underline M}=(x^{\underline m},\theta^{1\mu},\theta^{2\nu})$
include the space--time coordinates $x^{\underline m}$ $(\underline
m=0,1,\cdots,\, 9)$ and the fermionic coordinates
$\theta^{1\mu},\,\theta^{2\nu}$
$(\mu,\,\nu=1,\cdots,\,16)$\footnote{We use the conventions of the
superembedding approach proposed in \cite{Delduc:1992fk} such that
all underlined bosonic indices correspond to the target superspace
and the not underlined ones from different parts of the Latin
alphabet correspond to both the D--brane worldvolume and to
directions orthogonal to it.}. The D--brane interacts with fields of
the $D=10$ supergravity multiplet. The type $IIB$ supergravity
fields are contained in the dilaton superfield $\Phi(Z)$, the
Neveu--Schwarz--Neveu--Schwarz two--form superfield $B_2(Z)$, the
Ramond--Ramond superforms $C_{2n}(Z)$ $(n=0,1,2)$ and their duals
$C_{2n}(Z)$ $(n=4,3,2)$, and in the supervielbein
\begin{equation}\label{sv}
E^{A}\,(Z)=dZ^{M}\,E^{A}=(E^{\underline
a},\,E^{1\alpha},\,E^{2\beta})\,,
\end{equation}
where $\underline a=0,1,\cdots,\,9$ are $D=10$ vector tangent space
indices and $\alpha,\,\beta=1,\cdots,\,16$ are the Majorana--Weyl
spinor indices, $E^{1\alpha}$ and $E^{2\beta}$ being the two
Majorana--Weyl spinors of the same chirality.

The above listed superfields are subject to a certain set of
constraints (which are required to eliminate redundant and, in
particular, higher spin degrees of freedom) \cite{Howe+West=1983}.


\subsection{Type IIB supergravity constraints in the Einstein frame}
We shall deal with the following set of the type IIB supergravity
constraints in the Einstein frame. Below and in all places where it
will not cause confusion we shall skip the wedge product symbol. The
superspace torsion constraint is
\begin{eqnarray}\label{Ta=IIB}
{T}^{\underline{a}} =  {D}{E}^{\underline{a}} =
d{E}^{\underline{a}}-{E}^{\underline{b}}\,\Omega_{\underline
b}{}^{\underline a}\hspace{250pt}\nonumber\\
\\
= - i {E}^{{\alpha}1}\, {E}^{{\beta}1}
\gamma^{\underline{a}}_{\alpha\beta} - i {E}^{{\alpha}2}\,
{E}^{{\beta}2} \gamma^{\underline{a}}_{\alpha\beta} =- i {E}^{1}
\gamma^{\underline{a}} {E}^{1} - i {E}^{2}
\gamma^{\underline{a}} {E}^{2}  \nonumber
\end{eqnarray}
(where $\Omega_{\underline b}{}^{\underline a}$ is the $SO(1,9)$
spin connection) and the NS--NS field strength constraint has the
form
\begin{eqnarray}\label{H3=IIB}
H_{3} &=& dB_2 = - i e^{{1\over 2}{\Phi}} {E}^{\underline{a}}\,
({E}^{1}\,
\gamma_{\underline{a}} \,{E}^{1} -  {E}^{2}\,\gamma_{\underline{a}} \,{E}^{2})  -
\qquad \nonumber
\\ && - {1\over 2} e^{{1\over 2}{\Phi}} ({E}^{1}\, {\gamma}^{(2)}\,
{\nabla}_{1}\Phi - {E}^{2}\, {\gamma}^{(2)}\, {\nabla}_{2}\Phi)
+{1\over 3!} {E}^{\underline{c}_3}\,{E}^{\underline{c}_2}\,
{E}^{\underline{c}_1}
H_{\underline{c}_1\underline{c}_2\underline{c}_3}\; , \qquad
\end{eqnarray}
where the general structure of the first term is determined by the
superstring Wess--Zumino term in flat superspace, the dilaton
contribution $e^{{1\over 2}{\Phi}}$ can be recovered considering the
superstring in curved type IIB superspace, while the second term can
be then derived by studying Bianchi identities.

In eqs. (\ref{Ta=IIB}) and (\ref{H3=IIB})
$\nabla_{1,2\,\alpha}=E_{1,2\,\alpha}^{\underline M}\partial_{M}\,$,
$\gamma^{(2)}={1\over 2}\,E^{\underline b}\wedge E^{\underline
a}\,\gamma_{\underline a}\,\tilde\gamma_{\underline b} $ and
 $\gamma^{\underline a}_{\alpha\beta}$
and $\tilde\gamma^{\underline a\,\alpha\beta}$ are $16\times 16$
(symmetric) matrix counterparts of the Pauli matrices satisfying the
relations
\begin{equation}\label{gammas}
\gamma^{\underline
a}\,\tilde\gamma^{\underline b}+\gamma^{\underline
b}\,\tilde\gamma^{\underline a}=2\,\eta^{{\underline a}{\underline
b}}
\,,\qquad  \gamma^{[\underline a_1}\,\tilde\gamma^{\underline
a_2}\cdots \gamma^{\underline
a_5]}={1\over{5!}}\,\epsilon^{\underline a_1\cdots \underline
a_5\,\underline a_6\cdots \underline a_{10}}\,\gamma_{\underline
a_6}\,\tilde\gamma_{\underline a_7}\cdots \gamma_{\underline
a_{10}}\,.
\end{equation}
We use the mostly minus convention for the Minkowski metric
$\eta^{{\underline a}{\underline b}}=diag\,(+,-\cdots,\,-)$,
$\epsilon^{0123456789}=1$ and $\gamma^0\tilde\gamma^1\cdots
\tilde\gamma^9=-\tilde\gamma^0\gamma^1\cdots \gamma^9=1$. Note that
there is no ``charge conjugation" matrix in $D=10$ which would lower
or rise 16--component Majorana--Weyl spinor indices $\alpha$,
$\beta$. This is why one should distinguish between
$\gamma^{\underline a}_{\alpha\beta}$ and $\tilde\gamma^{\underline
a\,\alpha\beta}$.


The constraints on the field strengths of the RR superforms $C_{2n}$
($n=1,\cdots,5)$ are (see
\cite{c2})
\begin{eqnarray}\label{R=IIB}
 R_{2n+1} = &- 2i\, e^{{n-2\over 2}{\Phi}}  E^{2}\,{\gamma}^{(2n-1)}\,  E^{
 1}\,
  +
 {n-2\over 2}\,e^{{n-2\over 2}{\Phi}} \left(E^2  \,{\gamma}^{(2n)}\,\nabla_1\Phi
- (-)^n E^1 {\gamma}^{(2n)}\nabla_2\Phi\right) +
\nonumber \\
\\
& + {1 \over (2n+1)!} E^{\underline{a}_{2n+1}}
\cdots \, E^{\underline{a}_1} R_{\underline{a}_1\ldots
\;\underline{a}_{2n+1}}\;,\hspace{200pt}\nonumber
\end{eqnarray}
where
\begin{eqnarray}\label{g2n}
 \gamma^{(2n)}={1\over {2n!}}\,E^{\underline
 a_{2n}}\cdots E^{\underline a_{1}}\,\gamma_{\underline a_1}\tilde\gamma_{\underline
a_2}\,\gamma_{\underline a_{3}}\cdots\tilde\gamma_{\underline
a_{2n}}\,,\nonumber\\
\\
 \gamma^{(2n-1)}={1\over {(2n-1)!}}\,E^{\underline
a_{2n-1}}\cdots E^{\underline a_{1}}\,\gamma_{\underline
a_1}\tilde\gamma_{\underline a_2}\,\gamma_{\underline
a_{3}}\cdots\gamma_{\underline a_{2n-1}}\,.\nonumber
\end{eqnarray}

Note that RR 10--form $C_{10}$ does not carry any independent
degrees of freedom. Its 11--form field strength is non--trivial only
because of the presence of the Grassmann--odd directions in type IIB
D=10 superspace, while the purely bosonic part of its field strength
is identically zero.

\subsection{Type IIB supergravity constraints in the string
frame}\label{stringf}
 Let us now present how the type IIB
supergravity constraints look like in the string frame. This frame
is related to the Einstein frame by the following (conformal)
redefinition of the supervielbeins and the spin connection
\begin{eqnarray}\label{einstring}
&E_{str}^{\underline a}=e^{{1\over 4}\,\Phi}\,E^{\underline
a}\,,\qquad\qquad E_{str}^{I\alpha}=e^{{1\over 8}\,\Phi}\,(E^{I
\alpha}-{i\over 8}\,E^{\underline a}\,\gamma_{\underline
a}^{\alpha\beta}\,\nabla_{I\alpha}\,\Phi)\qquad (I=1,2)
\nonumber\\
&\nonumber\\
&\Omega^{\underline a\,\underline b}_{str}=\Omega^{\underline
a\,\underline b}+{1\over 2}\,E^{[\underline a}\,D^{\underline
b]}\,\Phi-{1\over 4}\,E^{I}\gamma^{\underline a\,\underline
b}\,\nabla_{I}\Phi+{i\over {64}}\,\nabla_{I}\Phi\,\gamma^{\underline
a\,\underline b\,\underline c}\,\nabla_{I}\Phi\,E_{\underline
c}\,,
\nonumber\\
&\\\
&{1\over{3!}}\,\left({E}^{\underline{c}_3}\,{E}^{\underline{c}_2}\,
{E}^{\underline{c}_1}\,H_{\underline{c}_1\underline{c}_2\underline{c}_3}\right)_{str}=
{1\over{3!}}\,{E}^{\underline{c}_3}\,{E}^{\underline{c}_2}\,
{E}^{\underline{c}_1}
\,H_{\underline{c}_1\underline{c}_2\underline{c}_3} -i\,{{
3!}\over{64}}\, e^{{\Phi}\over 2}\,(\nabla_{1}\Phi\,
{\gamma}^{(3)}\,
 {\nabla}_{1}\Phi
- {\nabla}_{2}\Phi\, {\gamma}^{(3)}\, {\nabla}_{2}\Phi)
\;,
\nonumber\\
&\nonumber\\
&{1 \over (2n+1)!}\,\left( E^{\underline{a}_{2n+1}}
\cdots \, E^{\underline{a}_1} R_{\underline{a}_1\ldots
\;\underline{a}_{2n+1}}\right)_{str}={1 \over (2n+1)!}\, E^{\underline{a}_{2n+1}}
\cdots \, E^{\underline{a}_1} R_{\underline{a}_1\ldots
\;\underline{a}_{2n+1}} \hspace{50pt}
\nonumber\\
&\nonumber\\
&\hspace{200pt}-i\,{{(n-4)\,(2n+1)}\over
{16}}\,e^{{(n-2)\,\Phi}\over 2}\,
\nabla_2\Phi
\,{\gamma}^{(2n+1)}\,\nabla_1\Phi\nonumber
\end{eqnarray}

With such a redefinition the torsion  constraint (\ref{Ta=IIB}) does
not change its form, while the constraints on the NS--NS field
strength (\ref{H3=IIB}) and on the RR field strengths (\ref{R=IIB})
take, respectively, the following form (for simplicity we skip the
subscript {\it string} in the notation of the string frame
supervielbeins):
\begin{eqnarray}\label{H3=IIBstring}
H_{3} = dB_2 = - i  {E}^{\underline{a}}\,({E}^{1}\,
\gamma_{\underline{a}}\, {E}^{1} -  {E}^{2}\, \gamma_{\underline{a}}\, {E}^{2}) +{1\over 3!}
{E}^{\underline{c}_3}\,{E}^{\underline{c}_2}\, {E}^{\underline{c}_1}
H_{\underline{c}_1\underline{c}_2\underline{c}_3}\; , \hspace{80pt}
\end{eqnarray}

\begin{eqnarray}\label{R=IIBstring}
 &R_{2n+1} = -2i\, e^{{- }{\Phi}}  E^{2}\,{\gamma}^{(2n-1)}\,  E^{
 1}\,
 -\,e^{{- }{\Phi}} \left(E^2  \,{\gamma}^{(2n)}\,\nabla_1\Phi
- (-)^n E^1 {\gamma}^{(2n)}\nabla_2\Phi\right) +\hspace{50pt}
\nonumber \\
&\\
&+ {1 \over (2n+1)!} E^{\underline{a}_{2n+1}}
\cdots \, E^{\underline{a}_1} R_{\underline{a}_1\ldots
\;\underline{a}_{2n+1}}\;,\hspace{150pt}
\nonumber
\end{eqnarray}

The superspace constraints imply the self--duality of the five--form
field strength and the duality relations between the lower and the
higher form field strengths
\begin{equation}
\label{R5=*R5}   R_{\underline a_1 \cdots \underline a_5} =
{1\over 5!} \epsilon_{ \underline a_1 \cdots \underline a_5
\underline b_1
\cdots \underline b_5} R^{\underline b_1
\cdots \underline b_5}\,,
\end{equation}
\begin{equation}
\label{Rq=*RD-q}  R_{\underline{a}_1 \cdots \underline{a}_{9-2n}} =
- (-)^{^{{n(n-1)\over 2}}} \; {1 \over (2n+1)!}\; \epsilon_{\;
\underline{a}_1 \cdots \underline{a}_{9-2n}\underline{b}_1 \cdots
\underline{b}_{2n+1}} R^{\underline{b}_1 \cdots
\underline{b}_{2n+1}}\, \quad (n=0,1,\cdots,\,4)\,.
\end{equation}
The proper action for the physical fields of type IIB supergravity
producing (self)--duality relations (\ref{R5=*R5}) and
(\ref{Rq=*RD-q}) as equations of motion was constructed in
\cite{Dall'Agata:1997ju}.

\subsection{The Dp--brane action}\label{framelike}

The super--Dp--brane action is constructed using the pullbacks to
the brane worldvolume of the dilaton superfield, the NS--NS
two--form, the RR $2n$--forms and the supervielbein vector
components (\ref{sv}). We shall denote the pullbacks of forms to the
worldvolume by the same letters as the forms themselves. This should
not produce any confusion since in this paper we mainly deal with
the pullbacks.

We shall consider the D--brane action in a form proposed in
\cite{Bandos:1997rq} which is equivalent to the original action
of \cite{Dp}--\cite{bt} but at the same time can be directly
uplifted to the superembedding approach via the generalized action
principle and similar techniques
\cite{bsv,howe,Bandos:1997rq}. We find this form of the action more
convenient for the derivation and the analysis of the D--brane
equations of motion.

In such a formulation the D--brane Lagrangian density is constructed
as a worldvolume differential $(p+1)$--form. In particular, we
should determine what is the worldvolume area density. To this end
we have to define a metric or a vielbein in the worldvolume induced
by its embedding into the target superspace. This can be done in the
following way. Let us take the pullback of the vector part of the
supervielbein (\ref{sv})
\begin{equation}\label{psv}
E^{\,\underline a}\,(Z(\xi))=d\xi^m\,\partial_m Z^{\underline
M}\,E_{M}^{\,\underline a}\,.
\end{equation}
 On the
$(p+1)$--dimensional non--degenerate surface there exist $p+1$
linearly independent vectors. So, using an appropriate $SO(1,9)$
transformation $u_{\underline b}^{~\underline a'}(\xi)$ in the
ten--dimensional tangent space with local parameters on the
worldvolume we can always transform the vielbein (\ref{psv}) in such
a way that $(p+1)$ of its components will be parallel to the
worldvolume and $(9-p)$ components will be orthogonal to the
worldvolume
\begin{equation}\label{rpsv}
E^{\underline a}\,(Z(\xi)) \quad \Rightarrow \quad E^{\underline
b}\,u_{\underline b}^{~\underline a'}(\xi):=(E^a,\,E^i) \quad
a'=(a,i)\,,\quad a=0,1,\cdots, p\,,\quad i=1,\cdots, 9-p
\end{equation}
such that
\begin{eqnarray}\label{hEi=0}  {E}^i = {E}^{\underline{a}}
u_{\underline{a}} ^{\; i} =0\; .
\end{eqnarray}
 Eq. (\ref{hEi=0}) is called the {\it embedding condition}
  (see {\it e.g.} \cite{se,bsv,Bandos:1997rq}).
It simply implies that the pullback onto the worldvolume of a vector
orthogonal to it is zero.

 In eqs. (\ref{rpsv}) and (\ref{hEi=0})
the ten--dimensional vector index $\underline a$ is split into the
$(p+1)$--dimensional worldvolume tangent space index $a$ and the
index $i$ labeling the $9-p$ directions orthogonal to the brane.
This splitting reflects the fact that the presence of the brane
spontaneously breaks the target space $SO(1,9)$ Lorentz symmetry
down to its subgroup $SO(1,p)\times SO(9-p)$. The components of the
local Lorentz matrix
\begin{equation}\label{lh}
u_{\underline b}^{~\underline a'}(\xi)=(u_{\underline b}^{~
a}(\xi)\,,u_{\underline b}^{~ i}(\xi))\,,
\end{equation}
satisfying the orthogonality conditions
\begin{eqnarray}\label{oc}
{u}^{\underline{c}\, a} u_{\underline{c}} ^{\; b} =\eta^{ab}=diag(+1
,\underbrace{- 1, \cdots , -1}_{p}\;) \; ,\quad {u}^{\underline{a}
\,a} u_{\underline{a}} ^{\; i} =-0\; ,
\quad {u}^{\underline{a}\, i} u_{\underline{a}} ^{\; j}
=-\delta^{ij}\,, &\nonumber\\
&\\
 u_{\underline{a}} ^{\; a} u_{\underline{b}} ^{\;
b}\eta_{ab}- u_{\underline{a}} ^{\; i}u_{\underline{b}} ^{\;
j}\delta_{ij}=\eta_{\underline{ab}}\,,  \hspace{100pt} &\nonumber
\end{eqnarray}
play the role of the Goldstone fields associated with this symmetry
breaking. They parameterize the coset space
${SO(1,9)}\over{SO(1,p)\times SO(9-p)}$ and are called {\it Lorentz
harmonics}. $u_{\underline b}^{~ a}(\xi)$ and  $u_{\underline b}^{~
i}(\xi)$ will be auxiliary fields of our construction (see
\cite{BZstr,BZ-p} for more details about Lorentz harmonics
techniques). The variation of $u_{\underline b}^{~ a}(\xi)$ in the
D--brane action will produce the embedding condition (\ref{hEi=0}).

Using the splitting (\ref{rpsv}) of the supervielbein components, we
can associate the pullback $E^a(Z(\xi))$ with the induced vielbein
of the worldvolume. Such a choice is in accordance with the
conventional definition of the induced metric on the brane
worldvolume
\begin{equation}\label{im}
g_{mn}=\partial_m\,Z^{M}\,E_{M}^{~\underline
a}\,\,\,\partial_n\,Z^{N}\,E_{N}^{~\underline
b}\,\eta_{\underline{ab}}\,\, \qquad (m,n=0,1,\cdots p)\,.
\end{equation}
It is easy to see, using the embedding equations
(\ref{psv})--(\ref{oc}), that the metric (\ref{im}) is expressed as
the product of the components of the induced worldvolume vielbein
$E^a$
\begin{equation}\label{im1}
g_{mn}=E_{m}^{~ a}\,E_{n}^{~b}\,\eta_{{ab}}\,, \qquad
E^a=d\xi^m\,E_{m}^{~ a}=\partial_m\,Z^{M}\,E_{M}^{~
\underline b}\,u_{\underline b}^{~ a}\,.
\end{equation}
In terms of the vielbein $E^a$ and the metric (\ref{im}),
(\ref{im1}) the worldvolume area density is
\cite{BZstr,BZ-p,se,bsv,Sorokin:1999jx}
\begin{equation}\label{vd}
{1\over {(p+1)!}}\,\epsilon_{a_1\cdots a_{p+1}}\,E^{a_1} \cdots
 E^{a_{p+1}}=d^{p+1}\,\xi\,\sqrt{|det\,g_{mn}|}\,.
\end{equation}

Let us now show following \cite{BZstr,se} that when the embedding
condition (\ref{hEi=0}) is satisfied
 the Lorentz harmonics can be expressed in terms of
components of the pullback of $E^{\underline a}$ and hence are
indeed auxiliary fields. As in the case of the frame formulation of
gravity, using local $SO(1,p)$ transformations in the worldvolume
tangent space (acting on the indices $a,b$) and the definition
(\ref{im}), (\ref{im1}) of the induced metric we can reduce
$E_{m}^{~ a}\eta_{ab}$ to a symmetric matrix. The components of this
matrix are completely determined by the components of the induced
metric (\ref{im}) and hence do not depend on the Lorentz harmonics.
Now notice that from eq. (\ref{hEi=0}) it follows that
\begin{equation}\label{im2}
E^{\underline a}=d\xi^m\,\partial_m\,Z^{M}\,E_{M}^{\;\underline a}
=d\xi^m\,E_m^{\;a}\,u_a^{\;\underline a}\quad \Rightarrow \quad
u_a^{\;\underline a}=E^{\;m}_a\,\partial_m\,Z^{\underline
M}\,E_{M}^{\;\underline a}\,,
\end{equation}
where $u_a^{\;\underline a}=\eta^{ab}\,u_{\underline b}^{~
b}\,\eta_{\underline{ab}}$ and $E^{\;m}_a$ is inverse of the induced
worldvolume vielbein $E_m^{\;a}$ associated with the induced metric
$g_{mn}$ in (\ref{im1}).

Therefore, the components of $u_{\underline b}^{~ a}$ are expressed
solely through components of the supervielbein pullback
$E_m^{\;\underline a}=\partial_m\,Z^{M}\,E_{ M}^{\;\underline a}$.
For instance, in the case of the bosonic brane in a flat background
one finds that $u_a^{\;\underline a}$ reduces to $u_a^{\;\underline
a}=E_a^m\,\partial_m\,x^{\underline a}$.

Finally, from the orthogonality conditions (\ref{oc}) it follows
that, up to local $SO(9-p)$  rotations in the directions orthogonal
to the brane, the components of $u_{\underline b}^{~ i}$ are
determined by the components of $u_{\underline b}^{~ a}$. This
completes the proof that the Lorentz harmonics are auxiliary
variables. Their use in the D--brane action simplifies the
derivation and the analysis of its equations of motion.

Using the induced  worldvolume vielbein $E^a=E^{\underline
a}\,u_{\underline a}^{\;a}$ and the pullbacks of the type IIB
supergravity fields {\bf in the Einstein frame} we construct the
Dp--brane action in the following form \cite{Bandos:1997rq}
\begin{eqnarray}\label{SDp-br}
S_{Dp} &=& \int_{{\cal{W}}^{p+1}} {\cal{L}} =
 \int_{{\cal{W}}^{p+1}} ({\cal{L}}^{DBI}_{p+1} +  {\cal{L}}^{LM}_{p+1} +
 {\cal{L}}^{WZ}_{p+1})= \nonumber
 \\
 \\
 &\!\!\!\!\!\!\!\! \!\!\!\!\!\!\!\!=&\!\!\!\!\!\!\!\! \!\!\!\!\!\!\!\!
 \int\limits_{{\cal{W}}^{p+1}}\,{{e^{{p-3\over 4} {\Phi}}}\over {(p+1)!}}
 \,\epsilon_{a_1\cdots a_{p+1}}\,E^{a_1} \cdots
  E^{a_{p+1}}\,
{\sqrt{|det\,(\eta_{ab}+F_{ab})|}} + Q_{p-1}  (e^{-{1\over 2}
{\Phi}} {\cal F}_2
 - F_{2}) + e^{{\cal F}_2} \wedge  {\mathbb{C}} \vert_{p+1} \; ,
 \nonumber
\end{eqnarray}
where in ${\cal{L}}^{WZ}_{p+1} = e^{{\cal F}_2} \wedge {\mathbb{C}}
\vert_{p+1}$, which is the conventional Wess--Zumino (WZ) term
\cite{Dp}, \cite{DbrWZ,dea},
 \hbox{ $\vert_{p+1}$ means} that we pick only the terms which are the $(p+1)$--forms
in the external product of the {\it formal} sums of the forms of
different order
$$
e^{{\cal F}_2}=1+ {\cal F}_2+ {1\over 2}\,{\cal F}_2\,{\cal F}_2 +
{1\over {3!}}\,{\cal F}_2\,{\cal F}_2 \,{\cal F}_2+{1\over{4!}
}\,{\cal F}_2\,{\cal F}_2 \,{\cal F}_2\,{\cal F}_2+{1\over
{5!}}\,{\cal F}_2\,{\cal F}_2 \,{\cal F}_2 \,{\cal F}_2\,{\cal
F}_2=\sum_{k=0}^{5}\,{1\over {n!}}\,({\mathcal F}_2)^n\,,
$$
\begin{equation}\label{C}
\mathbb{C}= C_0 + C_2 + C_4 + C_6 +
C_8 + C_{10}\,=\sum_{n=0}^{5}\,C_{2n}.
\end{equation}
The structure of the Wess--Zumino term tells us, in particular, that
the Dp--brane minimally couples to the RR field potential $C_{p+1}$
and also couples to lower order RR forms. Remember also that the
bosonic parts of the RR field strengths $R_{2k+1}$ and $R_{9-2k}$
$(k=0,1)$ are dual to each other and $R_5$ is self--dual
(\ref{Rq=*RD-q}) \footnote{Actually, the form of the action
(\ref{SDp-br}) is generic and also describes the Dp--branes in type
IIA supergravity. In that case one should consider the Dp--branes
with even values of $p$ coupled to ($2k+1$)--form RR fields.}.

The first term in (\ref{SDp-br}) is the Dirac--Born--Infeld--like
part of the D--brane action, where $F_{ab}(\xi)$ is an auxiliary
antisymmetric tensor field taking values in the worldvolume tangent
space. This field is related to the extended field strength of the
Born--Infeld (BI) gauge field $A=d\xi^m\,A_m(\xi)$
\begin{equation}\label{calF}
 {\cal F}_2  =  dA- {B}_2
\end{equation}
via the second term of the action (\ref{SDp-br})
\begin{eqnarray}\label{LLMp}
{\cal{L}}^{LM}_{p+1} &=&  Q_{p-1} \, \left(e^{-{1\over 2}
{\Phi}}{\cal F}_2
 - F_{2}\right)= Q_{p-1} \, \left(e^{-{1\over 2} {\Phi}}(dA-{B}_2) - {1
\over 2} {E}^{a} {E}^{b} F_{ba}\right)\;,
\end{eqnarray}
where $Q_{p-1}(\xi)$ is a ($p-1$)--form Lagrange multiplier.

Indeed, varying (\ref{LLMp}) with respect to $Q_{p-1}(\xi)$ we get
the algebraic relation
\begin{equation}\label{deltaQ}
{\delta {\cal{L}}^{LM}_{p+1} \over \delta Q_{p-1}} =0 \;\quad
\Rightarrow  \quad    F_2 = e^{-{1\over 2}{\Phi}} {\cal F}_2 \quad i.e. \quad {1
\over 2} {E}^{a} {E}^{b} F_{ba}= e^{-{1\over 2}{\Phi}} (dA -
 {B}_2)\; .
\end{equation}
Then substituting eq. (\ref{deltaQ}) into the first term of
(\ref{SDp-br}) and using the definition of the induced metric
(\ref{im}), (\ref{im1}) one can reduce the frame--like Dp--brane
action to the conventional form \cite{Dp}--\cite{bt} consisting of
the Dirac--Born--Infeld action and the WZ term
\begin{equation}\label{DBI}
{{S}}=
\int\,d^{p+1}\xi\,e^{{p-3\over 4}
{\Phi}}\sqrt{|g_{mn}+e^{-{1\over 2} {\Phi}}{\cal F}_{mn}|}+
\int\, e^{{\cal F}_2} \wedge  {\mathbb{C}} \vert_{p+1} \,,
\end{equation}
or {\bf in the string frame} (\ref{einstring})
\begin{equation}\label{DBIstring}
{{S}}_{str}=
\int\,d^{p+1}\xi\,e^{{ - }
{\Phi}}\sqrt{|g_{mn}+{\cal F}_{mn}|}+\int\, e^{{\cal F}_2} \wedge
{\mathbb{C}} \vert_{p+1} \,.
\end{equation}
Note that in the string frame relation (\ref{deltaQ}) takes the form
\begin{equation}\label{deltaQstring}
F_2 = {\cal F}_2 \quad ( = dA -
 {B}_2\;) .
 \end{equation}
 A reason why we introduced the auxiliary field
$F_{ab}(\xi)$ is that it is in some sense frame independent, that
simplifies a bit the derivation of the field equations and their
analysis.

By construction, the action (\ref{SDp-br}) is invariant under target
space superdiffeomorphisms and the local $SO(1,9)$ symmetry acting
on the background superfields, under the worldvolume diffeomorphisms
and under local worldvolume $SO(1,p-1)$ transformations acting on
the fields carrying worldvolume tangent space indices $a,b,\cdots$,
such as $E^a$ and $F_{ab}$. It is also invariant under local
worldvolume fermionic $\kappa$--symmetry whose form and properties
will be discussed in detail in Section
\ref{theta}.

Having constructed the D--brane action and described its properties
we now turn to the derivation and the consideration of the D--brane
equations of motion in an arbitrary type IIB supergravity
background.

\setcounter{equation}0
\section{Bosonic equations}

\bigskip

\subsection{Equations for the auxiliary fields and the embedding
condition}\label{aux}

 As we have already discussed, the variation of
the action (\ref{SDp-br}) with respect to the Lagrange multiplier
$Q_{p-1}$ expresses the auxiliary field $F_{ab}$ in terms of the BI
field strength (\ref{deltaQ})
\begin{equation}\label{deltaQ1}
 F_2 = e^{-{1\over 2}{\Phi}} {\cal F}_2 = e^{-{1\over 2}{\Phi}} (dA -
 {B}_2)\; .
\end{equation}

The variation with respect to $F_{ab}$ gives the expression for the
Lagrange multiplier in terms of other fields
\begin{eqnarray}\label{Q(p-1)=}
Q_{p-1} =  {e^{{p-3\over 4} {\Phi}}\over
{2\cdot(p-1)!}}\,\epsilon_{ab
\; c_1\cdots \; c_{p-1}}
 {E}^{c_1}  \cdots  {E}^{c_{p-1}}\; \, (\eta +
F)^{-1\, ab}
\sqrt{|\eta + F|} \; ,
\end{eqnarray}
where $(\eta + F)^{-1\, ab}$ is the inverse matrix  of $ \eta_{ab}+
F_{ab}$ and $|\eta + F|=|det\,(\eta_{ab}+ F_{ab})|$. To arrive at
eq. (\ref{Q(p-1)=}) we used the identity
$$
\epsilon_{ab
\; c_1\cdots \; c_{p-1}}
 {E}^{c_1}  \cdots  {E}^{c_{p-1}}\,
 {E}^c\,{E}^d = {2\over{p\,(p+1)}} \,\epsilon_{a_1\cdots a_{p+1}}\,E^{a_1} \cdots
 E^{a_{p+1}}\,\delta_{[a}{}^c\delta_{b]}{}^d \; .
$$
To perform the variation of the action (\ref{SDp-br}) with respect
to the auxiliary Lorentz harmonic variables $u_{\underline b}^{~a}$
contained in $E^a=E^{\underline b}\,u_{\underline b}^{~a}$ one
should take into account that they are constrained to be $SO(1,9)$
matrices (\ref{oc}). As discussed in detail {\it e.g.} in
\cite{BZstr,BZ-p,se,Bandos:1997rq} the independent variations of $u_{\underline
b}^{~a}$  subject to the orthogonality conditions (\ref{oc}) are
those projected along the directions orthogonal to the worldvolume,
i.e. $u_{\underline{b}}^{\; i}\, {\delta
\over \delta u_{\underline{b}}{}^a}$. With this in mind one gets the
following equations of motion of $u_{\underline b}^{~a}(\xi)$
\begin{equation}\label{Eqm-harm}
 u_{\underline{b}}^{\; i} {\delta S \over \delta u_{\underline{b}}{}^a} = 0
 \quad
 \Longrightarrow   \quad  {E}^i \, \wedge\left(
 {{e^{{p-3\over 4}  {\Phi}}}\over{p!}}\,\epsilon_{a
\; b_1\cdots \; b_{p}}
 {E}^{b_1}  \cdots  {E}^{b_{p}}  \,\sqrt{|\eta + F|}+Q_{p-1}
F_{ab}\,E^b  \right)  =0 \; ,
\end{equation}
where $Q_{p-1}$ is defined in (\ref{Q(p-1)=}). Substituting
(\ref{Q(p-1)=}) into (\ref{Eqm-harm}), and defining the worldvolume
form $E^i$ as $E^i=E^a\,E_a^{\,\,i}$ we get
\begin{equation}
E_b^{\,\,i}(\delta-FF)_a^{-1\,b} \,\sqrt{|\eta +
F|}\,(det\,{E_m^{\,\,a}})\, d^{p+1}\xi=0\,.
\end{equation}

Assuming that the brane worldvolume metric is non degenerate, i.e.
that $det\,{E_m^{\,\,a}}$ is non--zero, and that the matrix
$(\delta-FF)_a^{-1\,b}\sqrt{|\eta + F|}$ is also non--degenerate
\footnote{The singularity of $(\delta-FF)_a^{-1\,b}\sqrt{|\eta + F|}\sim (1-(F_{0m})^2)$
 occurs at the critical
value $(F_{0m})^2=1$ of the electric field $F_{0m}$.
 } we find that eq. (\ref{Eqm-harm}) reduces to the embedding
condition (\ref{hEi=0})
\begin{eqnarray}\label{hEi=01}
E_a^{\,\,i}=0\quad \Rightarrow \quad  {E}^i = {E}^{\underline{a}}\,
u_{\underline{a}} ^{\; i} =0\; .
\end{eqnarray}
Thus, the embedding condition (\ref{hEi=0}) naturally arises in our
formulation as the equation of motion (\ref{hEi=01}) of the
Lorentz--harmonic variables and ensures that the latter are purely
auxiliary fields, as demonstrated in the previous Section. The use
of eq. (\ref{hEi=01}) simplifies the derivation of the equations of
motion of the worldvolume fields $Z^{M}(\xi)$ which we shall
consider in Subsection \ref{ZM}.

\subsection{Equations of motion of the worldvolume BI gauge field}

The variation of the Lagrangian (\ref{SDp-br})  with respect to the
BI field $A=d\xi^m A_m(\xi)$ is
\begin{eqnarray}\label{vL(dQ=)}
\delta_{_A} {\cal L}_{p+1} =  \left[ d\left(e^{-{1\over 2}{\Phi}}\,Q_{p-1}\right) +
\left.  {\mathbb{R}} \, e^{(dA- {B}_2)}\right|_{p} \right] \,  \delta A
+ d\, \left[ e^{-{1\over 2} {\Phi}}\,Q_{p-1} \,  \delta A + \left.
 {\mathbb{C}}\,   e^{(dA- {B}_2)}\right|_{p-1}   \delta A  \right]\!,
\end{eqnarray}
where
\begin{equation}\label{R}
\mathbb{R} := \sum _{k=0}^{4} R_{2k+1} =  d \mathbb{C} - \mathbb{C}  \, H_3 \equiv
e^{B_2}   d\, (\mathbb{C}\, e^{-B_2} )\,, \qquad
\end{equation}
the form $\mathbb{C}$ was defined in (\ref{C}) and $|_{k}$ means
that only the $k$--form terms are retained in the product of formal
sums of the pullbacks of the differential forms.

Then, neglecting in (\ref{vL(dQ=)}) the second, total derivative
term ({\it i.e.} possible boundary contributions) and using eqs.
(\ref{deltaQ1}) and (\ref{Q(p-1)=}) one arrives at the BI field
equation in an arbitrary type IIB supergravity background {\bf in
the Einstein frame}
\begin{eqnarray}\label{DBI=d*F}
d\left({1\over {(p-1)!}}\,\epsilon_{ab
\; c_1\cdots \; c_{p-1}}
 {E}^{c_1}  \cdots  {E}^{c_{p-1}}\, (\eta + e^{-{1\over 2} {\Phi}}\,{\mathcal F})^{-1\, ab}\,
\sqrt{|\eta + e^{-{1\over 2}
{\Phi}}\,{\mathcal F}|} \,\,e^{{p-5\over 4} {\Phi}}\right) = -  2
\left.
 {\mathbb{R}} \, e^{{\mathcal F}_2}\right|_{p}\;
\end{eqnarray}
the right hand side of which can further be specified using the
definition of $\mathbb{R} $ (\ref{R}) and the superspace constraints
(\ref{R=IIB}).

In terms of the induced metric (\ref{im1}), the equation
(\ref{DBI=d*F}) takes the form
\begin{eqnarray}\label{DBI=d*F1}
{1\over{p!}} \epsilon_{m_1\cdots m_{p}l}\,d\xi^{m_1}\cdots
d\xi^{m_p}\,\partial_m
\left((g + e^{-{1\over 2} {\Phi}}\,{\mathcal F})^{-1\,[ml]}\,
\sqrt{|g_{ns} + e^{-{1\over 2}
{\Phi}}\,{\mathcal F}_{ns}|} \,\,e^{{p-5\over 4} {\Phi}}\right) =
\left.
 {\mathbb{R}} \, e^{{\mathcal F}_2}\right|_{p}
\end{eqnarray}
or, equivalently,
\begin{eqnarray}\label{DBI=d*F2}
\partial_m \left((g + e^{-{1\over 2} {\Phi}}\,{\mathcal F})^{-1\, [ml]}\,\sqrt{
|g + e^{-{1\over 2} {\Phi}}\,{\mathcal F}|}\,e^{{p-5\over 4} {\Phi}}
\,
 \right) =\hspace{200pt}\nonumber\\
 \\
  \quad  (-)^{^{p-1\over 2}}
 \sum\limits_{n=0}^{{p-1\over 2}} {{1} \over
2^{n} \,n!\,(p-2n)!}\; \epsilon^{l m_1\cdots m_{p-2n}r_1s_1 \cdots
r_{n}s_{n} } \; {R}_{m_1\ldots m_{p-2n}}\; {\mathcal F}_{r_1s_1}
\cdots {\mathcal F}_{r_ns_{n}}\,,  \nonumber
\end{eqnarray}
where ${R}_{m_1\ldots m_{n}}=E^{\,A_{n}}_{m_{1}} \cdots
E^{\,A_1}_{m_n}\,{R}_{A_1\ldots A_{n}}$ are the worldvolume
pullbacks of the RR field strengths (\ref{R=IIB}). Note that eqs.
(\ref{DBI=d*F1}) and (\ref{DBI=d*F2}) do not contain the Lorentz
harmonic variables.

{\bf In the string frame} the equation (\ref{DBI=d*F2}) simplifies a
bit
\begin{eqnarray}\label{DBI=d*F2string}
\partial_m \left((g + {\mathcal F})^{-1\, [ml]}\,\sqrt{
|g + {\mathcal F}|}\,e^{{-} {\Phi}}
\,
 \right) =\hspace{200pt}\nonumber\\
 \\
  \quad  (-)^{^{p-1\over 2}}
 \sum\limits_{n=0}^{{p-1\over 2}} {{1} \over
2^{n} \,n!\,(p-2n)!}\; \epsilon^{l m_1\cdots m_{p-2n}r_1s_1 \cdots
r_{n}s_{n} } \; {R}_{m_1\ldots m_{p-2n}}\; {\mathcal F}_{r_1s_1}
\cdots {\mathcal F}_{r_ns_{n}}\,,  \nonumber
\end{eqnarray}

The equations (\ref{DBI=d*F}) - (\ref{DBI=d*F2string}), which are
the differential equations for the BI field strength ${\mathcal
F}_2$, should be accompanied by the Bianchi identities which follow
from the definition (\ref{calF}) of ${\mathcal F}_2$
\begin{equation}\label{bianchi}
d{\mathcal F}_2=-H_3|_{p+1}\,.
\end{equation}

{}From the form of (\ref{bianchi}) we conclude that the worldvolume
flux ${\mathcal F}_2$ cannot be zero unless the pullback of the
NS--NS flux $H_{m_1 m_2m_3}$ is zero, while eqs. (\ref{DBI=d*F}),
(\ref{DBI=d*F1}) and (\ref{DBI=d*F2}) imply that ${\mathcal F}_2$
also cannot be zero if the pullback of the RR flux $R_{m_1\cdots
\,m_p}$ is non--zero. This should be taken into account in the
analysis of branes in the supergravity backgrounds with fluxes and,
in particular, of brane instantons wrapping compactified
submanifolds. Several examples will be considered in Subsection
\ref{warped}.

\subsection{Equations of motion of the coordinate fields
$Z(\xi)=(x(\xi),\theta(\xi))$. Equation for the bosonic fields
$x(\xi)$} \label{ZM} It is convenient to consider the variation of
the D--brane action (\ref{SDp-br}) with respect to the worldvolume
fields $Z^{M}(\xi)=(x^{\underline
m}(\xi),\theta^{1\mu}(\xi),\theta^{2\nu}(\xi))$ and the
corresponding field equations by projecting them to the tangent
space of the embedding superspace, {\it i.e.}
\begin{equation}\label{dL}
\delta \,Z^{M}\,{{\delta\,{\mathcal L}_{p+1}}\over{\delta\,Z^{
M}}}=\delta \,Z^{M}\,E_{M}^{\,A}\,\,E^{\,N}_{ A}{{\delta\,{\mathcal
L}_{p+1}}\over{\delta\,Z^{N}}}=(\delta
\,Z^{M}\,E_{M}^{\,
A})\,{{\delta\,{\mathcal L}_{p+1}}\over{\delta\,Z^{
N}\,E_{N}^{\,A}}}\,.
\end{equation}
The bosonic field equations are then identified with
\begin{equation}\label{bf}
\delta_{\underline a}\,{\mathcal L}_{p+1}={{\delta\,{\mathcal L}_{p+1}}\over{\delta\,Z^{N}\,E_{
N}^{\,\underline a}}}=0\,
\end{equation}
and the fermionic field equations are
\begin{equation}\label{ff}
\delta_{\hat\alpha}\,{\mathcal L}_{p+1}={{\delta\,{\mathcal L}_{p+1}}\over{\delta\,Z^{N}\,E_{
N}^{\,\hat\alpha}}}=0\,,
\end{equation}
where instead of two 16--component spinor indices $1\alpha $ and $
2\alpha$ we have introduced a single 32--component index
$\hat\alpha$, so that $E^{\hat\alpha}=(E^{1\alpha},E^{2\alpha})$.
 To get the explicit form of the $Z^{M}$ field equations it is
useful to recall that the variation of a generic $n$--{\it
super}\,form $G_n(Z)$ produced by a variation $Z^{M}$ is
\begin{eqnarray}\label{formv}
&\delta\, G_n:=d\,(i_\delta\, G_{n})+i_\delta \,dG_n\hspace{300pt}\nonumber\\
&\\
&={1\over{(n-1)!}}\,d(d Z^{M_n}\cdots \,dZ^{ M_{2}}\, \delta Z^{
M_{1}}\,G_{M_1\cdots\,M_n })+{1\over{n!}}\,d Z^{
M_{n+1}}\cdots\,dZ^{ M_{2}}\,\delta Z^{ M_{1}}\,\partial_{[M_1}G_{
M_2\cdots\,M_{n+1}\} }\,.\nonumber
\end{eqnarray}

However, the D--brane Lagrangian (\ref{SDp-br}) is a $(p+1)$--form
which depends not only on $Z^{M}(\xi)$ and their differentials but
also on the purely worldvolume fields $u_{\underline b}^{\,a}(\xi)$,
$F_{ab}(\xi)$, $Q_{p-1}(\xi)$ and $dA(\xi)$ \footnote{The
Born--Infeld field may become a target space field if one uses
enlarged superspaces
\cite{dea}.}. So to apply eq. (\ref{formv}) to the calculation of the
$Z^M$--variation of the Lagrangian (\ref{dL}) one takes into account
that the {\it formal} external differential $dG_n$ on an auxiliary
$(p+2)$--dimensional surface also contains terms of the form
$$
d Z^{ M_{n+1}}\,dZ^{ M_{n}}\cdots\,dZ^{ M_{1}} dw(\xi)\,
{{\partial\,G_{ M_1\cdots\,M_n M_{n+1}}}\over{\partial w(\xi) }}\,,
$$
where $w(\xi)$ stands for the worldvolume fields which are treated
as independent coordinates.

Thus, up to boundary terms and taking into account eqs.
(\ref{deltaQ1})--(\ref{hEi=01}),  the field equations (\ref{bf}) and
(\ref{ff}) can be given in the following generic form
\begin{eqnarray}\label{bf2}
\delta_{\underline a}\,{\mathcal L}_{p+1}
={1\over{(p+1)!}}\,E^{b_{p+1}}\cdots E^{b_1}\,(d{\mathcal
L}_{p+1})_{ {\underline a}\,b_1\cdots b_{p+1} }+
{1\over{p!}}\,\,E^{\hat\alpha}\,E^{b_{p}}\cdots E^{b_1}\,(d{\mathcal
L}_{p+1})_{{\underline
a}\,\hat\alpha \,b_1\cdots b_{p} }\nonumber\\
\\
+
{1\over{2(p-1)!}}\,E^{\hat\beta}\,E^{\hat\alpha}\,E^{b_{p-1}}\cdots
E^{b_1}\,(d{\mathcal L}_{p+1})_{ {\underline a}\,\hat\alpha\,
\hat\beta\,b_1\cdots b_{p-1} }=0\,,\qquad (b=0,1,\cdots, p)\nonumber
\end{eqnarray}
\begin{equation}\label{ff2}
\delta_{\hat\alpha}\,{\mathcal L}_{p+1}
={1\over{(p+1)!}}\,E^{b_{p+1}}\cdots E^{b_1}\,(d{\mathcal
L}_{p+1})_{ {\hat\alpha }\,b_1\cdots b_{p+1} }+{1\over{p !}}\,
E^{\hat\beta}\,E^{b_{p}}\cdots E^{b_1}\,(d{\mathcal L}_{p+1})_{
{\hat\alpha }\,\hat\beta\,b_1\cdots b_{p} }=0\,,
\end{equation}
where $E^{b}=E^{\underline c}\,u_{\underline c}{}^b$. Note that the
fermionic equation (\ref{ff2}) is linear in the fermionic
supervielbein $E^{\hat\alpha}=(E^{1\alpha},E^{2\beta})$ and the
bosonic equation (\ref{bf2}) is only a second order polynomial in
$E^{\hat\alpha}$. This is due to the structure of the D--brane
action (\ref{SDp-br}) and the form of the supergravity constraints
(\ref{Ta=IIB}), (\ref{H3=IIB}) and (\ref{R=IIB}).

Now let us remember that, since the theory is invariant under the
local worldvolume diffeomorphisms (in particular, $\delta_{diff}\,
Z^{M}=\delta
\xi^m\,\partial_m Z^{M})$,  among
$x^{\underline m}(\xi)$ there are only $9-p$ independent bosonic
degrees of freedom corresponding to the normal fluctuations of the
brane in target space. Therefore, among ten eqs. (\ref{bf}),
(\ref{bf2}) there are only $9-p$ independent bosonic field
equations. To identify them in a Lorentz covariant way it is again
convenient to use the Lorentz harmonics $u_{ \underline
a'}{}^{\underline b}(\xi)=(u_a{}^{\underline b},u_i{}^{\underline a}
)$ (\ref{lh}), (\ref{oc}).

One can notice that if (\ref{bf2}) is multiplied by
$u_a{}^{\underline a}$, the first term vanishes since the indices
$a$ and $b$ take $p+1$ values while  $(d{\mathcal L})_{ab_1\cdots
b_{p+1}}$ is an antisymmetric tensor of rank $p+2$. At the same time
the projection of the second and the third term of (\ref{bf2}) along
$u_a{}^{\underline a}$ are proportional to the Born--Infeld
(\ref{DBI=d*F2}) and fermionic equation (\ref{ff2}). This reflects
the worldvolume diffeomorphism invariance of the D--brane theory.

Thus, the $9-p$ independent bosonic equations are those obtained
from eq. (\ref{bf2}) multiplying it by $u_i{}^{\underline a}$
$(i=1,\cdots,9-p)$, {\it i.e.} projecting (\ref{bf2}) along the
directions orthogonal to the brane worldvolume
\begin{eqnarray}\label{bf3}
u_i{}^{\underline a}\,\Big[{1\over{(p+1)!}}\,E^{b_{p+1}}\cdots
E^{b_1}\,(d{\mathcal L}_{p+1})_{ {\underline a}\,b_1\cdots b_{p+1}
}+{1\over{p !}}\,
\,E^{\hat\alpha}\,E^{b_{p}}\cdots E^{b_1}\,(d{\mathcal
L}_{p+1})_{{\underline
a}\,\hat\alpha \,b_1\cdots b_{p} }\nonumber\\
\\
-{1\over{2\,(p-1)!}}\,
E^{\hat\beta}\,E^{\hat\alpha}\,E^{b_{p-1}}\cdots
E^{b_1}\,(d{\mathcal L}_{p+1})_{ {\underline a}\,\hat\alpha\,
\hat\beta\,b_1\cdots b_{p-1} }\Big]=0\,.\nonumber
\end{eqnarray}

{\bf In the Einstein frame}, in the conventional (induced metric)
form, the D--brane bosonic equation in an arbitrary type IIB
supergravity background look as follows
\begin{eqnarray}\label{cbf}
D_m\,\Big(e^{{p-3\over 4}\Phi} \,E_{n\underline
a}\,(g+F)^{-1(mn)}\sqrt{|g+F|}\Big)-{1\over 2}\,e^{{p-5\over
4}\Phi}\,H_{\underline
amn}\,(g+F)^{-1mn}\sqrt{|g+F|}-
\nonumber\\
\\
-e^{{p-3\over 4}\Phi}\,E^{\,M}_{\underline
a}\partial_M\,\Phi\Big({p-3\over 4}-{1\over
4}\,F_{nm}\,(g+F)^{-1mn}\Big)\,\sqrt{|g+F|}\nonumber\\
\nonumber\\
={{(-)^{{p+1}\over 2}}\over{(p+1)!}}\,\epsilon^{m_1\cdots m_{p+1}}\,
({\mathbb{R}} \, e^{{\mathcal F}_2})_{\underline a\,m_1\cdots
m_{p+1}}\,,\nonumber
\end{eqnarray}
where $d\xi^m\,D_m$ is the pullback of the target superspace
covariant derivative $D_{M}=\partial_M-\Omega_{M}$, $H_{\underline
amn}=E_m^{\,B}\,E_n^{\,A}\,H_{\underline a\,AB}$ is the pullback of
the NS--NS superform (\ref{H3=IIB}) and $F_2=e^{-{1\over
2}\Phi}\,{\mathcal F}_2$.

{\bf In the string frame}, in which the supervielbein is
$E_{str}^{\underline a}=e^{{1\over 4}\,\Phi}\,E^{\underline a}$ and
$F_2={\mathcal F}_2=dA-B_2$, the bosonic equation has even a bit
more compact form (in which we again skip the subscript {\it
string})
\begin{eqnarray}\label{cbfstring}
&D_m\,\Big(e^{-\Phi} \,E_{n\underline a}\,(g+{\mathcal
F})^{-1(mn)}\sqrt{|g+{\mathcal F}|}\Big)-{1\over 2}\,e^{ -
\Phi}\,H_{\underline
amn}\,(g+{\mathcal F})^{-1mn}\sqrt{|g+{\mathcal F}|}\hspace{50pt}\nonumber\\
\\
&+e^{ - \Phi}\,E^{\,M}_{\underline
a}\partial_M\,\Phi\,\sqrt{|g+{\mathcal F}|}=-{{(-)^{{p+1}\over
2}}\over{(p+1)!}}\,\epsilon^{m_1\cdots m_{p+1}}\, ({\mathbb{R}} \,
e^{{\mathcal F}_2})_{\underline a\,m_1\cdots m_{p+1}}
\,,\nonumber
\end{eqnarray}
where
\begin{eqnarray}\label{rhs}
& ({\mathbb{R}} \, e^{{\mathcal F}_2})_{\underline a\,m_1\cdots
m_{p+1}}=\sum_{n=0}^{{p+1}\over {2}}\,{{(p+1)!}\over{2^n\cdot
n!\,(p+1-2n)!}} \,R_{\underline a\,[m_1\cdots m_{p+1-2n}}\,{\mathcal
F}_{\cdot
\cdot}\cdots {\mathcal F}_{a_pa_{p+1}]}
\end{eqnarray}

One should keep in mind that in eq. (\ref{cbfstring}) the superforms
$H_3$ and ${\mathbb{R}}$ are subject to supergravity constraints in
the string frame (\ref{H3=IIBstring}) and (\ref{R=IIBstring}), which
are different from the Einstein frame constraints (\ref{H3=IIB}) and
(\ref{R=IIB}) used to derive (\ref{cbf}).

The linearly independent part of (\ref{cbfstring}) (or equivalently
of (\ref{cbf})) is singled out with the help of $u_i{}^{\underline
a}$ as in eq. (\ref{bf3}), while the projection of eq.
(\ref{cbfstring}) along $u_{a}^{\,\,\underline a}$ is zero modulo
the BI field equations (\ref{DBI=d*F2string}), the Bianchi
identities (\ref{bianchi}) and the fermionic equations considered in
Section \ref{theta}.

\subsection{The bosonic equations and the second fundamental form of the
embedding}\label{secondff}
 For further analysis to be carried out in
Sections \ref{conciseboson}-\ref{warped} we would like to present
the bosonic equation (\ref{cbfstring}) in a different but equivalent
form. This subsection is complimentary, so the reader who is not
interested in the details about the relation of the brane equations
to the geometric properties of the embedding such as the second
fundamental form (or the extrinsic curvature) can jump over to
Section \ref{theta} and return here lateron.

Let us first rewrite the {\bf string frame equation}
(\ref{cbfstring}) in the worldvolume tangent frame and single out
its independent components by projecting it along
$u_{i}^{\,\underline a}$
\begin{eqnarray}\label{cbfi}
&{1\over {\sqrt{|\eta+{\mathcal F}|}}}\,D_a\,\Big(e^{{-}\Phi}
\,E_{b}^{\,\,\underline a}\,(\eta+{\mathcal
F})^{-1(ab)}\sqrt{|\eta+{\mathcal F}|}\Big)\,u_{\underline ai}=\hspace{200pt}\nonumber\\
\\
&{1\over 2}\,e^{{ - }\Phi}\,u_{i}^{\,\underline a}\,H_{\underline
a\,ab}\,(\eta+{\mathcal F})^{-1ab} -e^{{ - }\Phi}\,u_i^{\,\underline
a}\,E_{\underline a}^{\,M}\,\partial_M\,\Phi \,-{{(-)^{{p+1}\over
2}}\over {(p+1)!\,\sqrt{|\eta+{\mathcal F}|}}}\,\epsilon^{a_1\cdots
a_{p+1}}\, ({\mathbb{R}} \, e^{{\mathcal {\mathcal
F}}_2})_{\underline a\,a_1\cdots a_{p+1}}\,u_i^{\,\underline
a}\,.\nonumber
\end{eqnarray}
We now notice that because of the embedding condition
$E_{b}^{\,\,\underline a}\,u_{\underline a\,i}=0$, $E_{b}^{\,\,\underline a}=u_{b}^{\,\,\underline a}$
(see Subsection
\ref{aux}) the left hand side of (\ref{cbfi})
reduces to
\begin{equation}\label{1str}
{1\over {\sqrt{|\eta+{\mathcal F}|}}}\,D_a\,\Big(e^{{ - }\Phi}
\,E_{b}^{\,\,\underline a}\,(\eta+{\mathcal
F})^{-1(ab)}\sqrt{|\eta+{\mathcal F}|}\Big)\,u_{\underline
a}^{\,i}=(D_a\,E_{b}^{\,\,\underline a})\,u_{\underline a}^{\,i}
\,(\eta+{\mathcal F})^{-1(ab)}\,e^{{ - }\Phi}\,.
\end{equation}
In the right hand side of (\ref{1str}) one can recognize the second
fundamental form
\begin{equation}\label{2ndf}
K_{b}^{\,\,\,i}=E^a\,K_{ab}^{\,\,\,i}=-D\,E_{b}^{\,\,\underline a}\,u_{\underline
a}^{\,i}=-D\,u_{b}^{\,\,\underline a}\,u_{\underline
a}^{\,\,i}=-d\,u_{b}^{\,\,\underline a}\,u_{\underline
a}^{\,\,i}+u^{\,\,\underline b}_{b}\,\Omega_{\underline
b}{}^{\underline a}\,u_{\underline a}^{\,\,i}
\end{equation}
characterizing the embedding of the (p+1)--dimensional worldvolume
into a curved target type IIB D=10 superspace (see
\cite{se,ser,Howe:1997wf,Sorokin:1999jx} for a review of geometrical grounds of
the embedding and superembedding).

Note that in the static gauge, in which $p+1$ components of the
$10D$ target space--time coordinates $x^{\underline
m}=(x^m,x^{\tilde i})$ are identified with the worldvolume
coordinates $x^m=\xi^m$ and the physical modes are associated with
$x^{\tilde i}(\xi^m)$ ($\tilde i=1,\cdots, 9-p$), the second
fundamental form looks as follows
\begin{equation}\label{2ndff}
K_{ab}^{\,\,\,i}=-E_a{}^m\,E_b{}^n(\partial_m\,\partial_n\,x^{\tilde
i}\,E_{\tilde i}{}^i+\Gamma_{mn}{}^{\underline l}\,E_{\underline
l}{}^i+\partial_m\,x^{\tilde i}\,\Gamma_{n\tilde i}{}^{\,\underline
l}\,E_{\underline l}{}^i)+ {fermion\,\,contributions},
\end{equation}
where $\Gamma_{mn}{}^{\underline l}$ and $\Gamma_{n\tilde
i}{}^{\,\underline l}$ are components of the target space
Christoffel symbol\linebreak $\Gamma_{\underline{mn}}{}^{\underline
l}=D_{(\underline m}\,E_{\underline n)}{}^{\underline
a}\,E_{\underline a}{}^{\underline l}$\,\,\, ($\underline
l,\underline m, \underline n =0,1,\cdots,9$\,;\,\, $m=0,1,\cdots
p$\,;\,\, $i=1,\cdots,9-p$).

We see that if in a pure bosonic background the brane does not
fluctuate in the transverse directions, {\it i.e.}
$\partial_m\,x^{\tilde i}=0$, its second fundamental form is
non--zero if the pullback of the target--space Christoffel symbol
components with one index corresponding to orthogonal directions is
non--zero. This occurs, {\it e.g.} in warped compactifications
considered in Section \ref{warped}.

In terms of the second fundamental form the Dp--brane bosonic equation takes the following form in the
{\bf string frame}
\begin{eqnarray}\label{cbfi1}
&K_{ba}^{\,\,\,\,\,i}\,(\eta+{\mathcal F})^{-1(ab)}= -{1\over
2}\,H^i{}_{ab}\,(\eta+{\mathcal F})^{-1ab} +D^i\,\Phi
+{{(-)^{{p+1}\over 2}\,e^{\Phi}}\over {(p+1)!\,\sqrt{|\eta+{\mathcal
F}|}}}\,\epsilon^{a_1\cdots a_{p+1}}\, ({\mathbb{R}} \, e^{{\mathcal
{\mathcal F}}_2})^i{}_{a_1\cdots a_{p+1}}\,\hspace{20pt}
\end{eqnarray}
where (as defined in Subsection \ref{framelike}) the index $i$ in
the right hand side denotes the component of the pullback of a
superform in the directions orthogonal to the brane, {\it e.g.}
$H_{i\,ab}=\,u_{i}^{\,\underline a}\,E^{\,M}_{\underline
a}\,H_{M\,ab}$, etc.

{}From the geometrical point of view the Dp--brane bosonic equation
(\ref{cbfi1}) is an embedding condition which tells us that its
``trace" $K_{ba}^{\,\,\,\,\,i}\,(\eta+F)^{-1(ba)}$ is expressed in
terms of the pullbacks of background fields. In particular, when the
worldvolume gauge field, the axion--dilaton, the background gauge
fields and all the fermionic fields are zero on the worldvolume the
trace of the second fundamental form (which is also called the
extrinsic curvature) is zero,
\begin{equation}\label{me}
K_{ba}^{\,\,\,\,i}\,\eta^{ba}=0\,,
\end{equation}
implying that the embedding of the surface in superspace is minimal
in the sense of its induced area. The presence of other fields on
the worldvolume and in the background modifies the minimal embedding
condition (\ref{me}) in accordance with eq. (\ref{cbfi1}).

Let us now reveal another property of the second fundamental form
(\ref{2ndf}) which follows from the embedding condition
(\ref{hEi=0})
\begin{equation}\label{anti}
E^i=0\,\quad \Rightarrow \quad 0=dE^i=(DE^{\,\,\underline
a})\,u_{\underline a}^{\,\,i}+E^{\,\,\underline a}\,D\,u_{\underline
a}^{\,\,i}=T^{\,\,\underline a}\,u_{\underline
a}^{\,\,i}-E^{\,a}\,K_a^{\,\,i}\,.
\end{equation}
In view of the torsion constraint (\ref{Ta=IIB}), from the form of
eq. (\ref{anti}) we conclude that the antisymmetric part of the
second fundamental form can only be non--zero due to the presence of
the worldvolume and/or background fermionic fields
\begin{equation}\label{anti1}
K_{[ab]}^{\,\,\,\,\,\,\,\,i} =  - i  {E}_a^1 \,\gamma^{i}\,{E}_b^1 -
i{E}_a^2\,
\gamma^{i}\,{E}_b^2\, .
\end{equation}

\setcounter{equation}0
\section{Fermionic equation, its concise form and $\kappa$--symmetry}\label{theta}

Let us now present the explicit form of the equation of motion
(\ref{ff2}) for the worldvolume fermions $\theta(\xi)$. After some
algebra and taking into account the bosonic equations one finds
\cite{Bandos:1997rq} the following variation of the D--brane
Lagrangian with respect to
$i_\delta\,E^{\,\hat\alpha}=\delta\,Z^{N}\,E_{ N}^{\,\hat\alpha}$
{\bf in the  Einstein frame}
\begin{eqnarray}\label{dL-(p+1)(G)}
& {{e^{-{p-3\over 4} \, {\Phi}}\over {\sqrt{|\eta + F|}
}}}\;\delta_{\hat\alpha}\,{\mathcal L}_{p+1}
 =
{{1}\over{(p+1)!}}\,\epsilon_{ b_1\cdots \; b_{p+1}}
 {E}^{b_1}  \cdots  {E}^{b_{p+1}}\, i_\delta {{E}}{}^{\hat{\alpha}}
(I-\bar{\Gamma})_{\hat{\alpha}}{}^{\hat{\beta}}
{{\nabla}_{\hat{\beta}}
\Phi}  \qquad \; \qquad \nonumber \\
\\
&   i (-)^p  {{1}\over{p!}}\,\epsilon_{a
\; b_1\cdots \; b_{p}}
 {E}^{b_1}  \cdots  {E}^{b_{p}} \,{i_\delta {E}} \,
\left(I-\bar{\Gamma} \right)\left(\gamma_b\, \otimes (\eta -\sigma_3 \, F)^{-1\, ba}
\right) \left({{E}} - {i\over 8}\,{{\tilde\Gamma}}{}^{(1)}
{{\nabla} \Phi} \right)\,.\nonumber
\end{eqnarray}
where $\bar{\Gamma}$ is the $\kappa$--symmetry projector
$(\bar{\Gamma}^2=1)$  whose properties will be discussed a bit
later,
\begin{eqnarray}
 \label{tE=E1E2}
& {{E}}^{\hat{\alpha}} ={{E}}^{{\alpha}I}= \left( E^{{\alpha}1},
E^{{\alpha}2}
\right)\; ,  \quad
{\nabla}_{\hat{\beta}} \Phi := \left(\matrix{{\nabla}_{{\beta}1}\Phi
\cr
{\nabla}_{{\beta}2} \Phi }\right)\; , \\
&\nonumber\\
& {\tilde\Gamma}^{(1)}:= E^{\underline{a}}\,{\tilde
\Gamma}_{\underline{a}} = \left(
\matrix{E^{\underline{a}}\tilde\gamma_{\underline{a}}^{\alpha\beta} & 0 \cr 0 &
E^{\underline{a}}\tilde\gamma_{\underline{a}}^{\alpha\beta} }
\right)\; , \qquad
\\
&\nonumber\\
& \left(\gamma_a\, \otimes (\eta -\sigma_3 \, F)^{-1\, ab} \right) =
\left(\matrix{\gamma_{a}\,(\eta -F)^{-1\, ab} \!\!\!\!\!\! & 0 \cr 0 & \!\!\!\!\!\!
\gamma_a\,(\eta +F)^{-1\, ab} }\right), \quad \gamma_a=\gamma^{\underline b} u_{\underline
b\,a}=E^{\,m}_{a}\,E^{\,\underline b}_m\,\gamma_{\underline
b},\hspace{10pt}\label{gtimes}
\end{eqnarray}
the Pauli matrix $\sigma_3$ acts on the indices $I=1,2$ of the
spinor $E^{\hat\alpha}=E^{I\alpha}$ and $E^{\,m}_{a}(\xi)$ is the
inverse induced worldvolume vielbein introduced in
(\ref{im1})--(\ref{im2}).

{}From the variation (\ref{dL-(p+1)(G)}) we find the concise form of
the fermionic equations for the D$p$--branes in an arbitrary type
IIB supergravity background {\bf in the  Einstein frame}
\cite{Bandos:1997rq} \footnote{In \cite{Bandos:1997rq} the fermionic
equation was given in the differential form $${1\over
p!}\,\epsilon_{ba_1\cdots a_p}\,E^{a_1}\cdots
E^{a_p}\,\left(I-\bar{\Gamma} \right)\,\left(\gamma_b\, \otimes
(\eta -\sigma_3 \, F)^{-1\, ba} \right) \left({{\cal E}}- {i\over
8}{\tilde {\Gamma}}^{(1)}  { {\nabla} \Phi} \right)
$$
$$
 +{i\over {2\cdot(p+1)!}}\,\,\epsilon_{a_0a_1\cdots
a_p}\,E^{a_0}\cdots E^{a_p}\,\left(I-\bar{\Gamma}
\right)\,{{\nabla} \Phi} =0\,.$$}
\begin{eqnarray}\label{Dirac(G)}
\left(I-\bar{\Gamma}
\right)\,\Big[\left(\gamma_b\, \otimes (\eta -\sigma_3 \, F)^{-1\, ba} \right)
\left({{\cal E}}_a - {i\over 8}{\tilde {\Gamma}}{}_a  { {\nabla}
\Phi} \right) +{i\over 2}\,{{\nabla} \Phi}\Big] =0\,,
  \, ,
\end{eqnarray}
where the calligraphic $\mathcal E_a$ denotes the pullback
\begin{equation}\label{pbg}
{\mathcal E}_a^{\,\hat\alpha}
=E_a^{\,m}(\xi)\,\partial_m\,Z^M\,E_M^{\,\hat\alpha}
\end{equation}
of the spinorial supervielbein to avoid the confusion with the
induced worldvolume  vielbein $E^a=dZ^M\,E_M^{\,\underline
b}\,u^{\,a}_{\underline b}=d\xi^m\,E_m^{\,a}$. Note that in the form
(\ref{Dirac(G)}) the D--brane fermionic equation does not contain
the Lorentz harmonics. Their role is taken up by the induced
worldvolume vielbein $E_m^{\,a}$ and its inverse.

{\bf In the string frame} the fermionic equation simplifies to
\begin{eqnarray}\label{Dirac(G)string}
\left(I-\bar{\Gamma}
\right)\,\Big[\left(\gamma_b\, \otimes (\eta -\sigma_3 \, F)^{-1\, ba} \right)
\,{{\cal E}}_a +{i\over 2}\,{{\nabla} \Phi}\Big] =0\, ,
\end{eqnarray}
where $\gamma_b\, \otimes (\eta -\sigma_3 \, F)^{-1\, ba}$ is
defined in (\ref{gtimes}) and $F_2=dA-B_2$.

\bigskip

\subsection{The $\kappa$--symmetry projector and the worldvolume Lorentz group}
The appearance of the projector ${1\over 2}\,(I-\bar{\Gamma} )$ in
the fermionic variation (\ref{dL-(p+1)(G)}) of the Dp--brane action
implies the well known fact that the actions of the superbranes
possess worldvolume fermionic $\kappa$--symmetry. This means that
the superbrane action is invariant under the following variation of
the worldvolume scalars $Z^M(\xi)$
\begin{equation}\label{kappa}
\delta_\kappa Z^M\,E_M^{\,\hat\alpha}=\kappa^{\hat\beta}(\xi)(I+\bar{\Gamma}
)_{\hat\beta}^{\,\,\,\hat\alpha}\,,
\qquad \delta_\kappa Z^M\,E_M^{\,\underline a}=0\,
\end{equation}
accompanied by an appropriate variation of the BI field $A_m(\xi)$
and of the auxiliary fields $u_{\underline b}^{\,a}(\xi)$ and
$F_{ab}(\xi)$ which we shall not present here. Note only that
$A_m(\xi)$ transforms in such a way ($\delta_\kappa
\,A=i_{\delta_\kappa}\,B_2$) that the $\kappa$--transformation of its extended
field strength ${\mathcal F}_2=dA-B_2$ is $\delta_\kappa\,{\mathcal
F}_2=-i_{\delta_\kappa} H_3$. If one substitutes the variation
(\ref{kappa}) into eq. (\ref{dL-(p+1)(G)}) one finds that it
vanishes identically since $(I-\bar{\Gamma} )\,(I+\bar{\Gamma} )=0$,
thus indicating the presence of a symmetry.

 In the case under consideration the $\kappa$--symmetry projector
$\bar{\Gamma}$ is a block--anti-diagonal $32\times 32$ matrix (see
\cite{bt,kallosh} and also \cite{abkz})
\begin{eqnarray}\label{barG=hh-1}
\bar{\Gamma}_{\hat{\beta}}{}^{\hat{\alpha}}= \left(\matrix{ 0 & h_\beta{}^\alpha \; \cr
(h^{-1})_\beta{}^\alpha & 0 } \right)\; .  \qquad
\end{eqnarray}
It can be written in the following differential form (see
\cite{c1,c2} and more recent \cite{Martucci:2005rb})
\\
\begin{eqnarray}\label{bGammaE=}
&{{1}\over {(p+1)!}}\,\epsilon_{a_1\cdots a_{p+1}}\,E^{a_1} \cdots
  E^{a_{p+1}}\,  \bar{\Gamma}  =  {1\over \sqrt{|\eta +F|}}\,
\sum\limits_{n=0}^{{p+1\over 2}} \,\left( \matrix{ 0 & (-)^n\,
{\gamma}^{(2n)}{}_\beta{}^\alpha \cr
{\gamma}^{(2n)}{}_\beta{}^\alpha & 0 }
\right)  \wedge  { F_2 ^{\,{p+1\over 2}-n }\over \left({p+1\over 2}-n
\right)! }\,\hspace{50pt}
\end{eqnarray}
\\
where the matrix differential forms ${\gamma}^{(2n)}$ have been
defined in (\ref{g2n}).

Eq. (\ref{bGammaE=}) implies that the matrix $h_\alpha{}^\beta$ in
(\ref{barG=hh-1}) is defined by the following relation
\\
\begin{eqnarray}\label{Vxh=}
&{{1}\over {(p+1)!}}\,\epsilon_{a_1\cdots a_{p+1}}\,E^{a_1} \cdots
  E^{a_{p+1}}\,  h_\beta{}^\alpha  =  {1\over \sqrt{|\eta +F|}}\,
\sum\limits_{n=0}^{{p+1\over 2}}   \,(-)^n \,{\gamma}^{(2n)}{}_\beta{}^\alpha \; \wedge
{F_2^{\,{p+1\over 2}-n }\over {\left({p+1\over 2}-n
\right)! }}\,\hspace{50pt}
\end{eqnarray}
\\
and its inverse is
\\
\begin{eqnarray}\label{Vxh-1=}
&{{1}\over {(p+1)!}}\,\epsilon_{a_1\cdots a_{p+1}}\,E^{a_1} \cdots
  E^{a_{p+1}}\,  (h^{-1})_\beta{}^\alpha   =  -{1\over \sqrt{|\eta +F|}}\,
\sum\limits_{n=0}^{{p+1\over 2}} \,{\gamma}^{(2n)}{}_\beta{}^\alpha \; \wedge
{F_2^{\,{p+1\over 2}-n }\over {\left({p+1\over 2}-n
\right)! }}\,.\hspace{50pt}
\end{eqnarray}
\\
The form of the matrix $h_\beta{}^\alpha$ drastically simplifies
when the worldvolume field strength $F_2$ vanishes. Then, it reduces
to the antisymmetrized product of the (p+1) gamma--matrices along
the brane worldvolume,
\begin{eqnarray}\label{h0=}
 \,h_\beta{}^\alpha
\Big|_{^{F_2=0}}   =-
  {{1}\over (p+1)! }\; \epsilon_{a_{_0}\cdots \, a_p} \,(\gamma^{a_{_0}}\tilde\gamma^{a_{_1}}
  \gamma^{a_{_2}}\cdots
  \tilde\gamma^{a_p})_\beta{}^\alpha =: \bar\gamma^{T\,-1}{}_\beta{}^\alpha\,,
   \qquad
\bar\gamma^{-1}= (-)^{{p-1\over 2}}\,\bar\gamma\,,\nonumber\\
\\
\bar\gamma={{1}\over (p+1)! }\; \epsilon_{a_{_0}\cdots \, a_p}
\,(\tilde\gamma^{a_{_0}}\cdots
  \gamma^{a_p})
 \; .\hspace{200pt}\nonumber
\end{eqnarray}
One can also check, using the properties (\ref{oc}) of the Lorentz
harmonics, that the following relations hold
\begin{eqnarray}\label{h0sa=sah0}
&&
\bar\gamma^T\,\gamma^a = (-)^{{p-1\over 2}} \gamma^a\,\bar\gamma
= {{(-)^{{p-1\over 2}}}\over p!}\, \epsilon^{a b_1\cdots \, b_p}\,
\gamma_{b_1} \cdots \, \gamma_{b_p}
\; ,
 \\
\nonumber\\
 \label{h0si=sih0} &&
 \bar\gamma^T\,\gamma^{i}
 =
 (-)^{{p+1\over 2}}\, \gamma^{i}\,
\bar\gamma
\,,
\qquad {\rm where}\qquad
\gamma^{i}=\gamma^{\underline a}\,u_{\underline a}^i\,,\quad
\gamma^{a}=\gamma^{\underline a}\,u_{\underline a}^a\,,
\end{eqnarray}
and hence
\begin{eqnarray}\label{h0sah0=sa}
\bar\gamma^T\,\gamma^a\,\bar\gamma
=  \gamma^a
\; , \qquad
 \bar\gamma^T\,\gamma^i \,
 \bar\gamma
 = -  \gamma^i
 \;\qquad (a=0,\cdots,\,p\,,\quad i=1,\cdots\,, D-p-1\,).
\end{eqnarray}
In the general case, in which the worldvolume field strength $F_2$
is non--zero, the relations (\ref{h0sah0=sa}) get modified and take
the following form
\begin{eqnarray}\label{hsah=sa}
{h}\,\gamma^a\, {h}^{T} =  \gamma^b\,k_b^{~a}=\gamma^b\,(\eta +
F)_{bc}(\eta - F )^{-1~ca}
\; , \qquad
 {h}\,\gamma^i \,{h}^T = -  \gamma^i  \;\,,
\end{eqnarray}
where the matrix $k_b^{~a} = (\eta + F)_{bc}(\eta - F )^{-1~ca}$ is
orthogonal
\begin{eqnarray}\label{kinin}
k_b^{~a} = (\eta + F)_{bc}(\eta - F )^{-1~ca}\qquad
\Rightarrow \qquad k_b^{~a}\,k_d^{~c}\,\eta_{ac}=\eta_{bd}, \quad
{\rm det}\, k=1\,.
\end{eqnarray}
(note that det $(\eta + F)_{bc}$=det $(\eta-F)_{bc}$).

Therefore, $k_b^{~a}$ belongs to the worldvolume Lorentz group
$SO(1,p)\subset SO(1,9)$ and is called the {\it Cayley image} of the
antisymmetric tensor $F_{ab}$. Because of the relations
(\ref{hsah=sa}) and of the fact that $k_b^{~a} \in SO(1,p)$ we
observe an interesting property of the Dp--brane $\kappa$--symmetry
projector: $h_\beta^{\,\alpha}$ belongs to a $16\times 16$ real
matrix representation of the group $Spin(1,p)\subset Spin(1,9)$
\footnote{The $\kappa$--symmetry projector and the BI field strength
of the type IIA D--branes have the same group--theoretical
meaning.}. The relations (\ref{hsah=sa}) and (\ref{kinin}) are very
useful for the analysis of the fermionic equations (\ref{Dirac(G)}),
(\ref{Dirac(G)string}). Note that the form of the $\kappa$--symmetry
projector  and all the above relations are the same in the Einstein
and the string frame.

If we substitute the form (\ref{barG=hh-1}) of the
$\kappa$--symmetry projector into the fermionic equation ({\it e.g.}
{\bf in the string frame} (\ref{Dirac(G)string})) we find the
following linearly independent 16--component equation
\begin{equation}\label{Dirac(Gh)string}
(\eta + F)^{-1\, ba}\,\gamma_b\,
\,({{\cal E}}^2_a-\,{{\cal E}}^1_a\,h) +{i\over 2}\,({\nabla}_2-h^{-1}\,{\nabla}_1\,)
\Phi=0\,.
\end{equation}
Notice that the fermionic supervielbeins $E^{1\alpha},\,E^{2\alpha}$
(remember the notation (\ref{pbg})) and their inverse (encoded in
$\nabla_{\alpha I}$) enter the equation ({\ref{Dirac(Gh)string})
only through the single 16--component combination $ E^2-\,E^1\,h$,
which manifests the fact that due to $\kappa$--symmetry only half of
the fermionic degrees of freedom of the superbrane are dynamical.
Note that so far we have not imposed any $\kappa$--symmetry gauge.
Therefore, in the above form the equation for the physical fermionic
modes on the superbrane is gauge independent up to the bosonic
equations, as will be explained in the next Subsection  and Section
\ref{conciseboson}.

\subsection{The linearized Dp--brane fermionic equation}\label{linear}
Let us now consider the fermionic equation of motion of the
Dp--branes in the linear approximation in $\theta(\xi)$. This will
allow us to obtain an explicit coupling of the worldvolume fermions
to the background and worldvolume fluxes, which may be useful for
various applications including brane instanton calculations.

To obtain the linearized fermionic equation we use the following
trick \cite{Bandos:2005ww}. It is based on the fact that the
Dp--brane action and equations of motion are invariant under the
target space superdiffeomorphisms $Z^{M}\,\rightarrow\,f^M(Z)$ and
that $\theta^{\underline \mu}(\xi)$ are the Goldstone fermions on
the brane associated with the spontaneous breaking of half of the
target space supersymmetry. Using the background diffeomorphisms, we
can choose a supercoordinate system in which the brane does not
fluctuate along the Grassmann--odd directions in superspace, {\it
i.e.} $\theta^{\underline \mu}(\xi)=0$ (see \cite{sbs} for a
discussion of dynamical supergravity--superbrane systems with the
target space superdiffeomorphisms being a gauge symmetry). Then the
fermionic equation, {\it e.g.} {\bf in the string frame}
(\ref{Dirac(G)string}), reduces to the algebraic relations on the
components of the pullbacks of the gravitino and the dilatino fields
$\psi_{\underline m}^{\,\hat\alpha}(x)=E_{\underline
m}^{\,\hat\alpha}|_{\theta=0}$ and $\lambda_{\hat\alpha}(x)=-{i\over
2}\,{{\nabla}_{\hat\alpha} \Phi}|_{\theta=0}$:
\begin{equation}\label{theta=0}
\left(I-\bar{\Gamma}
\right)\,\Big[\left(\gamma_b\, \otimes (\eta -\sigma_3 \, F)^{-1\, ba} \right)
\,\partial_a\,x^{\underline m}\,{\psi}_{\underline m}- \lambda \Big] =0\, ,
\end{equation}
 where the values of the
$\kappa$--symmetry projector $\bar\Gamma$ and of the pullback of the
bosonic vielbein $E_m^{\,\underline a}$ are taken at $\theta=0$.
{\it I.e.} in (\ref{theta=0}) $E_m^{\,\underline
a}=\partial_m\,x^{\underline n}\,E_{\underline n}^{\,\underline
a}(x)$ is just the pullback of the $D=10$ gravitational field.

Let us now recall that $\theta(\xi)$ are Goldstone fermions for the
broken target space supersymmetry. Therefore their presence in the
fermionic equation at linear order can be restored by performing in
(\ref{theta=0}) an infinitesimal local supersymmetry transformation
of the pullbacks of the type IIB supergravity fields $E_{\underline
n}^{\,\,\,\underline a}(x)$, $\psi_{\underline m}^{\,\hat\alpha}(x)$
and  $\lambda_{\hat\alpha}(x)$ whose supersymmetry parameter ,
$\epsilon^{\hat\alpha}(x(\xi))$, we identify (in the Wess--Zumino
gauge) with
$\theta^{\hat\alpha}(\xi)=\theta^{\mu}(\xi)\,E_\mu^{\,\hat{\underline
\alpha}}(x,\theta)$ so that
$d\,\epsilon^{\hat\alpha}(x(\xi))=d\,\theta^{\hat\alpha}(\xi)$. We
thus get
\begin{eqnarray}\label{deltae}
\left(I-\bar{\Gamma}
\right)\,\Big[\left(\gamma_b\, \otimes (\eta -\sigma_3 \, F)^{-1\, ba} \right)
\,\partial_a\,x^{\underline m}\,{\psi}_{\underline m}- \lambda \Big]
+\hspace{200pt}\nonumber
\\
\\
+
\delta\,\Big[(I-\bar{\Gamma})\,(\gamma_b\, \otimes (\eta -\sigma_3 \,
F)^{-1\, ba}\, E_a^{\,\,m}\Big]\,\psi_m+\delta
\bar{\Gamma}\,\lambda+\hspace{100pt}
\nonumber\\
\nonumber\\
+\left(I-\bar{\Gamma}
\right)\,\Big[\left(\gamma_b\, \otimes (\eta -\sigma_3 \, F)^{-1\, ba} \right)
\,E_a^{\,\,m}\,\delta\,{\psi}_{ m}- \delta\lambda \Big]=0\, ,\nonumber
\end{eqnarray}
where the second and the third term contains the supersymmetry
variation of the pullback of the gravitational field $E_{\underline
n}^{\,\,\,\underline a}(x)$ entering $\bar\Gamma$, $\gamma_b$ and
${\cal F}_2={1\over 2}\,E^b\,E^a\,F_{ab}=dA-B_2$.

Ignoring the three--fermion contributions, the pullbacks
$\delta\,{\psi}^{\hat\alpha}_{m}=\delta\,{\psi}^{I\alpha}_{m}$ and
$\delta\lambda_{\hat\alpha}=\delta\lambda_{I\alpha}$ of the local
supersymmetry variations of the type IIB gravitino and dilatino have
the following form {\bf in the string frame}
\begin{eqnarray}\label{susypsi}
\delta\psi_m&=& D_m \,\theta+ {1\over 4\cdot 2!}\,H_{m\underline{ab}}\,\tilde\gamma^{\underline{ab}}
\sigma_3\,\theta +\,\nonumber \\
\\
&+& {1\over 8}e^\Phi\big[R_{\underline a}\tilde\gamma^{\underline
a}\,(i\sigma_2) -{1\over 3!}{ R}_{\underline
{abc}}\tilde\gamma^{\underline {abc}}\,\sigma_1
+{1\over 2\cdot 5!}R_{\underline {abcde}}\,\tilde\gamma^{\underline {abcde}}(i\sigma_2)\big]\,\gamma_m\,\theta\,,\nonumber\\
\nonumber\\
 \label{susylambda}
\delta\lambda &=& {1\over 2}\,\big[D_{\underline
a}\Phi+e^\Phi\,R_{\underline a}\,(i\sigma_2)\, \big]
\,\gamma^{\underline a}\theta -{1\over {
4!}}\,\big[H_{\underline{abc}}-e^\Phi\,R_{\underline{abc}}\,(i\sigma_2)\,\big]\,\gamma^{\underline{abc}}\,\sigma_3\,\theta
\,,
\end{eqnarray}
where $\gamma_m=E_m^{\,\,\underline a}\,\gamma_{\underline a}$, all
$\tilde\gamma^{\underline {abc}\cdots}$ start with
$\tilde\gamma^{\underline a\,\alpha\beta}$, $D_m =
\partial_m - {1\over 4}\partial_m\,x^{\underline
n}\,\Omega_{\underline n}{}^{\underline{ab}}
\gamma_{\underline{ab}}$ is the covariant derivative with the spin connection
pulled back onto the worldvolume and the Pauli matrices act on the
index $I=1,2$ of the spinor $\theta^{\alpha I}=(\theta^{\alpha
1},\theta^{\alpha 2})$ \footnote{In eqs. (\ref{susypsi}),
(\ref{susylambda}) and in what follows we skip $\otimes$ product
symbol between $\gamma$ and $\sigma$ where this does not cause
confusion.}.

The type IIB supersymmetry variation of the graviton field
$E_{\underline m}^{\,\,\underline a}(x)$ is
\begin{equation}\label{susygraviton}
\delta\,E_{\underline m}^{\,\,\underline a}(x)=-2i\psi_{\underline
m}\,\gamma^{\underline a}\,\theta.
\end{equation}
Substituting eqs. (\ref{susypsi})--(\ref{susygraviton}) and using
the definition of $\bar\Gamma$ one can find the linear fermion
equations for any Dp--brane in an arbitrary type IIB supergravity
background with non--zero graviton, gravitino and dilatino. An
explicit form of such an equation for the M2--brane has been
obtained in \cite{Bandos:2005ww}.

If we are interested in purely bosonic supergravity backgrounds (in
which the gravitino and the dilatino are zero) the first, second and
third term of eq. (\ref{deltae}) vanish and the equation reduces to
the Dirac--like equation which can be derived from the quadratic
action of \cite{Martucci:2005rb}
$$
\left(I-\bar{\Gamma}
\right)\,\Big[\left(\gamma_b\, \otimes (\eta -\sigma_3 \, F)^{-1\, ba} \right)
\,E_a^{\,\,m}\,\delta\,{\psi}_{ m}- \delta\lambda \Big]=0\,,
$$
or
\begin{equation}\label{deltae0}
\left(I-\bar{\Gamma}
\right)\,\left(\gamma_b\, \otimes (\eta -\sigma_3 \, F)^{-1\, ba} \right)
\,{\mathbf D}_a\,\theta=0\,,
\end{equation}
where the generalized covariant derivative ${\mathbf D}_a$ contains
the pullback to the worldvolume of the $Spin(1,9)$ connection and
all contributions of the worldvolume and background gauge field
strengths in accordance with the form of the supersymmetry
transformations (\ref{susypsi}) and (\ref{susylambda})
\begin{eqnarray}\label{mathcalD}
{\mathbf D}_a&=D_a
+{1\over 4\cdot
2!}H_{a\,\underline{bc}}\,\tilde\gamma^{\underline{bc}}\,
\sigma_3+{1\over 8}e^\Phi\big(iR_{\underline a}\tilde\gamma^{\underline
a}\,\sigma_2 -{1\over 3!}{ R}_{\underline
{abc}}\tilde\gamma^{\underline {abc}}\,\sigma_1 +{i\over 2\cdot
5!}R_{\underline {abcde}}\,\tilde\gamma^{\underline
{abcde}}\,\sigma_2\big)\,\gamma_a\nonumber\\
\\
 &-{1\over {2\,(p+1)}}\,\tilde\gamma^b\, \otimes (\eta +\sigma_3 \, F)_{
ba}\,\Big[\big(D_{\underline a}\Phi+ie^\Phi\,R_{\underline
a}\,\sigma_2\, \big)
\,\gamma^{\underline a} -{1\over
2\cdot
3!}\,\big(H_{\underline{abc}}-ie^\Phi\,R_{\underline{abc}}\,\sigma_2\,\big)\,\gamma^{\underline{abc}}\,\sigma_3\Big]\nonumber\,,
\end{eqnarray}
where $ {1\over {p+1}}\,\tilde \gamma^b\, \otimes (\eta +\sigma_3 \,
F)_{ ba} $ is inverse of (\ref{gtimes}).

One can notice that the covariant derivative (\ref{mathcalD})
entering (\ref{deltae0}) is the covariant derivative of the
generalized holonomy which determines the gravitino supersymmetry
variation (\ref{susypsi}) extended by the terms from the
supersymmetry variation of the dilatino (\ref{susylambda}).

In the case of zero worldvolume field strength $F_2$ the equations
(\ref{deltae0}) and (\ref{mathcalD}) reduce, respectively, to
\begin{equation}\label{deltae0f0}
\left(I-\bar{\Gamma} \right)\,\gamma^a\, \,{\mathbf D}_a\,\theta=0\,
\end{equation}
and
\begin{eqnarray}\label{mathcalDf0}
{\mathbf D}_a&=D_a
+{1\over 4\cdot
2!}H_{a\,\underline{bc}}\,\tilde\gamma^{\underline{bc}}
\sigma_3\,+{1\over 8}e^\Phi\big(iR_{\underline a}\tilde\gamma^{\underline
a}\,\sigma_2 -{1\over 3!}{ R}_{\underline
{abc}}\tilde\gamma^{\underline {abc}}\,\sigma_1 +{i\over 2\cdot
5!}R_{\underline
{abcde}}\,\tilde\gamma^{\underline {abcde}}\,\sigma_2\big)\,\gamma_a\nonumber\\
\\
 &-{1\over {2\,(p+1)}}\,\tilde\gamma_a\, \,\Big[\big(D_{\underline a}\Phi+ie^\Phi\,R_{\underline
a}\,\sigma_2\, \big)
\,\gamma^{\underline a} -{1\over
2\cdot
3!}\,\big(H_{\underline{abc}}-ie^\Phi\,R_{\underline{abc}}\,\sigma_2\,\big)\,\gamma^{\underline{abc}}\,\sigma_3\Big]\nonumber\,.
\end{eqnarray}
They have been studied in detail for the D3 brane in particular
backgrounds with fluxes in
\cite{Tripathy:2005hv,Bergshoeff:2005yp,Park:2005hj}.

Let us now proceed with the analysis of the general structure of the
fermionic equation (\ref{deltae0}). Using the transposition identity
\begin{eqnarray}\label{id32-5}
 \left(I-\bar{\Gamma}\right) \, \left(\gamma_a\, \otimes (\eta -\sigma_3 \, F)^{-1\,
ab}\right) \equiv \left(\gamma_a\, \otimes (\eta -\sigma_3 \,
F)^{-1\, ab}\right) \,
\left(I-\bar{\Gamma}\right)^T \;
\end{eqnarray}
we can rewrite (\ref{deltae0}) in the form
\begin{eqnarray}\label{D2(G)Th}
{\mathbf D}_b  \,{\theta}\,  \left(I-\bar{\Gamma}\right)
\left(\gamma_a\,
\otimes (\eta -\sigma_3 \, F)^{-1\, ab}\right) \,=0 \; .
\end{eqnarray}
\\
Then using the identity $(I-\bar{\Gamma})\equiv {1\over
2}(I-\bar{\Gamma})(I-\bar{\Gamma})$, the Leibnitz rule and the
identity ${\cal D}\bar{\Gamma}\; (I-\bar{\Gamma})=
(I+\bar{\Gamma}){\cal D}\bar{\Gamma}$, one can equivalently write
eq. (\ref{D2(G)Th}) as
\begin{eqnarray}\label{D3(G)Th}
  {\mathbf D} _b \,( {\theta}\,  (I-\bar{\Gamma}))
(I-\bar{\Gamma})
\left(\gamma_a\, \otimes (\eta -\sigma_3 \, F)^{-1\, ab}\right)
+\hspace{200pt}
\nonumber \\
\\
\qquad + 2\,{\theta}(I+\bar{\Gamma})\,  {\mathbf D}_b
\bar{\Gamma}\, \left(\gamma_a\, \otimes (\eta -\sigma_3 \, F)^{-1\,
ab}\right) \,=0 \; .\nonumber
\end{eqnarray}
Now we observe that the first and the second terms should vanish
separately, {i.e.}
\begin{eqnarray}\label{D4(G)Th}
&  {\mathbf D}_b ( {\theta}\,  (I-\bar{\Gamma})) (I-\bar{\Gamma})
\left(\gamma_a\, \otimes (\eta -\sigma_3 \, F)^{-1\, ab}\right) =0 \;
\end{eqnarray}
and
\begin{eqnarray}\label{D5(G)Th}
 {\theta}\, (I+\bar{\Gamma}) \, {\mathbf D}_b \bar{\Gamma}\,
\left(\gamma_a\,
\otimes (\eta -\sigma_3 \, F)^{-1\, ab}\right) \,=0 \; .
\end{eqnarray}
The reason is that at the leading order in $\theta$ which we are
interested in at the moment, the $\kappa$--symmetry transformation
of $\theta$ which follows from eq. (\ref{kappa}) is
$$
\delta_\kappa\,\theta=\kappa(\xi)\,(1+\bar\Gamma)|_{\theta=0}\,.
$$
Therefore, ${\theta}(\xi)\, (I+\bar{\Gamma})$ is a pure gauge degree
of freedom. It can be put to zero as a $\kappa$--symmetry gauge
fixing condition. In this gauge the second term in (\ref{D3(G)Th})
vanishes. Now we notice that  eq. (\ref{D4(G)Th}) is
$\kappa$--invariant at the linear order in $\theta$, because
$\theta\, (I-\bar{\Gamma})$ is $\kappa$--invariant in this order.
Hence the second term (\ref{D5(G)Th}) of (\ref{D3(G)Th}) must vanish
in any gauge which is only possible if the following equation holds
\begin{eqnarray}\label{Beq(G)}
(I+\bar{\Gamma}) \, {\mathbf D}_b \bar{\Gamma}\,
\left(\gamma_a\, \otimes (\eta -\sigma_3 \, F)^{-1\, ab}\right) \,=0\;
\end{eqnarray}
at least for the purely bosonic supergravity backgrounds in which
all fermionic fields are set to zero. Eq. (\ref{Beq(G)}) on the
$\kappa$--symmetry projector is a differential equation for bosonic
fields on the Dp--brane. As such it cannot be an independent
equation but only a concise form of the bosonic equations which we
have already found by varying the Dp--brane action.

\setcounter{equation}0
\section{The $\kappa$--symmetry projector and the concise form of the Dp--brane bosonic
equations}\label{conciseboson}
 Let us study in detail eq. (\ref{Beq(G)}) and its
relation to the Dp--brane bosonic equations (\ref{DBI=d*F2}),
(\ref{DBI=d*F2string}), (\ref{cbf}) and (\ref{cbfstring}) and to the
Bianchi identity (\ref{bianchi}). Since eq. (\ref{Beq(G)}) contains
the projector $I+\bar{\Gamma}$ it is actually a $16\times 16$
component equation for the matrix $h_\beta^{\,\alpha}$ that
determines $\bar\Gamma$, eq. (\ref{barG=hh-1}). Upon some algebra
with the use of eqs. (\ref{hsah=sa}) it can be reduced to the
following form
\begin{eqnarray}\label{Beq(h)}
  {{\mathbf D}}_a\,h \, \gamma_b\, (\eta +F)^{-1\,
ba} = 0\quad
\Leftrightarrow  \quad     {{\mathbf D}}_a\,h \, h^{\!^{-1}} \gamma_b\,
(\eta -F)^{-1\, ba} = 0 \; .
\end{eqnarray}
Note that in (\ref{Beq(h)}) ${\mathbf D}_a\,h \, h^{\!^{-1}}$ is a
counterpart of the Cartan form $d\,h \, h^{\!^{-1}}$ for the group
$Spin(1,p)$. Eq. (\ref{Beq(h)}) can thus be regarded as a
Dirac--like equation on $h$ where the generalized  covariant
derivative ${{\mathbf D}}_a$ is the following $16\times 16$
counterpart of (\ref{mathcalD}) in the sense  that
\begin{equation}\label{mathcalDf016}
(1+\bar\Gamma)\,{{\mathbf D}}\,\bar\Gamma=\left(\matrix{h{\mathbf
D}h^{-1} & {\mathbf D}h \; \cr {\mathbf D}(h^{-1}) &h^{-1}{\mathbf
D}h } \right)\;
\end{equation}
(the explicit expression for ${\mathbf D}\,h$ is given in the
Appendix).

To reveal the contents of eq. (\ref{Beq(h)}) let us first simplify
things and consider the case when the dilaton and all the background
fluxes are zero. Then the generalized covariant derivative
${{\mathbf D}}_a$ (\ref{mathcalDf016}) reduces to the $Spin(1,9)$
covariant derivative $D_a$ and eq. (\ref{Beq(h)}) takes the form
\begin{eqnarray}\label{Beq(h)flux0}
 {{ D}}_a\,h \, h^{\!^{-1}} \gamma_b\,
(\eta -F)^{-1\, ba} = 0 \; .
\end{eqnarray}
  We now notice that eq. (\ref{Beq(h)flux0}) can be obtained by differentiating the relations
(\ref{hsah=sa}) as follows. Taking the covariant differential of
(\ref{hsah=sa}) we get
\begin{eqnarray}\label{Dhsa=}
 (Dh\, h^{-1})\gamma^{{a}} +  \gamma^{{a}} (Dh\, h^{-1})^{T}
 &=& \gamma^b (Dk\, k^{-1})_b{}^a + 2\gamma^i K_b^{\,\, i} (\eta
 +F)^{-1\, ba}
 \;  , \;\qquad \\ \label{Dhsi=}
 (Dh\, h^{-1})\gamma^{{i}} +  \gamma^{{i}} (Dh\, h^{-1})^{T}
 &= & 2K^{a\, i} (\eta +F)^{-1}{}_{ab} \gamma^b
 \;  ,
\end{eqnarray}
where $k_a{}^b=(\eta + F)_{bc}(\eta - F )^{-1~ca}$ is the Cayaley
image defined in (\ref{kinin}) and $K_a^{\,\, i}=E^b\,K_{ba}^{\,\,
i}$, with $K_{ba}^{\,\, i}$ being the second fundamental form
(\ref{2ndf}). To derive eqs. (\ref{Dhsa=}) and (\ref{Dhsi=}) we have
used that
$$
\gamma^a=\gamma^{\underline a}\,u_{\underline a}^{\,\,a}\,,\qquad
\gamma^i=\gamma^{\underline a}\,u_{\underline a}^{\,\,i}
$$
(see eqs. (\ref{lh}), (\ref{oc}) and (\ref{im2}) for the definition
of the harmonics $u_{\underline b}^{\,\,\underline
a'}=(u_{\underline a}^{\,\,a},u_{\underline a}^{\,\,i})$),  the
identities
\begin{eqnarray}\label{1+k=}
\delta_b{}^a + k_b{}^a \equiv 2\eta_{bc}(\eta -F)^{-1\, ca} \; , \quad \delta_b{}^a +
k^{-1}{}_b{}^a \equiv \delta_b{}^a + k^a{}_b \equiv  2\eta_{bc}(\eta
+F)^{-1\, ca} \; ,
\quad
\end{eqnarray}
and the definition of the second fundamental form (\ref{2ndf})
which, due to the harmonic relations (\ref{oc}), implies that
\begin{eqnarray}\label{Dua,Dui=}
Du^a_{\underline{a}}= u_{\underline{a}}^i K^{ai} \; , \qquad
Du^i_{\underline{a}}= u_{\underline{a}\, c} K^{ci} \qquad \\
\label{Dsa=Dsi} \Rightarrow \qquad   D\,\gamma^a = \gamma^i K^{ai}   \; ,
\qquad D\gamma^i = \gamma_a K^{ai}  \; .
   \;
\end{eqnarray}

The general solution of eqs. (\ref{Dhsa=}) and (\ref{Dhsi=}) is
\begin{eqnarray}\label{Dhh=}
 (Dh\, h^{-1})_{\beta}{}^{\alpha}= {1\over 4} (Dk\,
 k^{-1})^{ab}(\gamma_{{a}}\tilde{\gamma}_b)_{\beta}{}^{\alpha} - K^{a\, i} (\eta
 +F)^{-1}{}_{ab}(\gamma^{{b}}\tilde{\gamma}^i)_{\beta}{}^{\alpha}
 \;  ,
\end{eqnarray}
where the first term can be further written as follows (see
\cite{abkz} for the D9 brane case)
\begin{eqnarray}\label{Dkk=}
 (Dk\, k^{-1})^{ab} &=&  2 (\eta -F)^{-1\, ac} DF_{cd} (\eta+F)^{-1\, db}
=
2 DF_{cd}\,  (\eta +F)^{-1\, ca} (\eta+F)^{-1\, db}
  \;  .
 \qquad
\end{eqnarray}
Multiplying the left and the right hand side of (\ref{Dhh=}) with
$\gamma_a \,(\eta - F)^{-1\, ab}$ we get
\begin{eqnarray}\label{Dhs(1-F)}
  D_bh\, h^{-1} \gamma_{a}(\eta - F)^{-1\, ab} =\hspace{300pt}
& \nonumber \\
&\nonumber \\
  {} \qquad =  K_b{}^{ai}
(\delta-FF)^{-1}{}_a{}^b  \,  \gamma^i  - K_{[ab]}^{\;\;
\,\;  i} (\eta+F)^{-1\, ac}(\eta+F)^{-1\, bd} \;
 \gamma_{cd}\gamma^i  - \qquad \nonumber \\
&\nonumber \\
  {} \qquad {} \; -
D^aF_{bc}\, (\delta-FF)^{-1}{}_a{}^b(\eta+F)^{-1\, cd}\, \gamma_{d}
\qquad \nonumber &\\
&\nonumber \\
 {} \qquad {} \; + {1\over 2}D_{[a^\prime }F_{b^\prime c^\prime]}
(\eta+F)^{-1\, a^\prime a}(\eta+F)^{-1\, b^\prime b}(\eta+F)^{-1\,
c^\prime c} \, \gamma_{abc}
 \; .&
 \qquad
\end{eqnarray}
We see that the right hand side of eq. (\ref{Dhs(1-F)}) is expressed
as a polynomial in the linearly independent products of the matrices
$\gamma^a$ and $\gamma^i$. Therefore, if eq. (\ref{Beq(h)flux0})
holds, each term in the right hand side of (\ref{Dhs(1-F)}) must
vanish separately. But these are nothing but the bosonic equations
and the Bianchi identities of the Dp--brane in
 backgrounds with  vanishing dilaton and vanishing background fluxes.
Indeed, the first term in the {\it r.h.s.} of (\ref{Dhs(1-F)}) is
the {\it l.h.s.} of the equation of motion of the bosonic field
$x^{\underline m}$ written in terms of the (generalized) trace of
the second fundamental form (\ref{1str})--(\ref{cbfi1}). The second
term in (\ref{Dhs(1-F)}) contains the anti--symmetric part of the
second fundamental form $K_{[ab]}^{\;\; \,\;  i}$ (\ref{anti1})
which is zero in the pure bosonic background. The fourth term is the
l.h.s. of the Bianchi identity (\ref{bianchi}) which is zero when
the pullback of $H_3$ is zero and the third term contains the BI
equation (\ref{DBI=d*F2string}) (in which the dilaton and the
backgound fluxes are set to zero) and the Bianchi identity
(\ref{bianchi}). To see this one should use the identity
\begin{eqnarray}\label{D(1+F)-1}
D_b \left(  (\eta + F)^{-1\, [bc]\,}\sqrt{|\eta + F|}
\right)\,(\delta - F)_{c}{}^a
 & \equiv  -\sqrt{|\eta + F|} \left[ D^bF_{cd} (\delta - FF)^{-1}{}_b^{\;c} (\eta + F)^{-1}{}^{da} +
 \right. \hspace{20pt} \nonumber \\
\\
  &   \left. \quad + {3\over 2} D_{[b}F_{cd]}
 (\eta + F)^{-1 [bc]} (\eta - F)^{-1 [de]}\,(\delta -
 F)_{e}{}^a\right].\hspace{20pt}\nonumber
\end{eqnarray}

In the presence of the dilaton and the background fluxes the bosonic
equations (\ref{DBI=d*F2string}), (\ref{bianchi})  and (\ref{cbfi})
acquire the corresponding contributions. Hence, the {\it r.h.s.} of
(\ref{Dhs(1-F)}) is not zero anymore and eq. (\ref{Beq(h)flux0}) is
not valid. But, as one can show using the form of the bosonic
equations of motion (\ref{cbfi}) and (\ref{bianchi}), the dilaton
and background flux contributions arrange themselves in such a way
that the matrix $h$ satisfies the Dirac--like equation
(\ref{Beq(h)}) with the generalized covariant derivative ${\mathbf
D}_a$ (\ref{mathcalDf016}) (see the Appendix for its explicit form).
We have thus demonstrated that eqs. (\ref{Beq(G)}), (\ref{Beq(h)})
for the $\kappa$--symmetry projector contain in a compact form all
the bosonic field equations of the superbrane when the supergravity
fermionic fields are zero. When the supergravity background contains
the gravitino and the dilatino, the
 equations (\ref{Beq(G)}), (\ref{Beq(h)}) get modified by the
corresponding contributions of the fermionic fields. In this paper
we shall not elaborate this case in detail.

\setcounter{equation}0
\section{Linearized fermionic equation with couplings to worldvolume and background fluxes. Further analysis and
examples}\label{analysis}

We now go back to the consideration of the linearized fermionic
equations (\ref{D4(G)Th}) and rewrite them explicitly in terms of a
single physical 16--component  fermionic field
$\theta\,(1-\bar\Gamma)$ $\Rightarrow$
$\Theta^{\alpha}(\xi)=(\theta^2-\theta^1\,h)^\alpha$, similar to the
generic equation (\ref{Dirac(Gh)string}), without imposing any
$\kappa$--symmetry gauge fixing condition. Using the definition of
$\bar\Gamma$ (\ref{barG=hh-1}), the generalized covariant derivative
(\ref{mathcalDf0}) and the Dp--brane concise bosonic equations
(\ref{Beq(h)}) we get the following  Dirac--like equation for
16--component fermionic field {\bf in the Einstein frame}
\begin{eqnarray}\label{D(h)th=H,R,P}
D_b\,{\Theta}\,
 \gamma_a\, (\eta + F)^{-1\, ab}  =  - {e^{{{\Phi}}}\over 8} \;  {\Theta} \,
 \left[(p-3) \left( e^{- {\Phi}}\, {{D}\!\!\!\!/\, \Phi}- { {R}}\!\!\! /_{\,\, 1 }
h^{^{T\,-1}} \right) +    2 {R}_b\gamma_a (\eta - F)^{-1\,ab}
h^{^{T\,-1}}
\right. \, \nonumber\\
\nonumber\\
   \left. \qquad
 - e^{-
{\Phi}} \,{{D}\!\!\!\!/\, \Phi} \,\tilde{\gamma}_b
\,\gamma_a\, (F(\eta + F)^{-1})^{ ab} - { {R}}\!\!\! /_{\,\,1} \tilde{\gamma}_b
\,\gamma_a\, (F(\eta - F)^{-1})^{ ab} h^{^{T\,-1}} \right] \;  \nonumber\\
\\
 +
   {e^{{ {\Phi}\over 2}}\over 4} \;  {\Theta} \,
 \left[\left( e^{- {\Phi}} { {H}}\!\!\!\!/ - {p-1\over 2} { {R}}\!\!\!
 /_{\,\,3}\,
 h^{^{T\,-1}} \right) -
{1\over 2}\gamma^{\underline{c}\underline{d}}\, \gamma_a\, (\eta -
F)^{-1\, ab}
 \left( e^{- {\Phi}}  {H}_{b\underline{c}\underline{d}} -
 {R}_{b\underline{c}\underline{d}}h^{^{T\,-1}} \right)   - \right.\nonumber\\
\nonumber\\
\left. - {1\over 2} { {R}}\!\!\! /_{\,\,3} \tilde{\gamma}_b
\,\gamma_a\, (F(\eta - F)^{-1})^{ ab} h^{^{T\,-1}} \right]    \;
 - {1\over {8\cdot 4!}} \;  {\Theta} \,
 {R}_{b\underline{c}_1\ldots \underline{c}_4}{}\gamma^{\underline{c}_1\ldots
\underline{c}_4}\gamma_a
 ((\eta - F)^{-1})^{
ab} h^{^{T\,-1}}  \; , \nonumber
\end{eqnarray}
where $\Theta^{\alpha}(\xi)=(\theta^2-\theta^1\,h)^\alpha$ is a
single 16--component combination of the two worldvolume fermions and
the ``slashed'' objects are
\begin{equation}\label{slashed} {{D}\!\!\!\!/\,
\Phi}=\gamma^{\underline a}\,D_{\underline a}\,\Phi\,,\quad\quad {
{R}}\!\!\!  /_{\,\,1}=\gamma^{\underline a}\,R_{\underline a}\,,
\quad { {R}}\!\!\!  /_{\,\,3}={1\over {3!}}\,\gamma^{\underline a\,\underline b\,\underline
c}\,R_{\underline a\,\underline b\,\underline c}\,,\quad {
{H}}\!\!\!\!/ ={1\over {3!}}\,\gamma^{\underline a\,\underline
b\,\underline c}\,H_{\underline a\,\underline b\,\underline c}\,
\end{equation}
$\hspace{30pt}(\underline a\,,\underline b\,,\underline
c=0,1,\cdots,9$).

 {\bf In the string frame} (see Subsection \ref{stringf})
$\Theta|_{str}=e^{\Phi\over 8}\,\Theta$ and the fermionic equation
takes the form (where the overall dilaton factor $e^{-\Phi}$ is
brought to the left hand side) \footnote{\hbox{To arrive at
(\ref{D(h)th=H,R,P-string}) we used the relation} \linebreak
$[D_b\,{\Theta}\,
 \gamma_a\, (\eta + F)^{-1\, ab} ]_{str}=e^{-{\Phi\over
8}}\,\Big(D_b\,{\Theta}\,
 \gamma_a\, (\eta + F)^{-1\, ab}+{{p+1}\over 8}\,\Theta\, {{D}\!\!\!\!/\, \Phi}-
 {1\over 8}\,\Theta\,{{D}\!\!\!\!/\, \Phi}\tilde\gamma_b\,\gamma_a\,(F(\eta+F)^{-1})^{ab}\Big)$.}
\begin{eqnarray}\label{D(h)th=H,R,P-string}
&e^{ -{\Phi}}\,D_b\,{\Theta}\,  \gamma_a\, (\eta + F)^{-1\, ab}  = -
{1\over 8} \;  {\Theta} \,
 \left[ -4 \, e^{- {\Phi}}\, {{D}\!\!\!\!/\, \Phi}-(p-3)\, { {R}}\!\!\! /_{\,\, 1 }
h^{^{T\,-1}}  \right. \, \nonumber\\
&\nonumber\\
&   \left. \qquad +    2 {R}_b\gamma_a (\eta -
F)^{-1\,ab}\,h^{^{T\,-1}}
 - { {R}}\!\!\! /_{\,\,1} \tilde{\gamma}_b
\,\gamma_a\, (F(\eta - F)^{-1})^{ ab} h^{^{T\,-1}} \right] \;  \nonumber\\
\\
& +
   {1\over 4} \;  {\Theta} \,
 \left[\left( e^{- {\Phi}} { {H}}\!\!\!\!/ - {p-1\over 2} { {R}}\!\!\!
 /_{\,\,3}\,
 h^{^{T\,-1}} \right) -
{1\over 2}\gamma^{\underline{c}\underline{d}} \gamma_a\, (\eta -
F)^{-1\, ab}
 \left( e^{- {\Phi}}  {H}_{b\underline{c}\underline{d}} -
 {R}_{b\underline{c}\underline{d}}\,h^{^{T\,-1}} \right)   - \right.\nonumber\\
&\nonumber\\
&\left. - {1\over 2} { {R}}\!\!\! /_{\,\,3} \tilde{\gamma}_b
\,\gamma_a\, (F(\eta - F)^{-1})^{ ab} h^{^{T\,-1}} \right]    \;
 - {1\over {8\cdot 4!}} \;  {\Theta} \,
 {R}_{b\underline{c}_1\ldots \underline{c}_4}{}\gamma^{\underline{c}_1\ldots
\underline{c}_4}\gamma_a
 (\eta - F)^{-1\,
ab} h^{^{T\,-1}}  \; . \nonumber
\end{eqnarray}
Equations (\ref{D(h)th=H,R,P}) and (\ref{D(h)th=H,R,P-string})
contain all possible couplings of the Dp--brane fermions
 to the worldvolume and type IIB background fluxes linear in fermions.

This is however not the end of the story, because in its left hand
side the covariant derivative ${ D}_a=E_a^{\,\underline
b}\,\partial_{\underline b}-{1\over 4}\,\Omega_a^{\,\underline
b\underline c}\gamma_{\underline b\underline c}$ is not a proper
induced covariant derivative in the worldvolume. It contains the
pullback of the $Spin(1,9)$ connection rather than the induced
$Spin(1,p)\times Spin(9-p)$ connection of the worldvolume theory.
Therefore, to complete the derivation of the explicit form of the
Dirac equation for the brane fermions we should extract from ${
D}_a$ the proper worldvolume covariant derivative. To this end,
using the harmonic relations (\ref{lh}) and (\ref{oc}) we rewrite ${
D}$ as follows \footnote{We recall that in our mostly minus metric
convention $\eta_{\underline a\underline
b}=(\eta_{ab}\,,-\delta_{ij})$.}
\begin{eqnarray}\label{wwcd}
{  D}=d-{1\over 4}\,\Omega^{\,\underline b\underline
c}\gamma_{\underline b\underline c}=d-{1\over 4}\,\Omega^{\, b
c}\,\gamma_{ b  c}-{1\over 4}\,\Omega^{\,ij}\,\gamma_{ij}+{1\over
2}\,\Omega^{\,
b\,i}\gamma_{b}\,\tilde\gamma^{i}\nonumber\\
\\
={\mathcal D}+{1\over 2}\,K^{\,
b\,i}\,\gamma_{b}\,\tilde\gamma^{i}+{1\over 4}\,u^{\,\underline
a\,\underline b}\,d\,u_{\underline a}^{\,\,\underline
c}\,\gamma_{\underline b\underline c}
\,,\nonumber
\end{eqnarray}
where $K^{\,\, b\,i}=u^a\,D\,u^i=u^a\,du^i+\Omega^{\,\, b\,i}$ is
the second fundamental form defined in
 eq. (\ref{2ndf}),  $
\gamma^a=\gamma^{\underline a}\,u_{\underline a}^{\,\,a}\,,$
$\gamma^i=\gamma^{\underline a}\,u_{\underline a}^{\,\,i}$ and
\begin{equation}\label{wwcd1}
{\mathcal D}_a=E_a^{\,\underline b}\,\partial_{\underline b}-{1\over
4}\,\omega_a^{\, b  c}\gamma_{  b  c}-{1\over
4}\,A_a^{\,ij}\,\gamma_{ij}
\end{equation}
is the proper worldvolume covariant derivative with the $SO(1,p)$
spin connection  $\omega^{\,bc}$ and the R--symmetry $SO(9-p)$
connection $A^{\,ij}$ induced by the embedding. They are related to
the pullback of the $SO(1,9)$ connection as follows (see
\cite{se,Sorokin:1999jx} for details)
\begin{eqnarray}\label{omA}
\omega^{\, b  c}=\Omega^{\,\underline b\underline
c}\,u_{\underline b}^{\,b}\,u_{\underline c}^{\,c} -u^{\,\underline
a \,b} du_{\underline a}^{\,c}\,,\qquad
A^{\,ij}=\Omega^{\,\underline b\underline c}\,u_{\underline
b}^{\,i}\,u_{\underline c}^{\,j} -u^{\,\underline a
\,i} du_{\underline a}^{\,j}\,.
\end{eqnarray}

The inhomogeneous term $u\,d\,u$ in (\ref{wwcd}) is due to the
$SO(1,9)$ Lorentz transformation of the vector vielbein
$E^{\underline a}$ performed to adapt it to the brane worldvolume,
as was explained in Subsection \ref{framelike}. At the same time,
the spinorial objects like $E^{\alpha}$ and $\Theta^\alpha$ have not
yet been subject to the corresponding $Spin(1,9)$ transformation
associated with $u_{\underline a}^{\,\underline b'}$. Such a
transformation is performed by a $16\times 16$ matrix
$v_{\alpha}^{\,\beta'}$ of $Spin(1,9)$ (which defines spinor Lorentz
harmonic \cite{BZstr,BZ-p,se}) and which is determined by the
conventional relation between the vector and the spinor
representations of the Lorentz group
\begin{equation}\label{vs}
v\,\gamma^{\underline a}\,v^T\,u_{\underline a}^{\,\,\underline
b'}=\gamma^{\underline b'}=(\gamma^a,\,\gamma^i)\,.
\end{equation}
From this relation it follows, in particular, that
\begin{equation}\label{uduvdv}
{1\over 4}\,u^{\,\underline a\,\underline b}\,d\,u_{\underline
a}^{\,\,\underline c}\,\gamma_{\underline b\underline
c}=v\,d\,v^{-1}\,.
\end{equation}
Now, let us multiply both sides of eq. (\ref{D(h)th=H,R,P-string})
by $v^T$ from the right, use the relations (\ref{vs}),
(\ref{uduvdv}) and make the following redefinitions
$$
\Theta\,v^{-1}\,\,\Rightarrow \,\,\Theta\,,\qquad
 v\,\gamma^{\underline a}\,v^T\,u_{\underline a}^{\,\,a}=\gamma^a\,,
\qquad  v\,\gamma^{\underline a}\,v^T\,u_{\underline a}^{\,\,i}
=\gamma^i\,,\qquad v^{-1}hv \,\,\Rightarrow \,\,
h_\alpha{}^{\beta}\,,
$$
where upon the $Spin(1,9)$ rotation the matrices $\gamma^a$ and
$\gamma^i$ are {\it constant} as a consequence of (\ref{vs}). As a
result, with this new definitions, the right hand side of the
fermionic equation (\ref{D(h)th=H,R,P-string}) does not change its
form while its left hand side becomes
\begin{equation}\label{de}
e^{ -{\Phi}}\,D_b\,{\Theta}\,  \gamma_a\, (\eta + F)^{-1\, ab}
\quad \Rightarrow \quad e^{ -{\Phi}}\,{\cal D}_b\,{\Theta}\,  \gamma_a\, (\eta + F)^{-1\,
ab} -{1\over 2}\, e^{ -{\Phi}}\,K_b^{\, c\,i}\,{\Theta}
\,\gamma^{i}\tilde\gamma_{c}\,\gamma_a\,(\eta + F)^{-1\, ab}\,.
\end{equation}
Using the Clifford algebra, the second term in the {\it r.h.s.} of
eq. (\ref{de}) can be decomposed as follows
\begin{eqnarray}\label{decompose}
K_b^{\, c\,i}\,{\Theta}
\,\gamma^{i}\tilde\gamma_{c}\,\gamma_a\,(\eta + F)^{-1\, ab}=
\,
K_b{}^{c i}\;  {\Theta}\,
\gamma^i\tilde \gamma_{ca} (\eta + F)^{-1\,  a b} +  K_{ba}{}^{i} (\eta +
F)^{-1\, ab}\;  {\Theta}\, \gamma^i\nonumber\\
\\
= -K_b{}^{c i}\;  {\Theta}\,
\gamma^i\tilde \gamma_{ca} [F(\eta + F)^{-1}]^{a b} +  K_{(ba)}{}^{i} (\eta +
F)^{-1\, ab}\;  {\Theta}\, \gamma^i\,,\nonumber
\end{eqnarray}
where in the right hand side we split $K_{bc}{}^i$ into symmetric
and antisymmetric parts and took into account that $K_{[bc]}{}^i=0$
in a background with zero bulk fermions and in the leading order in
$\Theta$ (see eq. (\ref{anti1})). Note that the pair of {\it upper}
indices $i$ implies the contraction with the unit matrix
$\delta^{ij}$.

One can notice that the last term in (\ref{decompose}) is the left
hand side of the worldvolume scalar equation (\ref{cbfi1}) whose
right hand side contains the contribution of fluxes. Thus this term
brings additional couplings of the worldvolume fermion to fluxes.
Actually, as we shall see on examples these additional terms cancel
the non--Lagrangian flux terms of the fermion equation
(\ref{D(h)th=H,R,P-string}) in which the products of the
gamma--matrices are symmetric $S_{\alpha\beta}=S_{\beta\alpha}$,
{\it i.e.} the terms which (since $\Theta\,S\,\Theta\equiv 0$)
cannot be obtained from the quadratic Lagrangian of
\cite{Marolf:2003vf,Martucci:2005rb}, besides the terms with the
dilaton derivatives and those proportional to the BI field equation
and the Bianchi identity which result from integrating by parts the
kinetic term $L_{kin}=\sqrt{|\eta+F|}\,e^{-\Phi
}\,\Theta\,D_b\gamma_a\,\Theta\,(\eta + F)^{-1\, a b}$.

Finally, taking into account the bosonic equation  (\ref{cbfi1}),
the linear fermion Dirac equation which includes all possible
interactions of D--brane fermions with the worldvolume fields and
the background gravity and fluxes takes the following form
\begin{eqnarray}\label{finalf}
&\quad e^{ -{\Phi}}\,{\cal D}_b\,{\Theta}\,  \gamma_a\, (\eta +
F)^{-1\, ab} +{1\over 2}\, e^{ -{\Phi}}\,K_{(bc)}{}^{i}\; {\Theta}\,
\gamma^i\tilde \gamma^{c}{}_{a} [F(\eta + F)^{-1}]^{a b}=\nonumber\\
&\nonumber\\
&=- {1\over 8} \;  {\Theta} \,
 \left[ -4 \, e^{- {\Phi}}\, {\gamma^a\,D_a\, \Phi}-(p-3)\, { {R}}\!\!\! /_{\,\, 1 }
h^{^{T\,-1}}
\right. \,\nonumber\\
&\nonumber\\
&
  \left. 
+    2 {R}_b\gamma_a (\eta - F)^{-1\,ab} h^{^{T\,-1}}
  - { {R}}\!\!\! /_{\,\,1} \tilde{\gamma}_b
\,\gamma_a\, (F(\eta - F)^{-1})^{ ab} h^{^{T\,-1}} \right] \;  \nonumber\\
\\
& +
   {1\over 4} \;  {\Theta} \,
 \left[\left( e^{- {\Phi}} { {H}}\!\!\!\!/ - {p-1\over 2} { {R}}\!\!\!
 /_{\,\,3}\,
 h^{^{T\,-1}} \right) -
{1\over 2}\gamma^{\underline{c}\underline{d}} \gamma_a\, (\eta -
F)^{-1\, ab}
 \left( e^{- {\Phi}}  {H}_{b\underline{c}\underline{d}} -
 {R}_{b\underline{c}\underline{d}}\,h^{^{T\,-1}} \right)   - \right.\nonumber\\
&\nonumber\\
&\left. - {1\over 2} { {R}}\!\!\! /_{\,\,3} \tilde{\gamma}_b
\,\gamma_a\, (F(\eta - F)^{-1})^{ ab} h^{^{T\,-1}} \right]    \;
 - {1\over {8\cdot 4!}} \;  {\Theta} \,
 {R}_{b\underline{c}_1\ldots \underline{c}_4}{}\gamma^{\underline{c}_1\ldots
\underline{c}_4}\gamma_a
 (\eta - F)^{-1\,
ab} h^{^{T\,-1}}  \; \nonumber\\
&\nonumber\\
&-{1\over 4}\,e^{- {\Phi}}\,H_{i\,ab}\,(\eta+{F})^{-1ab}\;
{\Theta}\,
\gamma_i +{{(-)^{{p+1}\over
2}}\over {2\,(p+1)!\,\sqrt{|\eta+{\mathcal
F}|}}}\,\epsilon^{a_1\cdots a_{p+1}}\,
\;  {\Theta}\,
\gamma_i\,({\mathbb{R}} \, e^{{ {F}}_2})_{i\,a_1\cdots
a_{p+1}}\nonumber\,,
\end{eqnarray}
where the slashed quantities have been defined in (\ref{slashed}).

Let us repeat once again that the above Dirac equations for a single
Majorana--Weyl spinor $\Theta(\xi)$ are $\kappa$--gauge independent,
since they were derived without gauge fixing  $\kappa$--symmetry.

\subsection{On $\kappa$--symmetry gauge fixing}

Some comments about gauge fixing kappa--symmetry are now in order.
As we have already noted, the fermionic field
$\Theta^{\alpha}(\xi)=(\theta^2-\theta^1\,h)^\alpha$ is
$\kappa$--invariant in the leading order in $\theta$, while another
linearly independent combination of $\theta^1$ and $\theta^2$, {\it
e.g.} $\eta^{\alpha}(\xi)=(\theta^2+\theta^1\,h)^\alpha$ transforms
under the $\kappa$--symmetry (\ref{kappa}) as a Goldstone fermion
\begin{equation}\label{kappaline}
\delta_\kappa\, \eta^{\alpha}(\xi)=\delta_\kappa\,
(\theta^2+\theta^1\,h)^\alpha =
\kappa^\alpha(\xi):=2(\kappa^1+\kappa^2\,h)^\alpha\,+{\mathcal
O}(\theta\,\theta)\,.
\end{equation}
We can use (\ref{kappaline}) to gauge fix $\eta^{\alpha}(\xi)$ to
zero. This is the most natural choice of the $\kappa$--symmetry
gauge fixing condition which is always consistent with the
superbackground where the brane moves, since the properties of the
background are encoded in the form of the matrix
$h_\alpha^{~\beta}$, as we discussed in Section \ref{conciseboson}.
Such a gauge fixing was used for the analysis of the Dirac equation
for M5 and D3 branes with fluxes in
\cite{Kallosh:2005yu,Saulina:2005ve,Kallosh:2005gs,Bergshoeff:2005yp}.

One also often uses an alternative D=10 covariant gauge fixing
condition \cite{Aganagic:1996nn}, {\it e.g.}
$\theta^1=0\,\Rightarrow\,\Theta=\theta^2$, however, in this case
one should be careful and check whether this condition is consistent
with the background chosen, {\it e.g.} in
\cite{Bergshoeff:2005yp} it was shown that such a condition can be
inconsistent with orientifold compactifications.

We shall now analyze the structure of the Dp--brane fermion
equations in different cases. We shall work in the string frame
only. \label{withoutflux}

\subsection{Equations for all type IIB Dp--branes in the absence of the worldvolume flux. String frame}
Putting to zero the worldvolume field strength $F_2$ must be
compatible with the equation of motion of the BI field
(\ref{DBI=d*F2string}) and the Bianchi identity (\ref{bianchi}). The
Bianchi identity implies that ${\mathcal F}_2$ (which in the string
frame coincides with $F_2$) can be zero if the brane is placed into
the background in such a way  that the pullback of the NS--NS flux,
along the brane worldvolume is zero
\begin{equation}\label{Habc=0}
H_{abc}=0\,,
\end{equation}
while the BI equation of the Dp--brane implies that the pullback of
the RR p--form flux $R_{p}$ along the brane worldvolume, which is a
source for $F_{ab}$ must also be zero
\begin{equation}\label{Rp=0}
R_{a_1\cdots a_p}=0\,.
\end{equation}
In such a case the bosonic scalar field equation (\ref{cbfstring})
is
\begin{eqnarray}\label{cbf0}
{1\over{\sqrt{|g|}}}\,D_m\,\Big(e^{-\Phi}\,\sqrt{|g|} \,g^{mn}\,E_{n\underline
a}\Big) +e^{ -
\Phi}\,E^{\,M}_{\underline a}\partial_M\,\Phi\,=-{{(-)^{{p+1}\over
2}\,}\over{(p+1)!\,\sqrt{|g|}}}\,\epsilon^{m_1\cdots m_{p+1}}\,
{{R}}_{\underline a\,m_1\cdots m_{p+1}}\, \hspace{10pt}
\end{eqnarray}
or in terms of the second
fundamental form (\ref{2ndf})
$$
K_a{}^{ai}=D^i\Phi+ {{(-)^{{p+1}\over
2}\,}\over{(p+1)!}}\,\epsilon^{a_1\cdots a_{p+1}}\,
{{R}}^i{}_{a_1\cdots a_{p+1}}\,.
$$

Taking into account the simplification of the $\kappa$--symmetry
projector (\ref{h0=}), the Dirac equation (\ref{finalf})  reduces to
\begin{eqnarray}\label{DF0-string}
e^{ -{\Phi}}\,{\cal D}_a\,{\Theta}\,  \gamma^a\,  =  {1\over 2} \;
{\Theta} \,
 \left(  e^{- {\Phi}}\, {\gamma^a\,{D}_a \Phi}+{{p-3}\over 4}\, {
 {R}_{\underline a}\,\gamma^{\underline a}
 }\,
\bar\gamma   -  {1\over  2}\, {R}_a\gamma^a\, \bar\gamma\right)+
\nonumber\\
\\
 +
   {1\over 4} \;  {\Theta} \,
 \left[{1\over{3!}}\,\gamma^{\underline{abc}}\,( e^{- {\Phi}} \, {H}_{\underline{abc}}- {p-1\over 2}\,{R}_{\underline{abc}}\,
\bar\gamma) -
{1\over 2}\,\gamma^{\underline{c}\underline{d}\,a}
 ( e^{- {\Phi}}  {H}_{a\,\underline{c}\underline{d}} -
 {R}_{a\,\underline{c}\underline{d}}\,\bar\gamma )  \right]    \;
 \nonumber\\
\nonumber\\
-{1\over {8\cdot 4!}} \;  {\Theta} \,
 {R}_{a\,\underline{c}_1\cdots \underline{c}_4}{}\gamma^{a\,\underline{c}_1\cdots
\underline{c}_4}\, \bar\gamma  \;
+{{(-)^{{p+1}\over 2}}\over {2\cdot (p+1)!}}\;
{\Theta}\,\epsilon^{a_1\cdots a_{p+1}}\, {R}_{i\,a_1\cdots
a_{p+1}}\gamma_i\,.\nonumber
\end{eqnarray}
We recall that $\bar\gamma= {1\over (p+1)!}\, \epsilon_{a_0a_1\cdots
\, a_p} \tilde\gamma^{a_0} \cdots \, \gamma^{a_p}$, as defined in
(\ref{h0=}), $\underline a=(a,i)$, the indices $a$ are those of the
worldvolume and the index $i$ corresponds to the directions
orthogonal to the brane. The pair of indices $i$ which are both
lower (or upper) ones denotes the contraction with the unit matrix
$\delta_{ij}$, as in the last term of (\ref{DF0-string}).

Note that in agreement with our arguments below eq.
(\ref{decompose}), except for the dilaton derivative term, all
products of the gamma matrices in the right hand side of
(\ref{DF0-string}) are antisymmetric, as we shall now see on the
examples.

\subsubsection{D1 brane}
In the case of the D--string (see {\it e.g.} \cite{B00}) the
worldvolume field strength has only one component
$F_{ab}=\epsilon_{ab}\,F$ and the Born--Infeld equation simplifies
to
\begin{equation}\label{d1bi}
{\partial}_b\,\left( e^{-\Phi}\,{F\over{\sqrt{1-F^2}}}\right)=R_b =
{\partial}_b\,C_0\quad
\Rightarrow \qquad  F=e^\Phi\,{{c+C_0}\over\sqrt{1+(c+C_0)^2\,e^{2\Phi}}}\,,
\end{equation}
where $C_0(x)$ is the axion and $c$ is an integration constant.

 The worldvolume scalar field equation (\ref{1str}), (\ref{cbfi1}) is
$$
2 e^{-\Phi}\,K_a{}^{ai}=2\,e^{-\Phi}\,(1-F^2)\,{
D}^i\,\Phi+e^{-\Phi}\,F\,H^i{}_{ab}\,\epsilon^{ab}-\sqrt{1-F^2}\,R^i{}_{ab}\,\epsilon^{ab}+2F\,\sqrt{1-F^2}\,R^i\,
\quad (i=1,\cdots 8)\,.
$$
One can see that $F=0$ requires that $R_a=0$.

 The Dirac equation for a D1 brane with $F_{ab}=0$ is
\begin{eqnarray}\label{D1}
e^{ -{\Phi}}\,{\cal D}_a\,{\Theta}\,  \gamma^a\,  =  {1\over 2} \;
{\Theta} \,
 \left(  e^{- {\Phi}}\, {\gamma^a\,{D}_a \Phi}+{{1}\over 2}\, {
 {R}_{i}\,\gamma_{i}
 }\,
\bar\gamma  \right)-{1\over {2\cdot 4!}} \;  {\Theta} \,
 \epsilon^{ab}\,{R}_{ab\,i_1i_2 i_3}{}\gamma_{\,i_1i_2
i_3} \;
\nonumber\\
\\
-
   {1\over 8} \;  {\Theta} \,
 \left(
\,e^{- {\Phi}} \, \epsilon^{ab}\,{H}_{{abi}}\,\gamma_i\bar\gamma
+{1\over{3}}\,e^{- {\Phi}} \,
{H}_{{ijk}}\,\gamma_{{ijk}}-{R}_{a\,ij}\,\epsilon^{ab}\,\gamma_{b}\,\gamma^{ij}\,
  \right)    \;,\qquad \bar\gamma={1\over 2}\epsilon_{ab}\,\tilde\gamma^{ab}\,.\nonumber
\end{eqnarray}

\subsubsection{D3 brane (without and with a topological Born--Infeld flux on K3)}\label{D3i}
The D3 brane worldvolume field strength can be zero if $H_{abc}$ and
$R_{abc}$ are zero along the brane. The Dirac equation for the D3
brane,
 which in the case of the constant dilaton--axion
coincides with the equation derived in
\cite{Tripathy:2005hv,Bergshoeff:2005yp}, is
\begin{eqnarray}\label{D3}
e^{ -{\Phi}}\,{\cal D}_a\,{\Theta}\,  \gamma^a\,  =  {1\over 2} \;
{\Theta} \,
    \gamma^a\,\left(e^{- {\Phi}}\,{D}_a \Phi -  {1\over  2}\, {R}_a \bar\gamma\right)
   -{1\over {4\cdot 4!}} \;  {\Theta} \,
 {R}_{a\,i_1\cdots i_4}{}\gamma^{a\,i_1\cdots
i_4}\, \bar\gamma  \;
\nonumber\\
\\
 +
   \;  {\Theta} \,
 \left({1\over 8}\,\gamma^{ab}\,\gamma_i\,
 ( e^{- {\Phi}}  {H}_{abi} -
 {R}_{abi}\,\bar\gamma )  -{1\over {4!}}\gamma_{ijk}\,( e^{- {\Phi}} \, {H}_{ijk}
 - \,{R}_{ijk}\,
\bar\gamma)
 \right)    \;
 ,\qquad \bar\gamma={1\over 4!}\,\epsilon_{abcd}\,\tilde\gamma^{abcd} \,.\nonumber
\end{eqnarray}

Though, even if the background sources for the Born--Infeld field
{\it i.e.} $H_{abc}$ and $R_{abc}$ are zero, in general, the BI
field strength $F_2$ may acquire non--zero values when D3 branes
wrap a topologically nontrivial manyfold.

Consider, {\it e.g.} a D3--brane instanton wrapping the
4--dimensional manifold $K3$ in type IIB orientifold
compactification on $M_4\times K3\times T^2/Z^2$. Without the
worldvolume flux this example was considered in
\cite{Bergshoeff:2005yp,Park:2005hj}. The Dirac equation derived
therein has (in a constant dilaton--axion background and in the
absence of the $R_5$ flux) the form
\begin{equation}\label{d3fl}
{\cal D}_a\,{\Theta}\,  \gamma^a\,  = {1\over
8}\,\Theta\,{\tilde G}_{ab I}\,\gamma^{ab}\,\gamma_{I}\,,
\end{equation}
where $I=1,2$ is the index corresponding to $T^2$ and $a,b=1,2,3,4$
are the indices of the D3 brane worldvolume which coincides with
$K3$ (we consider the case of a static D3 which does not oscillate
in the directions of the physical space $M_4$) and
\begin{equation}\label{G}
{\tilde G}_{abI} \equiv H_{abI} - i  R_{abI}\,\bar\gamma\,.
\end{equation}
The imaginary unit $i$ appeared in (\ref{G}) because of the Wick
rotation to Euclidean space \footnote{Note that in contrast to
\cite{Bergshoeff:2005yp} in eq. (\ref{G}) $R_3$ enters with the
minus sign. This is because in our present convention for the Wick
rotation to Euclidean space $\epsilon^{abcd} \rightarrow
i\epsilon^{abcd}$, $\bar\gamma \rightarrow i\bar\gamma$ and
Euclidean $\gamma_a$ satisfy the anti--commutation relations
$\gamma_a\tilde
\gamma_b+\gamma_b\tilde \gamma_a=-\delta_{ab}$. }.

In a complex coordinate system, the flux on $K3\times T^2/Z^2$ has
the following form \cite{Tripathy:2002qw}
\begin{equation}
  G_3= R_3-iH_3 = c \,\Omega_2\wedge d\ov z \ +G_z\wedge d z,
\label{fluxD3}
\end{equation}
where  $c$ is a constant, $\Omega_2$ is the holomorphic self--dual
(2,0)--form on $K3$, $G_z$ is a closed  anti--self--dual (1,1)--form
on $K3$ and ($z,\ov z$) are complex coordinates on $T^2$. In
\cite{Tripathy:2002qw} it was shown that for the compactification of
$K3\times T^2/Z^2$ with the flux $G_3$ to preserve supersymmetry,
either $G_z=0$ or $c=0$, so that $G_3$ is either self--dual or
anti--self--dual. This prompts us also to consider a Born--Infeld
instanton field on the D3--brane instanton wrapping $K3$. This BI
instanton can be probably related to the presence of a worldvolume
field on D7 branes (filling $M_4$ and wrapping $K3$) which
participate in the tadpole cancelation and gaugino condensation
\cite{Tripathy:2002qw,luis}.

In the presence of a non--zero worldvolume field $F_2$ the D3--brane
worldvolume scalar field equation (\ref{cbfi1}) is
\begin{eqnarray}\label{D3K3}
&K_{b}^{\,\,\,\,\,ai}\,(\delta+{ F})_a^{-1\,b}=
\delta^{i}_I\,\left({1\over 2}\,H^I{}_{ab}\,(\delta-{  F})^{-1ab}
+{i\over {4\,\sqrt{|\delta+{ F}|}}}\,\epsilon^{a_1\cdots a_{4}}
R^I{}_{a_1a_2}\, F_{a_3a_4}\right)\,,
\end{eqnarray}
where  the imaginary unit $i$ appeared because of the Wick rotation,
the indices  $i=1,\cdots,6$ are those of the directions orthogonal
to D3 (and $K3$) and $I=1,2$ are those of $T^2$.

Remember that in the static gauge in which $p+1$  coordinates of the
target space are identified with those of the worldvolume,
$x^m=\xi^m$,  the second fundamental form has the form (\ref{2ndff})
\begin{equation}\label{2ndff1}
K_{ab}^{\,\,\,i}=-E_a{}^m\,E_b{}^n(\partial_m\,\partial_n\,x^{\tilde
i}\,E_{\tilde i}{}^i+\Gamma_{mn}{}^{\underline l}\,E_{\underline
l}{}^i+\partial_m\,x^{\tilde i}\,\Gamma_{n\tilde i}{}^{\,\underline
l}\,E_{\underline l}{}^i),
\end{equation}
where $\Gamma_{mn}{}^{\underline l}$ and $\Gamma_{n\tilde
i}{}^{\,\underline l}$ are components of the target space
Christoffel symbol and \linebreak \hbox{ $dx^{\underline
l}\,E_{\underline l}{}^i=(dx^m E_{m}{}^i,\,dx^{\tilde i}\,E_{\tilde
i}{}^i)$} are components of the target space vielbein in the
$i$--directions orthogonal to the brane worldvolume.

If we assume that, like in the case of $F_2=0$, the D3--brane wraps
$K3$ smoothly, {\it i.e.} it is static and does not fluctuate in
transverse directions, $\partial_m\,x^{\tilde i}=0$, the second
fundamental form of the $D3$ worldvolume is zero (since
$\Gamma_{mn}{}^{\underline l}\,E_{\underline l}{}^i=0$ for
$M_4\times K3\times T^2/Z^2$)
 so
the left hand side of (\ref{D3K3}) must vanish, and we are left with
the condition
\begin{equation}\label{f2sd}
{i\over {2\,\sqrt{|\delta+{ F}|}}}\,\epsilon^{a_1\cdots a_{4}}
R_{a_1a_2I}\,  F_{a_3a_4}+\,H_{abI}\,(\delta-{ F})^{-1ab} =0\,.
\end{equation}
In the case of a weak $F_2$ field we can restrict eq. (\ref{f2sd})
to the linear order in $F_2$
\begin{equation}\label{f2sdl}
{1\over 2}\,\epsilon^{a_1\cdots a_{4}} R_{a_1a_2I}\,
F_{a_3a_4}-i\,H_{abI}\, { F}^{ab} =0\,.
\end{equation}
Consider now the case of the compactification with the flux
(\ref{fluxD3}) be self--dual in $K3$, {\it i.e.} $G_z=0$. Then both
$R_3$ and $H_3$ are {\it self--dual}, and eq. (\ref{f2sdl}) takes
the form
\begin{equation}\label{f2sdlsd}
(R_{abI}-i\,H_{ abI})\, { F}^{ab} = G_{abI}\,{ F}^{ab}=0\,,
\end{equation}
from which it follows that for eq. (\ref{f2sdlsd}) to be satisfied,
$F_2$ must be {\it anti--self--dual},\linebreak
 \hbox{$
F_{ab}=-{1\over 2}\,\epsilon_{abcd}\,F^{cd}$}.  On the other hand,
if $G_3$ in (\ref{fluxD3}) is  {\it anti--self--dual} in $K_3$, {\it
i.e.} $c=0$, the equation (\ref{f2sdl}) takes the form
\begin{equation}\label{f2sdlasd}
(R_{abI}+i\,H_{abI})\, { F}^{ab} = {\bar G}_{abI}\,{ F}^{ab}=0\,.
\end{equation}
It is satisfied if $F_2$ is {\it self--dual}, $ F_{ab}={1\over
2}\,\epsilon_{abcd}\,F^{cd}. $ In both of the cases, because of
(anti)--self--duality of $F_2$ and of the Bianchi identity $dF_2=0$,
the linearized BI equation on $K3$ is identically satisfied, ${\cal
D}_a\,F^{ab}=0$.

Away from the linear approximation, the (anti)--self--duality
condition on $F_2$ seems to become non--linear
\begin{equation}\label{nlsd}
\mp{1\over 2}\,\epsilon^{abcd}\,F_{cd}=(\delta+{
F})^{-1[ab]}\,\sqrt{|\delta+{F}|}\,,
\end{equation}
\hbox{which  in view of the Bianchi identity $dF_2=0$ implies the BI
equation}\linebreak
 \hbox{ ${\mathcal D}_a\,(\sqrt{|\delta+{
F}|}\,(\delta+{ F})^{-1[ab]})=0$}.

However surprisingly, one can verify that the non--linear
self--duality relation (\ref{nlsd}) actually reduces to the linear
one $ F_{ab}=\pm {1\over 2}\,\epsilon_{abcd}\,F^{cd}.$ The simplest
way to see this is to choose in a given point $\xi$ on the D3 brane
worldvolume a special $SO(4)$ frame in which the only non--zero
components of the {\it Euclidean} $F_{ab}$ are $F_{12}=-F_{21}$ and
$F_{34}=-F_{43}$. In this frame the {\it r.h.s.} of eq. (\ref{nlsd})
takes the form
$$
(\delta+{ F})^{-1[ab]}\,\sqrt{|\delta+{
F}|}=-{{F^{ab}+F_{12}\,F_{34}\,F^{*ab}}\over
{\sqrt{1+F^2_{12}}\,\sqrt{1+F^2_{34}}}}\,.
$$
Substituting this expression into (\ref{nlsd}) we get the
conventional linear self--duality condition
$$
\pm F^*_{ab}={{F_{ab}+F_{12}\,F_{34}\,F^*_{ab}}\over
{\sqrt{1+F^2_{12}}\,\sqrt{1+F^2_{34}}}}\,
\quad\Rightarrow\quad
{{F_{12}}\over{\sqrt{1+F^2_{12}}}}=\pm{{F_{34}}\over{\sqrt{1+F^2_{34}}}}
$$
\begin{equation}\label{sd}
\quad\Rightarrow\quad F_{12}=\pm F_{34} \quad\Rightarrow\quad
F_{ab}=\pm{1\over 2}\,\epsilon_{abcd}\,F^{cd}\,.
\end{equation}

 In the case of a non--zero generic BI field and in the
setup under consideration, eq. (\ref{f2sd}) is not satisfied. This
happens, for instance, when {\it both}, the $G_3=R_3-iH_3$ and
$F_{ab}$ are either {\it self--dual} or {\it anti--self--dual} on
$K3$. The situation may be cured if in the compactification
background there is also a non--zero $R_5$ flux.

The presence of $R_5$, in general, leads to a warping of the
compactified space. We shall discuss more examples of this setup in
Section \ref{warped}. A generic form of the part of $R_5$ along the
tangent space of the compactified manifold $X_6$ (which, by duality,
preserves Lorentz--invariance in effective four--dimensional
space--time) is
\begin{equation}\label{R5wick}
R_{q_1\cdots q_5}=-\epsilon_{q_1\cdots q_5q_6}\,\partial_{q_6} \ln
Z(y)\,,
\end{equation}
where $Z(y^q)$ is a function of $X_6$ coordinates whose square root
is the warp (conformal) factor $Z^{1\over 2}(y)$ of the compactified
space metric (see eq. (\ref{warpedm}) of Section \ref{warped}). In
the Wick rotated space time the self--dual $R_5$ is complex. We can
 make a natural assumption that upon Wick rotation the metric warp factor $Z(y^q)$
and $R_5$ remain real in the {\it a priori} Euclidean $X_6$. If so,
by duality, $R_5$ becomes pure imaginary in the effective
four--dimensional space $M_4$
\begin{equation}\label{R5wick1}
R_{r_1r_2r_3r_4q}=-i\epsilon_{r_1r_2r_3r_4}\,\partial_{q} \ln
Z(y)\quad(r=1,2,3,4).
\end{equation}
 In other words, the imaginary unit appears
in the above expression because of the Wick rotation of the time coordinate in
$M_4$.

In the warped $K3\times T^2/Z^2$ background, the non--fluctuating
(i.e. $\partial_a x^i=0$) D3 brane instanton wrapping the
conformally warped $K3$ has a non--zero second fundamental form
(\ref{2ndff1}) with
the indices $I,J=1,2$ corresponding to $T^2$ 
\begin{equation}\label{2ndnot0}
K_{a}{}^{bI}={1\over 4}\,\delta_{a}^{b}\,\partial^I\,\ln Z\,.
\end{equation}
Then, in accordance with the generic equation (\ref{cbfi1}) the D3
 brane instanton scalar field equation acquires the contribution from $R_5$  and takes the form
\begin{eqnarray}\label{f2sdr5}
{1\over 2}\,\partial_I\,\ln Z\,\delta_{b}^{a}\,
(\delta+F)_a^{-1\,b}\,\sqrt{|\delta+F|}=
\nonumber\\
\\
={i\over {12}}\,\epsilon^{a_1\cdots a_{4}} R_{a_1\cdots
a_4I}+{i\over {2}}\,\epsilon^{a_1\cdots a_{4}} R_{a_1a_2I}\,
F_{a_3a_4}+\,H_{abI}\,(\delta-{ F})^{-1ab}\,\sqrt{|\delta+{ F}|}
\,.\nonumber
\end{eqnarray}
The assumption about the reality of Wick rotated bosonic fields
helps us to find particular solutions of the D3 brane bosonic field
equations. Thus, equation (\ref{f2sdr5}) imposes additional
conditions which relate the warp factor and generic bulk and
worldvolume fluxes
\begin{eqnarray}\label{agf}
{1\over 2}\,\partial_I\,\ln Z(y)\,\delta_{b}^{a}\,
(\delta+F)_a^{-1\,b}\,\sqrt{|\delta+F|}=\,H_{abI}\,(\delta-{
F})^{-1ab}\,\sqrt{|\delta+F|}\,,\nonumber\\
\\
2\,\epsilon_{IJ}\partial_J\,\ln Z(y)={1\over 2}\,\epsilon^{a_1\cdots
a_{4}} R_{a_1a_2I}\, F_{a_3a_4}\,.\nonumber
\end{eqnarray}
We should note that eqs. (\ref{agf}) {\it cannot} be regarded as
differential equations on $\ln Z$, since these equations are valid
on the D3 brane worldvolume at a certain point in $T^2/Z_2$, while
$\partial_I$ is the derivative in $T^2/Z_2$ directions orthogonal to
the worldvolume.

 Consider now the case when $F_{ab}$ is either
self--dual or anti--self--dual, but let us for a moment do not
specify $H_3$ and $R_3$. Now take into account that for a self--dual
or anti--self--dual $F_{ab}$ the equation (\ref{nlsd}) and the
following relations hold
$$
\sqrt{|\delta+F|}=1+{1\over 4} F^2 \,,\qquad
F_{ac}\,F_{b}{}^c={1\over 4}\delta_{ab}\,F^2\,,\qquad (\delta\pm
F)^{-1ab}={{\delta^{ab}\mp F^{ab}}\over{1+{1\over 4} F^2}}
$$
$$
 (\delta+F)^{-1(ab)}\,(1+{1\over 4} F^2)=
\delta^{ab}\,\quad\Rightarrow
\quad \delta_{ab}\,(\delta+F)^{-1ab}\,\sqrt{|\delta+F|}=4\,,
$$
we reduce eqs. (\ref{agf})  to
\begin{equation}\label{agf1}
2\,\partial_I\,\ln Z(y)\,= H_{abI}\,F^{ab}\,,
\qquad
2\,\epsilon_{IJ}\,\partial_J\,\ln Z(y)={1\over
2}\,\epsilon^{a_1\cdots a_{4}} R_{a_1a_2I}\, F_{a_3a_4}\,.
\end{equation}
If $F_{ab}$ is anti--self--dual, while $\Omega_2$ of $G_3=R_3-iH_3$
is self--dual, then the right hand sides of eqs. (\ref{agf1}) vanish
identically, so their left hand sides must also be zero,
$\partial_I\,Z=0$, and hence the warp factor should not depend on
the coordinates on $T^2$.

On the other hand, if $F_{ab}$ is self--dual, e.g. proportional to
$\Omega_2$, then the r.h.s. of (\ref{agf1}) do not vanish, and eqs.
(\ref{agf1}) give a relation between the warp factor and the
topological background and worldvolume fluxes. In addition, if
$F_{ab}$ is self--dual, the conditions (\ref{agf1}) require the
3--form flux $G_3=R_3-iH_3$ to be imaginary self--dual on the whole
six--dimensional manifold (being warped $K3\times T^2/Z^2$ in the
case under consideration), i.e. $G_3=i^*G_3$ or $H_{abI}=-{1\over
2}\,\epsilon_{IJ}\,\epsilon_{abcd} R_{cdJ}$, which is indeed the
case when in (\ref{fluxD3}) the term $G_z$ is zero.

One can turn around the above argument of the necessity to have a
non--zero $R_5$ flux when a generic worldvolume gauge field is
induced. Namely, if one starts from the consideration of the D3
brane instanton on $K3$ without the worldvolume field and switches
on the $R_5$ flux, the letter contributes to the {\it r.h.s.} of the
D3 brane scalar field equation (\ref{cbfstring}), (\ref{cbfi1}).
Since the Wick rotated $R_5$ is complex and the {\it l.h.s.} of eqs.
(\ref{cbfstring}), (\ref{cbfi1}) are assumed to remain real, for
these equations to have a solution in general case one should also
excite a worldvolume gauge field on D3 as in eq. (\ref{f2sdr5}). The
message is that if one considers a D3 brane instanton wrapping a
four--fold of a Calabi--Yau manifold in warped compactifications
with non--zero $R_5$ flux, one should in certain cases take into
account the worldvolume gauge field modes of such an instanton.

It is of interest to analyze (elsewhere) in detail whether the above
instanton configurations, and in particular, the fluxes defined by
eqs. (\ref{agf}) are compatible with Wick rotated supergravity
solutions determining Calabi--Yau compactification backgrounds.


Finally, let us present the Dirac equation for the fermionic modes
on the D3--brane instanton in the presence of the BI instanton or
anti--instanton, $F_2=\pm ^*F_2$, in the case without warping
\begin{eqnarray}\label{deb3ibi}
&-{\cal D}_b\,{\Theta}\,  \gamma_a\, (\delta^{ab} + F^{ ab} )=
   {1\over 8} \;  {\Theta} \,\gamma^{ab}\,\gamma_I\,
 \left((1-{1\over 4}\,F^2)\,H_{abI} - i\,(1+{1\over 4}\,F^2)\,R_{abI}\,
 h^{^{T\,-1}}\right)- \nonumber\\
\\
& -{1\over 4} \;  {\Theta}
 \,\gamma_I\,\gamma^{ab}\,F_{ca}\,
{H}_{bcI} +{1\over 4} \;  {\Theta}
 \,\gamma_I\,H_{abI}\,F^{ab}- {i\over 8} \;  {\Theta}
 \,\gamma_I\,R_{abI}\,F^{ab}\, h^{^{T\,-1}}-{i\over 8} \;  {\Theta}
 \,\gamma_I\,R_{abI}\,^*F^{ab}\, \bar\gamma\,h^{^{T\,-1}}\,.
 &\nonumber
\end{eqnarray}
In the cases in which $H_{abI}$ and $R_{abI}$ have opposite K3
duality in comparison with $F_2=\pm ^*F_2$, {\it i.e.} $H_{abI}=\mp
{1\over 2}\,\epsilon_{abcd}\,H^{cd}{}_{I}$ and $R_{abI}=\mp {1\over
2}\,\epsilon_{abcd}\,R^{cd}{}_{I},$ the last two terms in eq.
(\ref{deb3ibi}) vanish and it reduces to
\begin{eqnarray}\label{deb3ibi1}
&{\cal D}_b\,{\Theta}\,  \gamma_a\, (\delta^{ab} +
F^{ ab} )=\hspace{300pt} \nonumber\\
\\
& =
  - {1\over 8} \;  {\Theta} \,\gamma^{ab}\,\gamma_I\,
 \left({ {H}}_{abI} - i{ {R}}_{abI}\,
 h^{^{T\,-1}}  -
2F_{c[a}\,
 {H}_{b]cI}  -{1\over 4}\,F^2\,({H}_{abI} +i{R}_{abI} \,
 h^{^{T\,-1}})\right)
\;\,,\nonumber
\end{eqnarray}
where $h^{^{T\,-1}}={1\over{1+{1\over 4}\,F^2}}\,(\bar\gamma-{1\over
4}\,\epsilon^{abcd}\,F_{ab}\,\tilde\gamma_{cd}+ {1\over
8}\,\epsilon^{abcd}\,F_{ab}\,F_{cd}\,\cdot{\mathbf 1})$,
$\bar\gamma={1\over 4!}\,\epsilon_{abcd}\,\tilde\gamma^{abcd}$ and
\linebreak
\hbox{$\gamma_a\tilde\gamma_b+\gamma_b\tilde \gamma_a=-\delta_{ab}$}.

\subsubsection{D5 brane}
$F_{ab}$ on the D5 brane can be zero if $H_{abc}$ and $R_{a_1\cdots
a_5}$ are zero. Thus, the D5 brane Dirac equation reduces to
\begin{eqnarray}\label{D5}
e^{ -{\Phi}}\,{\cal D}_a\,{\Theta}\,  \gamma^a\,  =  {1\over 2} \;
{\Theta} \,
   e^{- {\Phi}}\,\left( \gamma^a\,{D}_a \Phi -{1\over 2}  \gamma_i\, {R}_i \bar\gamma\right)
   +{1\over {4!}} \epsilon^{a_1\cdots a_4 bc}\;  {\Theta} \,
 {R}_{i\,a_1\cdots a_4}{}\,\gamma_i\,\gamma_{bc}\,
\nonumber\\
\\
 +
   \; {1\over 8}\, {\Theta} \,
 \left(
  e^{- {\Phi}}  {H}_{abi}\,\gamma^{ab}\,\gamma_i -{1\over {3}}\,e^{- {\Phi}} \,
  {H}_{ijk}\,
 \gamma_{ijk} -
 \gamma^{ij}\,\gamma_a\,{R}_{ija}\,\bar\gamma
 \right)    \;,
 \qquad \bar\gamma={1\over 6!}\,\epsilon_{a_0\cdots a_5}\,\tilde\gamma^{a_0\cdots a_5}\,.\nonumber
\end{eqnarray}

\subsubsection{D7 brane}
In this case for the BI field strength $F_{ab}$ to be zero, it is
necessary that $H_{abc}=0$ and $R_{a_1\cdots a_7}=-{R^*_{ija}}=0$,
and the D7 brane Dirac equation takes the form
\begin{eqnarray}\label{D7}
e^{ -{\Phi}}\,{\cal D}_a\,{\Theta}\,  \gamma^a\,  =  {1\over 2} \;
{\Theta} \,
   e^{- {\Phi}}\,\left( \gamma^a\,{D}_a \Phi+ {1\over 2} \gamma^a\, {R}_a \bar\gamma\right)
+{1\over {2\cdot 4!}}\,{\Theta} \,
\gamma^{abc}\,\epsilon_{ij}R_{abcij}
\;
\nonumber\\
\\
 +
   \; {1\over 8}\, {\Theta} \,
 \gamma^{ab}\,\gamma_i\,
  \left(e^{- {\Phi}}  {H}_{abi} 
-\epsilon_{ij}\,{R}_{jab}
 \right)    \;,
 \qquad \bar\gamma={1\over 8!}\,\epsilon_{a_0\cdots a_7}\,\tilde\gamma^{a_0\cdots a_7}\, .\nonumber
\end{eqnarray}

\subsubsection{D9 brane (with and without worldvolume field)}
The D9  brane is space filling, so there are no orthogonal
directions and the worldvolume indices $a,b,\ldots$ coincide with
target space indices $\underline a, \underline b,\ldots$. The D9
brane always carry a non--zero worldvolume field in the backgrounds
with a non--zero $H_3$ and/or the \hbox{axion $R_1$.} Its Dirac
equation is
\begin{eqnarray}\label{finalfD9}
&\quad e^{ -{\Phi}}\,{\cal D}_b\,{\Theta}\,  \gamma_a\, (\eta +
F)^{-1\, ab}= - {1\over 8} \;  {\Theta} \,
 \left[ -4 \, e^{- {\Phi}}\, {\gamma^a\,D_a\, \Phi}-6\, \gamma^a\,{
 {R}}_a\,
h^{^{T\,-1}}
\right. \,\nonumber\\
&\nonumber\\
&
  \left. 
+    2 {R}_b\gamma_a (\eta - F)^{-1\,ab} h^{^{T\,-1}}
  - \gamma^c\,{
 {R}}_c\,\tilde{\gamma}_b
\,\gamma_a\, (F(\eta - F)^{-1})^{ ab} h^{^{T\,-1}} \right] +\;  \nonumber\\
\\
& +
   {1\over 4} \;  {\Theta} \,
 \left[{1\over 3!}\,\gamma^{abc}\,\left( e^{- {\Phi}} { {H}}_{abc} - 4 { {R}}_{abc}\,
 h^{^{T\,-1}} \right) -
{1\over 2}\gamma^{{c}{d}} \gamma_a\, (\eta - F)^{-1\, ab}
 \left( e^{- {\Phi}}  {H}_{b{c}{d}} -
 {R}_{b{c}{d}}\,h^{^{T\,-1}} \right)   - \right.\nonumber\\
&\nonumber\\
&\left. - {1\over 2\cdot {3!}}\,\gamma^{c_1c_2c_3}\, {
{R}}_{c_1c_2c_3}\, \tilde{\gamma}_b \,\gamma_a\, (F(\eta -
F)^{-1})^{ ab} h^{^{T\,-1}} \right]    \;
 - {1\over {8\cdot 4!}} \;  {\Theta} \,
 {R}_{b{c}_1\ldots {c}_4}{}\gamma^{{c}_1\ldots
{c}_4}\gamma_a
 (\eta - F)^{-1\,
ab} h^{^{T\,-1}} \,\nonumber
\end{eqnarray}
with $h^{^{T\,-1}}$ defined in (\ref{Vxh-1=}).

 The Dirac
equation for the D9  brane  with $F_{ab}=0$ (which requires
$H_{abc}=0$ and R$^*_a=R_{a_1\cdots a_9}=0)$ has the following form
\begin{eqnarray}\label{D9}
e^{ -{\Phi}}\,{\cal D}_a\,{\Theta}\,  \gamma^a\,  =  {1\over 2} \;
{\Theta} \,
   e^{- {\Phi}}\, \gamma^a\,{D}_a \Phi -\; {1\over{4!}}\, {\Theta} \,
 \gamma^{abc} \, {R}_{abc}\; .\nonumber
\end{eqnarray}
Note that in this case $\bar\gamma=  \textbf{1} $ and the term with
$R_5$ vanishes because of the self--duality.

\setcounter{equation}0
\section{Dp--branes with intrinsically non--zero worldvolume flux}\label{warped}

 Let us now consider examples of backgrounds
with fluxes in which the BI field on the brane cannot be put to zero
and thus should be taken into account in the solutions of the
Dp--brane bosonic and fermionic equations of motion. As we have
already mentioned, this happens when $H_{abc}$ and/or
$R_{a_1\cdots\,a_p}$ are non--zero along the brane worldvolume.

 Such a situation may occur in (warped) compactifications
of type IIB string theory with $R_5$ and/or $H_3$ and $R_3$ fluxes
turned on. These include {\it e.g.} the classical $AdS_5
\times X_5$ backgrounds \cite{Schwarz:1983qr,Romans:1984an}
with $X_5$ being a sphere $S^5$ or an Einstein manifold such as the
Sasaki--Einstein manifolds of $S^2\times S^3$ topology. In a certain
coordinate system the $AdS_5
\times X_5$ metric has the following form (in our mostly minus
metric signature)
\begin{equation}\label{adsx}
ds^2=\left({r\over
\rho}\right)^2\,(dx_0^2-dx^idx^i)-\left({\rho\over
r}\right)^2\,dr^2-\rho^2\,ds^2_{X_5}\,, \qquad i=1,2,3
\end{equation}
where $\rho$ is a characteristic radius of $AdS_5$ and $X_5$ (the
square root of the inverse cosmological constant), $r$ is the radial
coordinate of $AdS_5$ and $dx_0^2-dx^idx^i$ is the metric of the
Minkowski space boundary of $AdS_5$ corresponding to a stack of N
near horizon D3 branes.

The $AdS_5\times X_5$ compactifications are triggered by the
non--zero $R_5$ flux which in the local Lorentz frame is\footnote{In
our mostly minus metric convention $\epsilon^{r_1\cdots
r_5\,q_1\cdots q_5}=\epsilon^{r_1\cdots r_5}\,\epsilon^{q_1\cdots
q_5}$, $\epsilon_{r_1\cdots r_5\,q_1\cdots q_5}=-\epsilon_{r_1\cdots
r_5}\,\epsilon_{q_1\cdots q_5}$ and $\epsilon^{1\cdots
5}=\epsilon_{1\cdots 5}=1$.}
\begin{equation}\label{R5on5}
R^{\,AdS_5}_{r_1\cdots r_5}={4\over \rho}\,\epsilon_{r_1\cdots
r_5}\,,\qquad R^{\,X_5}_{q_1\cdots q_5}={4\over
\rho}\,\epsilon_{q_1\cdots q_5}\,.
\end{equation}

In addition to  the $R_5$ flux, $R_1$, $R_3$ and $H_3$ fluxes are
switched on in more general type IIB supersymmetry configurations
with general fluxes (see \cite{Dall'Agata:2004dk,Grana:2005jc} for
review and references therein).

The generic setup is the following. The D=10 space--time is a warped
$M_4
\times_w X_6$ (where $\times_w$
means that because of warping the D=10 manifold is not really the
direct product) manifold with a metric (of mostly minus signature)
\begin{equation}\label{warpedm}
ds^2=Z(y)^{-1/2}\,ds^2_{M_4}-Z(y)^{1/2}\,ds^2_{X_6}
\end{equation}
and with fluxes $H_3$ and $R_3$ turned on in the compactified
manifold $X_6$ (usually taken to be a Calabi--Yau) and with the
self--dual $R_5$ flux of the following form in a local Lorentz frame
\begin{equation}\label{R5}
R_{r_1\cdots r_4\,q}=-\epsilon_{r_1\cdots r_4}\,\partial_{q}\,\ln
Z\,\qquad R_{q_1\cdots q_5}=\epsilon_{q_1\cdots
q_6}\,\partial^{\,q_6}\,\ln Z\,,
\end{equation}
where $Z(y)$ is a warp factor depending on the coordinates $y^q$ of
$X_6$ $(q=1,\cdots\, 6)$ and the indices $r=0,1,2,3$ are those of
the space--time $M_4$ in a local Lorentz frame.  When
$Z(y)={\rho^4\over r^4}$,  $r^2=y_q\,y_q$, the metric
(\ref{warpedm}) reduces to that of $AdS_5\times X_5$.
The explicit form of $H_3$, $R_3$ and of the axion--dilaton depends
on the compactification solution  and will not be given here (see
\cite{Grana:2005jc} for a review and references).

In what follows we shall work in the static gauge $\xi^m=x^m$ for
the brane worldvolume coordinates. We shall restrict ourselves to
situations which allow for static brane configurations, {\it i.e.}
those whose oscillating modes do not depend on time variable
\hbox{$\partial_0\,x^i(\xi)=0$}.

\subsection{D5 branes in $AdS_5\times S^5$ and $AdS_5\times X_5$}\label{d5ads5}
Let us consider  the example of a probe D5 brane wrapping $S^5$ in
the conventional $AdS_5 \times S^5$ background, or  a
(Sasaki--)Einstein space $X_5$ in $AdS_5\times X_5$, and having the
time direction in $AdS_5$. For an $AdS_5$ observer it thus looks
like a particle. The D5 brane wrapping $S^5$ has been under
extensive study from various perspectives
\cite{Imamura:1998gk}--\cite{Gomis:1999xs}, in particular, as a
manifestation of the Hanany--Witten effect \cite{Hanany:1996ie} and
as a baryon vertex of an effective $D=4$ gauge theory
\cite{baryon,Gross:1998gk}.

We add to these results the derivation of the explicit form of the
linearized Dirac equation for the fermionic modes on this
``baryonic" D5 brane.

Since $R_5$ is given by eq. (\ref{R5on5}) and other fluxes are zero,
and if the D5 brane is static and if it does not have modes excited
in transverse directions, {\it i.e.} $\partial_m x^i\,(\xi)=0$, the
D5 brane scalar field equation (\ref{cbfi}), (\ref{cbfi1}) reduces
to
$$
K_{ab}{}^i\,(\eta+F)^{-1ba}=0\,,
$$
where the second fundamental form $K_{ab}{}^i$ describing the
embedding of the $D5$ brane worldvoulme into target space--time has
been defined in (\ref{2ndff}).

The BI equation (\ref{DBI=d*F2string}) takes the form
\begin{eqnarray}\label{DBIADS}
\partial_m \left((g + { F})^{-1\, [ml]}\,\sqrt{ |g + {F}|}
\,
 \right) =
  {{1} \over
5!}\; \epsilon^{l\,m_1\cdots m_{5}} \; {R}_{m_1\ldots
m_{5}}\;,
\end{eqnarray}
where $g_{mn}$ is the (almost minus signature) metric in the D5
brane worldvolume induced by embedding into $AdS_5\times X_5$
(\ref{adsx}) which for the static D5 wrapping $X_5$ is
\begin{equation}\label{d5x5}
ds^2_{D5}=\left({r\over
\rho}\right)^2\,dx_0^2-\rho^2\,ds^2_{X_5}
\end{equation}

Taking into account (\ref{d5x5}) one gets the following BI equation
for the static D5 brane
\begin{eqnarray}\label{DBIADSf}
{\mathcal D}_m \left((g + { F})^{-1\, [ml]}\,\sqrt{ |\eta + {F}|}
\,
 \right) =
  {{1} \over
5!\,\sqrt{|g_{D5}|}} \epsilon^{l\,m_1\cdots m_{5}} \; {R}_{m_1\ldots
m_{5}}={4\over \rho}\, {\rho\over r}\,\delta^{l0}={4\over
r}\, \delta^{l0}\;,
\end{eqnarray}
where $r$ is the radial coordinate of $AdS^5$ (see (\ref{adsx})).
For the static D5 brane $r$ takes a constant value on its
worldvolume, i.e. $\partial{_m}\,r=0$.

We notice that the $R_5$ flux (\ref{R5}) produces the source term to
the electric part of the BI field strength which thus cannot be
zero.

In the linear BI field approximation eq. (\ref{DBIADSf}) reduces to
\begin{equation}\label{DBIstatic}
{\mathcal D}^q\,{  F}^{q0}={4\over r}\,,\qquad {\mathcal D}_{q_2}\,{
F}^{q_2q_1}-\partial_0\,{ F}^{q_1\,0}=0\,,\qquad q=1,\cdots,\,5\,,
\end{equation}
where ${\mathcal D}_q$ is the covariant derivative on $X_5$. We thus
get the Maxwell equation for the electric field $F^{q0}$ on the
$X_5$ surface of the D5 brane which is electrically charged with the
constant charge density $1/r$.

In the case in which the magnetic part $F_{q_1q_2}$ of the
worldvolume field strength is zero \footnote{Effects of a non--zero
self--dual magnetic field $F_{q_1 q_2}$ have been recently studied
in \cite{Yolanda}.} the equations (\ref{DBIstatic}) reduce to the
Poincare equation for the static electric potential $A^0$ on $X_5$
with the constant source
\begin{equation}\label{DBIstatic1}
{\mathcal D}_q\,{\mathcal  D}_{q}\,A^0=\triangle\,A^0={4\over r}\,.
\end{equation}
This equation, as well as the Born--Infeld equation (\ref{DBIADSf})
and the Maxwell equation (\ref{DBIstatic}) (which imply that the
vector $V^q=(\eta + { F})^{-1\, q0}\,\sqrt{ |\eta + {F}|}$ must have
a constant divergence ${\mathcal D}_q\,V^q={1\over r}$) do not have
solutions on the compact manifold $X_5$ and in particular on $S^5$.
The physical explanation of this is that since the D5 brane carries
an effective electric charge there should be an electric flow in a
direction orthogonal to $S^5$. This flow can be associated with N
strings attached to $S^5$, extended along the radial direction of
$AdS_5$ and ended on its Minkowski space boundary \cite{baryon},
which is formed at the near horizon limit by N coincident D3 branes.
The charge of these N strings compensates N unites of the $R_5$ flux
on $S_5$.

This points to the well known fact that the brane cannot smoothly
wrap $X_5$, {\it i.e.} the condition $\partial_m\,x^i(\xi)=0$ is not
consistent and should be relaxed. In particular, if a brane mode is
excited along the radial coordinate $r$ of $AdS_5$ (along which the
N strings joining D5 with the Minkowski boundary are stretched),
$r(\xi)$ becomes a function of the D5 worldvolume coordinates of
$S^5$ and does not depend on $x_0$ ($\partial_0\,r=0$), the source
in the BI equation becomes non--constant and the BI equation with a
properly behaved $r(\xi)$ has (supersymmetric) solutions
\cite{Imamura:1998gk}--\cite{Gomis:1999xs}.

Indeed, the D5 brane worldvolume metric becomes
\begin{equation}\label{d5x5r}
ds^2_{D5}=\left({r\over
\rho}\right)^2\,dx_0^2-\left({\rho\over
r}\right)^2\, \partial_{q_1} r \, \partial
_{q_2}r\,d\xi^{q_1}\,d\xi^{q_2}-\rho^2\,ds^2_{X_5}
\end{equation}
(where $\xi^q$ are the coordinates of $X_5$) and the BI equation
takes the following form
\begin{eqnarray}\label{DBIADSfn}
&{\mathcal D}_m \left((g + { F})^{-1\, [ml]}\,\sqrt{ |\eta + {F}|}
\,
 \right) =
  {{1} \over
5!}\; {\rho\over r}\,{1\over{\sqrt{1+ \left({\rho\over
r}\right)^2\,\partial_q r\,\partial^qr}}}\,
\delta^{l0}\epsilon^{q_1\cdots q_{5}} \; {R}^{X_5}_{q_1\ldots q_{5}}
\nonumber\\
&\\
&\hspace{90pt}={4\over{ r\,\sqrt{1+ \left({\rho\over
r}\right)^2\,\partial_qr\,\partial^q r}}}\,\delta^{l0}\;.\nonumber
\end{eqnarray}
(the indices $q$ in $\partial_qr\,\partial^q r$ are contracted with
the metric ${1\over{\rho^2}}\,g^{q_1q_2}$ on $X_5$). The right hand
side of this equation is non--constant and allows for solutions in
$X_5$.

For such a D5--brane configuration the Dirac equation (\ref{finalf})
takes the form
\begin{eqnarray}\label{D5fAdS}
&  {\cal D}_b\,{\Theta}\,  \gamma_a\, (\eta + F)^{-1\, ab} +{1\over
2}\,K_{(bc)}{}^{i}\; {\Theta}\,
\gamma^i\tilde \gamma^{c}{}_{a} [F(\eta + F)^{-1}]^{a b}=\\
\nonumber\\
& - {1\over {8\cdot 4!}} \;  {\Theta} \,
 {R}_{a\underline b_1\cdots \underline b_4}{}\gamma^{\underline b_1\ldots
\underline b_4}\gamma_c
 (\eta- F)^{-1\,
ca} h^{^{T\,-1}}\,=\nonumber\\
\nonumber\\&=  - {1\over {2\,\rho\,{\sqrt{1+ \left({\rho\over
r}\right)^2\,\partial_qr\,\partial^q r}}}} \; {\Theta}\left[
\,(\eta-F)^{-1b0} +{\rho\over
r}\,\gamma_0\,\tilde\gamma_q\,(\eta-F)^{-1bq}\right]\gamma_b\,\,\bar\gamma\,
 h^{^{T\,-1}}
+\nonumber\\
\nonumber\\
&+{1\over{2\,r}}{\Theta}\, \partial_{\tilde q}\, r\,e^{\,\tilde
q}_a(\xi)\,\, \left[\gamma_b\,(\eta- F)^{-1\, ba}\, - \gamma^{a}
\,\tilde\gamma_{0}\gamma_{b}\,(\eta- F)^{-1\,b0} +
{r\over\rho}\,\gamma^{aq}\,\gamma_0\,\tilde\gamma^b \,(\eta-
F)^{-1\,}_ {bq}\right]\,\gamma^r\,\bar\gamma\,
 h^{^{T\,-1}}
\nonumber\,,
\end{eqnarray}
where $\underline b=(a,i)$, with $a=(0,q)$ being tangent space
worldvolume indices and $q=1,\cdots ,5$ corresponding to $X_5$,
$i=1,2,3,r$ denote the directions transverse to the D5 brane with
the index $r$ being of the radial $AdS_5$ direction,
$e^{\,m}_a(\xi)=(e^{\,\tilde q}_a,e^{\,0}_a)$, $e^{\,\tilde
q}_0=e^{\,0}_q=0$, is the inverse vielbein associated with the
induced worldvolume metric (\ref{d5x5r}), $\bar\gamma={1\over
6!}\,\epsilon_{a_1\cdots a_6}\,\tilde\gamma^{a_1}\cdots\gamma^{a_6}$
and
\begin{eqnarray}
&\bar\gamma\,h^{^{T\,-1}}={1\over{\sqrt{|\eta+F|}}}\, \left({\mathbf
1}-{1\over{2}}\,\tilde\gamma^{a_1
a_2}\,F_{a_1a_2}-{1\over{8}}\,\tilde\gamma^{a_1a_2a_3a_4}\,F_{a_1a_2}\,F_{a_3a_4}
-{1\over{2\cdot 4!}}\,\epsilon^{a_1\cdots
a_6}\,F_{a_1a_2}\,F_{a_3a_4}\,F_{a_5a_6}\,\bar\gamma
\,\right)\,.\nonumber
\end{eqnarray}

The equation (\ref{D5fAdS}) tells us how the worldvolume fermions
interact with the non--zero worldvolume flux $F_2$ induced by the
background $R_5$ flux.

Analogously, one can consider the dual case, {\it i.e.} a D5 brane
filling $AdS_5$ and wrapping a cycle in $X_5$. The form of the
equations remain the same with only difference that the time
component gets replaced with the component corresponding to the
cycle in $X_5$ wrapped by the \hbox{D5 brane}. A warp
compactification counterpart of this situation will be considered in
Subsection \ref{D5st}.

\subsection{D5 and D7 branes in warped compactification backgrounds}
Let us consider D5 and D7 branes in a more generic, warped $M_4
\times_w X_6$ compactification background  introduced in the
beginning of Subsection \ref{warped}. To simplify things we shall
mainly restrict ourselves to backgrounds with a constant
dilaton--axion except for the example of a D5--brane instanton.
Recall that, to preserve Lorentz invariance in $M_4$ the three form
fluxes $H_3$ and $R_3$ along $M_4$ are assumed to be zero.

\subsubsection{Space--time filling D5 brane}\label{D5st}

Consider a D5 brane that fills the effective four--dimensional
space--time and wraps a two--cycle in $X_6$. This brane
configuration participates {\it e.g.} in the tadpole cancelation
(see
\cite{Grana:2005jc} for a review and references). The D5 brane
equations take the following form.

The worldvolume scalar field equation is (\ref{cbfi}),
(\ref{2ndff}), (\ref{cbfi1})
\begin{eqnarray}\label{bsx4}
e^{-\Phi}\,K_{ba}^{\,\,\,\,\,i}\,(\eta+{ F})^{-1(ab)}=
e^{-\Phi}\,D^i\,\Phi-{{e^{-\Phi}}\over 2}\,
H^i{}_{ab}\,(\eta+{F})^{-1ab} -{1\over{6!}}\, {1\over {\sqrt{|\eta+{
F}|}}}\,\epsilon^{a_1\cdots a_6}\, {R}^i{}_{a_1\cdots a_6}
\nonumber\\
\\
- {1\over {2\cdot 4!}}\, {1\over {\sqrt{|\eta+{
F}|}}}\,\epsilon^{a_1\cdots a_6}\, {R}^i{}_{a_1\cdots a_{4}}
\, { F}_{a_5a_6}\,- {1\over {16}}\,{1\over {\sqrt{|\eta+{ F}|}}}
\,\epsilon^{a_1\cdots a_6}\, {R}^i{}_{a_1a_{2}}
\, { F}_{a_3a_4}\, { F}_{a_5a_6}
\nonumber\\
\nonumber\\
- {1\over {2\cdot 4!}}\,{1\over {\sqrt{|\eta+{ F}|}}}
\,\epsilon^{a_1\cdots a_6}\, {F}_{a_1a_{2}}\,
\, { F}_{a_3a_4}\, { F}_{a_5a_6}\,{R}^{i}\,,\nonumber
\end{eqnarray}
where the first and the last term vanish in the constant
axion--dilaton background.

In the case when the brane is static and does not fluctuate in
orthogonal directions, and taking into account that $R_7=-R^*_3$ and
the form of $R_5$ (\ref{R5}), the equation (\ref{bsx4}) in the
linear BI field approximation reduces to
\begin{equation}\label{sb}
e^{-\Phi}\,K_{a}{}^{ai}=e^{-\Phi}\,D^i\,\Phi+{{e^{-\Phi}}\over 2}\,
H^i{}_{q_1q_2}\,{ F}^{q_1q_2} +{1\over
6}\,\,\epsilon^i{}_{i_2i_3i_4}\, {R}_{i_2i_3i_4} -{1\over
2}\,\epsilon^{q_1q_2} \, { F}_{q_1q_{2}}\,\partial^i\, \ln
Z(y)+{\mathcal O}(FF)\,.
\end{equation}
Recall that the index $i=1,2,3,4$ corresponds to the four directions
in $X_6$ transversal to the brane, $q_1,q_2=1,2$ correspond to the
directions of $X_6$ along the brane and the indices $r$ are those of
$M_4$ which is part of the worldvolume, {\it i.e.} $a=(r_1,\cdots,
r_4,q_1,q_2)$.

The BI equation for the D5 brane  is
\begin{eqnarray}\label{DBI=d*F2stringx4}
{\mathcal D}_b \left(e^{-\Phi}\,(\eta + { F})^{-1\, [ba]}\,\sqrt{
|\eta + { F}|} \,
 \right) = {1\over 5!}\,
 \epsilon^{a b_1\cdots b_5}  {R}_{b_1\cdots b_5}
 + {{1} \over
{2\cdot 3!}}\;\epsilon^{a b_1\cdots b_5} \, {R}_{b_1b_2 b_3}\,
F_{b_4b_5}
\\
\nonumber\\
+{1\over 8}\,\epsilon^{ac b_1\cdots b_4} \, {R}_c\,{F}_{b_1b_2}\,
F_{b_3b_4}\,\nonumber
\,,
\end{eqnarray}
where again the last term vanishes in the constant axion--dilaton
background $R_c=0=\Phi$, which we shall further assume.

If the space--time $M_4$ filling D5 brane does not fluctuate in
$X_6$, the second term of the right hand side of
(\ref{DBI=d*F2stringx4}) also vanishes (since $R_3$ with any number
of indices in $M_4$ is zero) and in the linear approximation eq.
(\ref{DBI=d*F2stringx4}) reduces to the Maxwell equation
\begin{eqnarray}\label{DBI=d*F2stringx4l}
{\mathcal D}_b \,{ F}^{ ba} = -{1\over 5!}\,
 \epsilon^{a b_1\cdots b_5} \; {R}_{b_1\cdots b_5}=- \delta^a_{q_1}\,\epsilon^{q_1q_2}
\partial_{q_2}\, \ln Z\;\quad
 \Rightarrow
\nonumber\\
\\
\Rightarrow \quad {\mathcal D}_b \,{ F}^{ br}=0,\quad {\mathcal D}_b \,{ F}^{ bq_1} =
   -\epsilon^{q_1q_2}\,\partial_{q_2}\,\ln
 Z(y)\,,\nonumber
\end{eqnarray}
We observe that in the directions of $M_4$ the worldvolume gauge
field equation is source--less, while in the directions of $X_6$
along the brane it acquires the non--constant source term. We can
thus put to zero all components of $F_2$ along $M_4$ (in which case
the D5--brane configuration preserves $M_4$ Lorentz invariance).
Remaining Maxwell equation for $F_{q_1q_2}$ is
$$
{\mathcal D}_{q_2} \,{ F}^{ q_2q_1} =
  -\epsilon^{q_1q_2}\,\partial_{q_2}\,\ln
 Z(y)\,,\quad \Rightarrow \quad { F}_{ q_1q_2}=\epsilon_{q_1q_2}\,(c+\ln
 Z(y))\,,
$$
where $c$ is an integration constant.

In the full non--linear case, the Born--Infeld equation for the
worldvolume field on the $X_6$ part of the D5--brane worldvolume is
similar to the $D1$--brane equation (\ref{d1bi}) (the difference is
only in the sign in the square root because of the Euclidean metric
on $X_6$) and is also exactly solvable
\begin{equation}\label{d5bi}
{\partial}_q\,\left({F\over{\sqrt{1+F^2}}}\right)=
{\partial}_q\,\ln Z\quad
\Rightarrow \qquad  F= {{c+\ln Z}\over\sqrt{1- (c+\ln Z)^2}}\,,\quad
F_{q_1q_2}=\epsilon_{q_1q_2}\,F\,.
\end{equation}

We see that the worldvolume gauge field may have a non--trivial flux
on a two--cycle of $X_6$ wrapped by the D5 brane. The constant part
of its field strength can be regarded as the magnetic field of a
monopole located in the center of the compact two--cycle of $X_6$.

The bosonic equation (\ref{bsx4}) for the static space--time filling
D5 brane with the BI field (\ref{d5bi}) in the constant
axion--dilaton background takes the following form
\begin{equation}\label{bsx41}
{1\over {1+F^2}}\,K_q{}^{qi}={F\over {2(1+F^2)}}\;
H^i{}_{q_1q_2}\,{\epsilon}^{q_1q_2} -{1\over {6\,\sqrt{1+{
F}^2}}}\,\epsilon^{i_1i_2i_3i_4}\,R_{i_2i_3i_4} -\, (\,c-1+\ln
Z)\,\partial^i\, \ln Z(y)\,.
\end{equation}
When deriving (\ref{bsx41}) we took into account that for the static
($\partial_m\,x^i=0$) D3 brane under consideration  the trace
$K_r{}^{ri}$  of the components  of the second fundamental form in
$M_4$ is
 \begin{equation}\label{km4tr}K_r{}^{ri}=-\partial^i\,\ln Z\,
 \end{equation}
 which follows from the
definition (\ref{2ndff}) of the second fundamental form and the form
of the warped $M_4 \times_w X_6$ metric and connection. Note also
that the components
 $K_{qr}{}^i$ of the second
fundamental form with the index $r$ in $M_4$ and the index $q$ in
$X_6$ are zero.

Finally, for this D5 brane configuration the Dirac equation is
\begin{eqnarray}\label{finalfx4}
& {\cal D}_r\,{\Theta}\,\gamma^r+{1\over{1+F^2}}\,\left({\cal
D}_q\,{\Theta}\,  \gamma^q\, -{\cal
D}_{q_2}\,{\Theta}\,\gamma_{q_1}\,\epsilon^{q_1q_2}\,F\right)+
 =
{F\over 4}\,{\Theta}\,
\gamma_i\,\tilde\gamma^{q_1q_2}\,\epsilon_{q_1q_2}\,\partial_i\ln Z\;  \nonumber\\
\nonumber\\
&
  + {1\over 4} \;  {\Theta} \,\gamma^{\underline{ab}}\,\gamma^{\underline c}\,
 \left[{1\over{3!}}\,\left(   { {H}}_{\underline{abc}}
  - {{2+F^2}\over{1+F^2}}\, { {R}}_{\underline{abc}}\,
 h^{^{T\,-1}} \right) -
{1\over 2} \eta_{\underline c\, a}\, (\eta - F)^{-1\, ad}
 \left(   {H}_{d\underline{a}\underline{b}} -
 {R}_{d\underline{a}\underline{b}}\,h^{^{T\,-1}} \right)   +\right.\nonumber\\
&\\
&\left. +{F\over 2\cdot {3!}\,(1+F^2)} \,{ {R}}_{\underline{abc}}
\tilde{\gamma}_{q_1q_2} \,\epsilon^{q_1q_2} \,h^{^{T\,-1}} \right]
\;
 - {1\over {8\cdot 4!}} \;  {\Theta} \,
 {R}_{b\underline{c}_1\ldots \underline{c}_4}{}\gamma^{\underline{c}_1\ldots
\underline{c}_4}\gamma_a
 (\eta - F)^{-1\,
ab} h^{^{T\,-1}}+
\nonumber\\
&\nonumber\\
&+{1\over 4}\, {\Theta}\, \gamma_i\,(1+{F\over
2}\,\tilde\gamma^{q_1q_2}\,\epsilon_{q_1q_2})\,\left({F\over
{(1+F^2)}}\; H_{i\,q_1q_2}\,{\epsilon}^{q_1q_2} +{1\over
{3\,\sqrt{1+{ F}^2}}}\,\epsilon_{ii_2i_3i_4}\,R_{i_2i_3i_4}
-\partial_i\,(\,c+\ln Z)^2\,\right)\,\nonumber\,,
\end{eqnarray}
where the underlined indices run (as usual) from 0 to 9; $a=(r,q)$;
$r=0,1,2,3$ are in $M_4$, $q=1,2$ and transverse $i=1,2,3,4$ are in
$X_6$,
$$
 h^{^{T\,-1}}=-{{1\over {\sqrt{1+{ F}^2}}}}\,(F-{1\over
2}\,\epsilon_{q_1q_2}\,\tilde\gamma^{q_1q_2})\,\tilde\gamma^5 \quad
{\rm and} \quad \tilde\gamma^5={1\over 4!}\,\epsilon_{r_1\cdots
r_4}\,\tilde\gamma^{r_1\cdots r_4}.$$

It seems of interest to study in detail this D5 brane configuration
in concrete compactification schemes (see \cite{Grana:2005jc} and
references therein).


\subsubsection{D5 brane instanton wrapping Calabi--Yau six--fold}

Consider now a Euclidean D5 brane (instanton) wrapping $X_6$
\footnote{NS5 brane instantons wrapping a Calabi--Yau manifold in
type IIA theory have been studied in \cite{stefan} and references
therein.} . Then $H_3$, $R_3$
 and $R_5$, which in the general case are non--zero in $X_6$, can
induce a worldvolume flux in the D5 brane worldvolume in virtue of
the BI equations.

For the  D5--brane instanton on $X_6$,  which does not fluctuate in
transverse directions, and the background fluxes are non--zero only
on $X_6$, the both sides of the worldvolume scalar field equation
(\ref{bsx4}) are identically zero separately.

The D5 worldvolume field strength must satisfy the Bianchi identity
$ d\,{ F}_2=-H_3|_{X_6} $ and obey the BI field equation
(\ref{DBI=d*F2string}). In view of our ansatz (\ref{R5wick}),
(\ref{R5wick1}) for the Wick rotated $R_5$ flux in $X_6$ the BI
equation takes the form
\begin{eqnarray}\label{DBId5y61}
&-{\mathcal D}_b \left((\delta-{ F})^{-1\, [ba]}\,\sqrt{ |\delta + {
F}|} \,e^{-\Phi}\,
 \right) =
\nonumber\\
&\\
& = -i\; \partial^{\,a}\,\ln Z\; +
  {{i} \over
{2\cdot 3!}}\; \epsilon^{a b_1b_2 b_3b_4b_5}
\; {R}_{b_1b_2 b_3}\; { F}_{b_4b_5}+{{i} \over
{8}}\; \epsilon^{a b_1b_2 b_3b_4b_5}
\; {R}_{b_1}\,F_{b_2 b_3}\; { F}_{b_4b_5}
\,.  \nonumber
\end{eqnarray}
Since with our assumption about the Wick rotation, the left hand
side of (\ref{DBId5y61}) is real, while its right hand side is
imaginary they should be equal to zero separately. Such an ansatz
allows us to find particular solutions of the D5 brane equations of
motion. {\it E.g.} this imposes the relation between the warp factor
and the $R_1$, $R_3$ and $F_2$ fluxes similar to that for the D3
brane instanton of Subsection \ref{D3i} \footnote{Eric Bergshoeff
pointed our attention to the fact that since the axion field $C_0$
is  a pseudoscalar, it  should be replaced with $iC_0$ upon Wick
rotation. This, in particular, is required for getting D--instanton
solutions in Euclidean type IIB supergravity.  In our case this will
result in additional factor of $i$ in the term with the $R_1$ flux
and hence in a different splitting of eq. (\ref{DBId5y61}) into the
real and imaginary part, namely
$$
{\mathcal D}_b \left((\delta-{ F})^{-1\, [ba]}\,\sqrt{ |\delta + {
F}|} \,e^{-\Phi}\right)={{1} \over {8}}\; \epsilon^{a b_1b_2
b_3b_4b_5} \; {R}_{b_1}\,F_{b_2 b_3}\; { F}_{b_4b_5}\,,
$$
$$
\partial^{\,a}\,\ln Z\; =
  {{1} \over
{2\cdot 3!}}\; \epsilon^{a b_1b_2 b_3b_4b_5} \; {R}_{b_1b_2 b_3}\; {
F}_{b_4b_5}.
$$}
$$
\;- \partial^{\,a}\,\ln Z\; +
  {{1} \over
{2\cdot 3!}}\; \epsilon^{a b_1b_2 b_3b_4b_5}
\; {R}_{b_1b_2 b_3}\; { F}_{b_4b_5}+{{1} \over
{8}}\; \epsilon^{a b_1b_2 b_3b_4b_5}
\; {R}_{b_1}\,F_{b_2 b_3}\; { F}_{b_4b_5}=0
\,.
$$

The Dirac equation (\ref{finalf}) for the D5 brane fermionic modes
is (where under the Wick rotation $h^{^{T\,-1}} \rightarrow
i\,h^{^{T\,-1}}$)
\begin{eqnarray}\label{finalfd5y6}
&-e^{-\Phi}\,{\cal D}_b\,{\Theta}\,  \gamma_a\, (\delta - F)^{-1\,
ab} =- {1\over 8} \;  {\Theta} \,
 \left[ -4 \, e^{- {\Phi}}\, {\gamma^a\,D_a\, \Phi}-2i\, {
 {R}}_a\,\gamma^a
h^{^{T\,-1}}
\right. \,\nonumber\\
&\nonumber\\
&
  \left. 
-  2 i{R}_b\gamma_a (\delta + F)^{-1\,ab} h^{^{T\,-1}}
  + iR_c\,\gamma^c\, \tilde{\gamma}_b
\,\gamma_a\, (F(\delta +F)^{-1})^{ ab} h^{^{T\,-1}} \right] \;  \nonumber\\
\\
& +
   {1\over 4} \;  {\Theta} \,
 \left[{1\over{3!}}\, \gamma^{abc}\,\left(e^{-\Phi}\,H_{abc} - 2i\,  R_{abc}\,
 h^{^{T\,-1}} \right) +
{1\over 2}\gamma^{{c}{d}} \gamma_a\, (\delta + F)^{-1\, ab}
 \left( e^{-\Phi}\, {H}_{bcd} -
 i\,{R}_{bcd}\,h^{^{T\,-1}} \right)   - \right.\nonumber\\
&\nonumber\\
&\left. + {i\over {2\cdot 3!}} \,R_{cdf}\,\gamma^{cdf}
\tilde{\gamma}_b \,\gamma_a\, (F(\delta + F)^{-1})^{ ab}
h^{^{T\,-1}} \right]    \;
\nonumber\\
&\nonumber\\
& +{1\over {8}}\; {\Theta} \gamma^5\,\gamma_a \,\partial_b\,\ln Z\;
 (\delta + F)^{-1\,
ab} h^{^{T\,-1}}+ {i\over {8}} \;
  {\Theta} \,
\gamma_{cb}\,\gamma_a
\partial^c\ln Z\; (\delta + F)^{-1\,
ab} \bar\gamma\, h^{^{T\,-1}}   \;, \nonumber
\end{eqnarray}
where $\gamma^5={1\over4!}\,\epsilon_{i_1\cdots
i_4}\,\gamma^{i_1}\cdots\gamma^{i_4}$ is the gamma--five matrix of
$M_4$ orthogonal to the D5 brane worldvolume, $\bar\gamma={1\over
6!}\,\epsilon_{a_1\cdots a_6}\,\tilde\gamma^{a_1}\cdots\gamma^{a_6}$
and
\begin{eqnarray}
&h^{^{T\,-1}}={1\over{\sqrt{|\delta+F|}}}\,\left(\bar\gamma-{1\over{2\cdot\,4!}}\,\epsilon^{a_1\cdots
a_6}\,\tilde\gamma_{a_1\cdots
a_4}\,F_{a_5a_6}+{1\over{16}}\,\epsilon^{a_1\cdots
a_6}\,\tilde\gamma_{a_1a_2}\,F_{a_3a_4}\,F_{a_5a_6}\right.\nonumber\\
\nonumber\\
&\left.-{1\over{2\cdot 4!}}\,\epsilon^{a_1\cdots
a_6}\,F_{a_1a_2}\,F_{a_3a_4}\,F_{a_5a_6}\,{\mathbf
1}\right)\,.\nonumber
\end{eqnarray}

Note that the $D5$--brane instanton acquires the worldvolume flux
also in compactifications on $M_4 \times X_6$ with non--zero fluxes
$H_3$ and $R_3$ in $X_6$ but with zero $R_5$ and $R_1$.

In such cases the D5 worldvolume flux is defined by the relations
$$
dF_2=-H_3|_{X_6}, \qquad {\mathcal D}_b \left((\delta - { F})^{-1\,
[ba]}\,\sqrt{ |\delta + { F}|} \,e^{-\Phi}
 \right) =  - {{i} \over {2\cdot 3!}}\; \epsilon^{a b_1b_2 b_3b_4b_5}
\; {R}_{b_1b_2 b_3}\; { F}_{b_4b_5}\,.
\,
$$
Since, as we assume (upon the Wick rotation) the {\it l.h.s.} of the
latter equation is real and the {\it r.h.s.} is imaginary, they must
vanish separately which imposes further condition on $F_{ab}$
\begin{equation}\label{rf}
 {R}_3\; { F}_2=0 \quad \Rightarrow \quad d\,({R}_3\; {
 F}_2)=-R_3\,H_3=0\,.
\end{equation}
The simplest situation when the above equation is satisfied is
$R_3=0$. This takes place in the compactification solutions of Type
A in the Table 3.4 of \cite{Grana:2005jc}.


In such a background the Dirac equation of the D5 brane instanton is
\begin{eqnarray}\label{finalfd5y6z0}
&\quad \,{\cal D}_b\,{\Theta}\,  \gamma_a\, (\delta - F)^{-1\, ab}
=-
   {1\over 8} \;  {\Theta} \,
 \left(4 \, {\gamma^a\,D_a\, \Phi}+{1\over{3}}\, \gamma^{abc}\,H_{abc}
 +
\gamma^{{c}{d}} \gamma_a\, (\delta +F)^{-1\, ab}
 {H}_{bcd}\right).\hspace{30pt}
\end{eqnarray}
It will be of interest to analyze non--perturbative effects due to
the D5 brane instantons in effective four--dimensional theory.

\subsubsection{Compact space filling D7 brane}

A D7  brane  wraps $X_6$ and its worldvolume has time and one space
 direction along $M_4$. So it looks like a string for an observer
living in $M_4$. The D7  brane scalar field equation is
(\ref{cbfi}),  (\ref{cbfi1})
\begin{eqnarray}\label{cbfid7} &e^{-\Phi}\,K_{ba}^{\,\,\,\,\,i}\,(\eta+{
F})^{-1(ab)}=e^{-\Phi}\,D^i\Phi-{e^{-\Phi}\over
2}\,H^i{}_{ab}\,(\eta+{F})^{-1ab} \hspace{150pt}\nonumber\\
\nonumber\\ & \hspace{100pt} +{1\over
{\sqrt{|\eta+F|}}}\,\left(-\epsilon^{ij}\,R^j+{1\over{16\cdot
4!}}\,\epsilon^{a_1\cdots
a_{8}}\,F_{a_1a_2}\,F_{a_3a_4}\,F_{a_5a_6}\,F_{a_7a_8}\,R^i
\right.\nonumber\\
\\ & \left. \hspace{100pt}-{1\over
2}\,\epsilon^{ij}\,R^{jab}\,F_{ab}+{1\over{16\cdot
3!}}\,\epsilon^{a_1\cdots
a_{8}}\,R^i{}_{a_1a_2}\,F_{a_3a_4}\,F_{a_5a_6}\,F_{a_7a_8}
\right.\nonumber\\
\nonumber\\ & \left.\hspace{150pt} +{1\over{8\cdot
4!}}\,\epsilon^{a_1\cdots
a_{8}}\,R^i{}_{a_1a_2a_3a_4}\,F_{a_5a_6}\,F_{a_7a_8}\right)
 \,.\nonumber
\end{eqnarray}
The second fundamental form $K_{ba}^{\,\,\,\,\,i}$ of the embedded
D7 brane worldvolume is defined in (\ref{2ndff}).

If the brane is static and does not fluctuate along the transverse
space directions, for a given $M_4$ Lorentz preserving configuration
of non--zero fluxes $H_3$, $R_3$ and $R_5$ in $X_6$ and a constant
axion--dilaton, the right hand side of (\ref{cbfid7}) vanishes. Also
$K_{ba}^{\,\,\,\,\,i}\,=0$ for such a static embedding into the
warped background, due to the definition (\ref{2ndff}) of the second
fundamental form and because $x^i$ are transverse directions in
$M_4$ on which the warp factor $Z(y)$ does not depend. In this case
the BI equation takes the form
\begin{eqnarray}\label{DBI=d*F2stringd7}
{\mathcal D}_b \left((\eta + { F})^{-1\, [ba]}\,\sqrt{ |\eta + {
F}|}
\,
 \right) =
 -{1 \over {2}}\; \delta^{a}_ {q}\;\epsilon^{r_1r_2}\, {
F}_{r_1 r_2}\,\partial_{q}\,\ln Z -\; \delta^{a}_
{r_1}\;\epsilon^{r_1r_2}\, { F}_{r_2q}\,\partial_{q}\,\ln Z
\nonumber\\
\\
-{1\over
4!}\,\delta^{a}_q\,\epsilon^{qq_1q_2q_3q_4q_{5}}\,\epsilon^{r_1r_2 }
\;(
 {R}_{q_1q_2q_3} { F}_{q_4q_{5}}\; {
F}_{r_1r_2} -2{R}_{q_1q_2q_3} { F}_{r_1q_4}\; { F}_{r_2q_{5}})
\nonumber\\
\nonumber\\
-{1\over 3!}\,\delta^{a}_{r_1}\,\epsilon^{r_1r_2
}\,\epsilon^{q_1q_2q_3q_4q_{5}q_6}\,{R}_{q_1q_2q_3} { F}_{r_2q_4}\;
{ F}_{q_5q_6} \nonumber
\end{eqnarray}
where $a=(r_1,r_2,q)$ is the D7 worldvolume index, the index $q$
corresponds to the $X_6$ part of the D7  brane, the indices
$r_1,r_2$ label the two directions of the D7
 worldvolume along $M_4$ and the indices $i,j=1,2$ correspond to the $M_4$ orthogonal directions.
The BI field strength satisfies the Bianchi identity $d{
F}_2=-H_3|_{X_6}$ and thus at least $F_{q_1q_2}$ in $X_6$ is
intrinsically non--zero if the pullback of the NS--NS flux is
non--zero.

Finally the static D7  brane Dirac equation is
\begin{eqnarray}\label{finalfd7}
& {\cal D}_b\,{\Theta}\,  \gamma_a\, (\eta + F)^{-1\, ab} =
\hspace{250pt}
\nonumber\\
&\\
& -
   {1\over 8} \;  {\Theta} \,\gamma_{q_1q_2}\,\gamma_{a}\,
 \left[{1\over{3}}\,\delta^{a}_{q_3}\,\left(H_{q_1q_2q_3} - 3  R_{q_1q_2q_3}\,
 h^{^{T\,-1}} \right) -
 (\delta - F)^{-1\, a}{}_{q_3}
 \left( {H}_{q_1q_2q_3} -
 {R}_{q_1q_2q_3}\,h^{^{T\,-1}} \right)    \right.\nonumber\\
&\nonumber\\
&\left. -  {1\over {  3!}} \,\delta^{a}_{q_3}\,R_{q_1q_2q_3} \,
\tilde{\gamma}_b \,\gamma_c\, (F(\eta - F)^{-1})^{ cb} h^{^{T\,-1}}
\right]    \;
 - {1\over {8\cdot 4!}} \;  {\Theta} \,
 {\epsilon}^{q_1\cdots q_6}\,\gamma_{q_1\cdots q_4}\partial_{q_5}\,\ln
 Z\,\gamma^a
 (\eta - F)^{-1\,
}{}_{aq_6}\, \,h^{^{T\,-1}} \nonumber\\
&\nonumber\\
&-{1\over {16}} \;  {\Theta} \,
 {\epsilon}_{ij}\,\gamma^{ij}\,\gamma^{r_1}\,{\epsilon}_{r_1r_2}\,
 (\eta + F)^{-1\,
r_2a}\,\partial_q{}\,\ln
 Z\,\tilde\gamma_q\,\gamma_a \,h^{^{T\,-1}}  \;, \nonumber
\end{eqnarray}
where
\begin{eqnarray}
 &h^{^{T\,-1}}={1\over{\sqrt{|\eta+F|}}}\,\left(\bar\gamma-{1\over{2 \cdot 6!}}\,\epsilon^{a_1\cdots
a_8}\,\tilde\gamma_{a_1\cdots a_6}\,F_{a_7a_8}+{1\over{8\cdot
4!}}\,\epsilon^{a_1\cdots
a_8}\,\tilde\gamma_{a_1\cdots a_4}\,F_{a_5a_6}\,F_{a_7a_8}\right.\nonumber\\
\nonumber\\
&\left.-{1\over{16\cdot 3!}}\,\epsilon^{a_1\cdots
a_8}\,\tilde\gamma_{a_1a_2}\,F_{a_3a_4}\,F_{a_5a_6}\,F_{a_5a_6}\,
+{1\over{16\cdot 4!}}\,\epsilon^{a_1\cdots
a_8}\,F_{a_1a_2}\,F_{a_3a_4}\,F_{a_5a_6}\,F_{a_5a_6}\,{\mathbf 1}
\right)\,\nonumber
\end{eqnarray}
and $\bar\gamma={1\over{8!}}\,\epsilon^{a_1\cdots
a_8}\,\tilde\gamma_{a_1\cdots a_8}$.

\subsubsection{Space--time filling D7 brane}

A D7  brane  fills $M_4$ and wraps a four--fold in $X_6$. Such a
D7--brane takes part in the tadpole cancelation and gaugino
condensation (see \cite{Grana:2005jc} for a review and references).
The role of D7 brane worldvolume fluxes in the generation of
supersymmetry breaking terms in effective four--dimensional theory
has been discussed in \cite{luis,Martucci:2006ij}.

 If the D7 brane is static and
does not fluctuate in the transverse directions of $X_6$ its scalar
field equation is similar to (\ref{cbfid7}) and in the constant
axion--dilaton background takes the form
\begin{eqnarray}\label{d7m4}
&K_{ba}^{\,\,\,\,\,i}\,(\eta+{ F})^{-1(ab)}=-{1\over
2}\,H^i{}_{q_1q_2}\,(\eta+F)^{-1q_1q_2}
-{1\over{2\sqrt{|\eta+F|}}}\,\epsilon^{ij}\,R^j{}_{q_1q_2}\,F^{q_1q_2}\nonumber\\
\\
&+{1\over{8\sqrt{|\eta+F|}}}\,\partial^{i}\,\ln
Z\,F_{q_1q_2}\,F_{q_3q_4}\,\epsilon^{q_1\cdots q_4}
-{1\over{8\sqrt{|\eta+F|}}}\,\epsilon^{ij}\,\partial^j\,\ln
Z\,F_{r_1r_2}\,F_{r_3r_4}\,\epsilon^{r_1\cdots r_4}
\nonumber\\
\nonumber\\
&+{1\over{4\cdot 4!\sqrt{|\eta+F|}}}\,\epsilon^{a_1\cdots
a_8}\,R^{i}{}_{a_1a_2}\,F_{a_3a_4}\,F_{a_5a_6}\,F_{a_7a_8}
\,,\nonumber
\end{eqnarray}
where the second fundamental form $K_{ba}^{\,\,\,\,\,i}$ of the
embedded worldvolume has been defined in (\ref{2ndff}),  $a=(r,q)$
is the D7 worldvolume index, the indices $r$ correspond to the $M_4$
part of the D7 brane
 and the indices $q=1,2,3,4$ label the four directions of D7
along $X_6$. The indices $i$ denote two directions of $X_6$
orthogonal to D7. $Z(y)$ is the warp factor defining the $R_5$ flux
(\ref{R5}).

 The equation of motion of the D7--brane BI field
(satisfying the Bianchi identity $dF_2=-H_3$) is
\begin{eqnarray}\label{DBI=d*F2stringd7x4}
&{\mathcal D}_b \left((\eta + { F})^{-1\, [ba]}\,\sqrt{ |\eta + {
F}|} \,
 \right) ={1\over
 2} \,\delta^a_{q}\,\epsilon_{i_1i_2}\,R^{q i_1i_2}
 -{1\over
 {2}}\,\delta^a_{q_1}\,\epsilon^{q_1q_2q_3q_4}
 \,F_{q_2q_3}\,\partial_{q_4}\,\ln Z\nonumber\\
 \\
&-{1\over
 {2 \cdot
 3!}} \,\delta^a_{r_1}\epsilon^{r_1r_2r_3r_4}\,
 \epsilon^{q_1q_2q_3q_4}\,R_{q_1q_2q_3}\,F_{q_4r_2}\,F_{r_3r_4}
  -{1\over
 {2 \cdot
 4!}}\,\delta^a_{q_1}\,\epsilon^{q_1q_2q_3q_4}\,R_{q_2q_3q_4}\,F_{r_1r_2}\,F_{r_3r_4}
 \,\epsilon^{r_1r_2r_3r_4}\,.
 \nonumber
\end{eqnarray}

We can further assume that for the space--time filling D7--brane
configuration to preserve $M_4$ Lorentz invariance, the vacuum value
of its Born--Infeld field in $M_4$ is zero, namely
\hbox{$F_{r_1r_2}=F_{rq}=0$}. Then, taking into account that the
trace of the components of the second fundamental form along warped
$M_4$ is given by eq. (\ref{km4tr}), the bosonic equations
(\ref{d7m4}) and (\ref{DBI=d*F2stringd7x4}) reduce, respectively, to
\begin{eqnarray}\label{d7m41}
&K_{q_2q_1}{}^{i}\,(\delta+{ F})^{-1(q_1q_2)}=-\partial^i\,\ln
Z-{1\over 2}\,H^i{}_{q_1q_2}\,(\delta-F)^{-1q_1q_2}
+{1\over{2\sqrt{|\delta+F|}}}\,\epsilon^{ij}\,R^j{}_{q_1q_2}\,F^{q_1q_2}
\nonumber\\
\\
&-{1\over{8\sqrt{|\delta+F|}}}\,\partial^i \ln
Z\,F_{q_1q_2}\,F_{q_3q_4}\,\epsilon^{q_1\cdots q_4}\, \nonumber
\end{eqnarray}
 and
\begin{eqnarray}\label{DBI=d*F2stringd7x41}
{\mathcal D}_{q_2} \left((\delta - { F})^{-1\, [q_2q_1]}\,\sqrt{
|\delta + { F}|} \,
 \right)
  =-{1\over
 2} \,\epsilon_{i_1i_2}\,R^{q_1 i_1i_2}
 +{1\over
 {2 }} \,\epsilon^{q_1q_2q_3q_4}
 \,F_{q_2q_3}\,\partial_{q_4}\,\ln Z
 \,.
\end{eqnarray}
For $F_{r_1r_2}=F_{rq}=0$ in $M_4$, the Dirac equation has the
following form
\begin{eqnarray}\label{finald7m4}
& {\cal D}_r\,{\Theta}\,  \gamma^r -{\cal D}_{q_1}\,{\Theta}\,
\gamma_{q_2}\, (\delta- F)^{-1\, q_1q_2}-{1\over
2}\,K_{(q_2q_1)}{}^{i}\; {\Theta}\, \gamma^i\tilde
\gamma^{q_1}{}_{q_3} [F(\delta -F)^{-1}]^{q_3q_2}= 
\nonumber\\
\nonumber\\
& -
   {1\over 8} \;  {\Theta} \,\gamma_{q_1q_2}\,\gamma_{q_3}\,
 \left[{1\over{3}}\,\left(H_{q_1q_2q_3} - 3  R_{q_1q_2q_3}\,
 h^{^{T\,-1}} \right) -
 (\delta - F)^{-1\,q_3}{}_{q_4}
 \left( {H}_{q_1q_2q_4} -
 {R}_{q_1q_2q_3}\,h^{^{T\,-1}} \right)    \right.\nonumber\\
&\nonumber\\
&\left. + {1\over {  3!}} \,R_{q_1q_2q_3}\, \tilde{\gamma}_{q_4}
\,\gamma_{q_5}\, (F(\delta + F)^{-1})^{ q_5q_4} h^{^{T\,-1}} \right]
\;-{1\over {2}} \;  {\Theta}\,\gamma^5 \,
 (\gamma_{q}\,\partial_q\,\ln Z+\gamma_{i}\,\partial_i\,\ln Z)
 \, h^{^{T\,-1}}
\nonumber\\
\\
& +{1\over {16\cdot 3!}} \;  {\Theta}
 \,\gamma^i\,\epsilon_{ij}\,\partial^j\,\ln Z\,
 {\epsilon}_{q_2q_3q_4q_5}{}\tilde\gamma^{q_2q_3q_4}\,\gamma_{q_1}
 (\delta- F)^{-1\,
q_5q_1}\, h^{^{T\,-1}}
\nonumber\\
\nonumber\\
&-{1\over {64}} \;  {\Theta}
 \,\gamma^{ij}\,\epsilon_{ij}\,
 {\epsilon}_{q_2q_3q_4q_5}\partial^{q_4}\,\ln Z\,\tilde\gamma^{q_2q_3}\,\tilde\gamma_{q_1}
 (\delta - F)^{-1\,
q_5q_1}\,h^{^{T\,-1}}
\nonumber\\
\nonumber\\
 &-{1\over {8}} \;  {\Theta} \,\gamma^5\,
 \partial_{q_1}\,\ln Z\,(\delta-F)^{-1q_1q_2}\,\gamma_{q_2}
 \, h^{^{T\,-1}}
 +{1\over 4}\,\Theta\,\gamma_i\,H_{iq_1q_2}\,(\delta-F)^{-1q_1q_2}
\nonumber\\
\nonumber\\
&-{1\over{4\sqrt{|\delta+F|}}}\,\Theta\,\gamma_i\,\epsilon^{ij}\,R_{jq_1q_2}\,F^{q_1q_2}
+{1\over{16\,\sqrt{|\delta+F|}}}\,\Theta\,\gamma_i\,\partial_i \ln
Z\,F_{q_1q_2}\,F_{q_3q_4}\,\epsilon^{q_1q_2q_3q_4}\,\;, \nonumber
\end{eqnarray}
 where $\gamma^5={1\over 4!}\,\epsilon_{r_1r_2r_3r_4}\,\gamma^{r_1r_2r_3r_4}$
 is the gamma--five matrix on $M_4$ and
$$
h^{^{T\,-1}}={1\over{\sqrt{|\delta+F|}}}\,\epsilon^{q_1q_2q_3q_4}\,
\left({1\over 4!}\,\tilde\gamma_{q_1q_2q_3q_4}-{1\over
4}\,F_{q_1q_2}(\tilde\gamma_{q_3q_4}-{1\over 2}F_{q_3q_4}\,{\mathbf
1})\right)\,\tilde\gamma^5\,.
$$
In the case when $F_2$ is self--dual or anti--self--dual on the
four--cycle of $X_6$, the Dirac equation can be simplified using the
relations for self--dual $F_2$ given in Subsection \ref{D3i}.

\subsubsection{D3 branes in $AdS_3\times S^3\times X_4$  and $M_4\times S^3\times S^3$}
In the variety of brane configurations and compactification setups
considered in the literature one can easily find other examples when
the BI worldvolume field on the brane is non--zero. For instance,
consider a compactification of type IIB string theory on
$AdS_3\times S^3\times X_4$ or $M_4\times S^3\times S^3$ with
non--zero $H_3$ and $R_3$ fluxes in $S^3$ and/or $AdS_3$. Then a D3
brane wrapping $S^3$ or $AdS_3$ will carry a non--zero BI field and
a corresponding effective charge, like the \hbox{D5 branes} in the
examples of Subsection \ref{d5ads5}.

\section{Conclusion}
We have carried out a detailed analysis of the bosonic and fermionic
equations of motion of type IIB D branes in generic supergravity
backgrounds with fluxes taking into account the worldvolume BI
field. We have presented the explicit form of these equations for
each of the Dp branes $(p=1,3,5,7,9)$ in the cases in which the
$H_3$ and $R_p$ fluxes along the brane worldvolume are zero and the
supergravity backgrounds do not induce the worldvolume BI gauge
field. We then gave several examples of brane configurations in
which the worldvolume flux is intrinsically non--zero and thus must
be taken into account in studying the problems in which such branes
are involved.

It would be of interest to analyze further the brane configurations
with background and worldvolume fluxes discussed in the above
Subsections and other ones and to study their effects in concrete
compactification setups, supersymmetry etc., {\it e.g.} within the
lines of recent papers considering properties and effects of D3, D5
and D7 branes \cite{Martucci:2006ij} and Euclidean D3 branes
\cite{maldacena} on cycles of warped Calabi--Yau manifolds.

\section*{Acknowledgements}
The authors are grateful to Itzhak Bars, Eric Bergshoeff, Gianguido
Dall'Agata, Jos\'e de Azc\'arraga, Paolo Pasti, Mario Tonin and Paul
Townsend for fruitful and stimulating discussion at various stages
of this work. This work was partially supported by research grants
of the Ministerio de Educaci\'on y Ciencia (FIS2005-02761) and EU
FEDER funds, the Generalitat Valenciana (ACOMP06/187, GV05/102), the
EU MRTN-CT-2004-005104 grant `Forces Universe', by the INTAS
Research Grant 2005--2006--7928, by the MIUR-PRIN-2005 contract and
by INFN Special Initiative TS11 (D.S.), and by the Ukrainian State
Fund for Fundamental Research n.383 (I.B.). D.S. thanks the MEC of
Spain for the sabbatical grant and is grateful to Jose de
Azc\'arraga for warm hospitality extended to him in Valencia.

\section*{Appendix}

The underlined indices $\underline a, \underline b,
\ldots=0,1,\cdots, 9$ are $D=10$ target space indices. The indices
$a,b,c,d=0,1,\cdots p$ are those of ($p+1$)--dimensional Dp--brane
worldvolume and the indices $i,j,k=1,\cdots, 9-p$ are of the
directions orthogonal to the worldvolume.
\\
\\
We use the mostly minus convention for the Minkowski metric
$\eta^{{\underline a}{\underline b}}=diag\,(+,-\cdots,\,-)$. The
unit antisymmetric tensor $\epsilon^{{\underline a}_0\cdots
{\underline a}_9}$ is defind as follows $\epsilon^{0123456789}=1$
and $\gamma^0\tilde\gamma^1\cdots
\tilde\gamma^9=-\tilde\gamma^0\gamma^1\cdots \gamma^9=1$, where
$\gamma^{\underline a}_{\alpha\beta}$ and $\tilde\gamma^{\underline
a\,\alpha\beta}$ are $16\times 16$ (symmetric) matrix counterparts
of the Pauli matrices satisfying the relations
\begin{equation}\nonumber
\gamma^{\underline a}\,\tilde\gamma^{\underline
b}+\gamma^{\underline b}\,\tilde\gamma^{\underline
a}=2\,\eta^{{\underline a}{\underline b}} \,,\qquad
\gamma^{[\underline a_1}\,\tilde\gamma^{\underline a_2}\cdots
\gamma^{\underline a_5]}={1\over{5!}}\,\epsilon^{\underline
a_1\cdots \underline a_5\,\underline a_6\cdots \underline
a_{10}}\,\gamma_{\underline a_6}\,\tilde\gamma_{\underline
a_7}\cdots \gamma_{\underline a_{10}}\,.
\end{equation}
Note that there is no ``charge conjugation'' matrix in $D=10$ which
would lower or rise 16--component Majorana--Weyl spinor indices
$\alpha$, $\beta$. This is why one should distinguish between
$\gamma^{\underline a}_{\alpha\beta}$ and $\tilde\gamma^{\underline
a\,\alpha\beta}$.
\\
\\
The explicit expression for the generalized covariant derivative
acting on the $\kappa$--symmetry projector matrix $h$ in the concise
form of the bosonic equations (\ref{Beq(h)})  in {\bf the Einstein
frame} is
\begin{eqnarray}\label{bbDh:=}
&{\mathbf D}_b h
  = D_bh   + {1\over 8} \left\{{H}\!\!\!\!/\,\tilde{\gamma}_{{b}} +
{\gamma}_{{b}}\,{\tilde{H}}\!\!\!\!/ \, , \, h \right\}   - {1\over
4(p+1)} \,\left\{{H}\!\!\!\!/\,\tilde{\gamma}^{c}  \, , \, h
\right\} \,(\eta +F)_{cb} - \qquad \nonumber \\ &   - {1\over
4}\left[ \nabla\!\!\!\!/ \Phi\, \tilde{\gamma}_{{b}} \, , \, h
\right] - {1\over 2(p+1)}\, \left[ \nabla\!\!\!\!/ \Phi\,
\tilde{\gamma}^{c} \, , \, h \right]\, (\eta +F)_{cb} - \qquad
\nonumber \\ &  - \; {1\over 8}\, e^{\Phi} \,\left( \gamma_{{b}}\,
{\tilde{R}}\!\!\!/_{1}  + h\gamma_{{b}}\, {\tilde{R}}\!\!\!/_{1}\, h
\right) - {1\over 2(p+1)} \,e^{\Phi} \,\left(  { {R}}\!\!\!/_{1}\,
\tilde{\gamma}^c \,(\eta -F)_{cb} - h\, { {R}}\!\!\!/_{1}
\,\tilde{\gamma}^c\,h (\eta +F)_{cb}  \right)+ \qquad \nonumber \\
&  + \; {1\over 8}\, e^{{1\over 2}\Phi} \,\left( \gamma_{{b}}\,
{\tilde{R}}\!\!\!/_{3}  - h\gamma_{{b}}\, {\tilde{R}}\!\!\!/_{3}\, h
\right) + {1\over 4(p+1)}\, e^{{1\over 2}\Phi} \left(  {
{R}}\!\!\!/_{3}\, \tilde{\gamma}^c \,(\eta -F)_{cb} + h\, {
{R}}\!\!\!/_{3} \,\tilde{\gamma}^c\,h \,(\eta +F)_{cb}  \right)-
\qquad \nonumber \\ &   - {1\over 16} \,
\left(\gamma_{\underline{b}}\, {\tilde{R}}\!\!\!/_{5}  +
h\,\gamma_{\underline{b}}\, {\tilde{R}}\!\!\!/_{5} \,h\right) \; \,
\qquad\nonumber
\end{eqnarray}
where the slashed objects imply the contraction of the components of
$D=10$ n--forms $W_n$ $(n=1,3,5)$ with $\gamma^{\underline
a}_{\alpha\beta}$ and $\tilde\gamma^{\underline a\,\alpha\beta}$
such that ${ {W}}\!\!\!\!\!/_{n}:={1\over n!}\,\gamma^{\underline
a_1\cdots\underline a_n}\,W_{\underline a_1\cdots\underline a_n}$
and $\widetilde{{W}\!\!\!\!\!/}_{n}:={1\over
n!}\,\tilde\gamma^{\underline a_1\cdots\underline
a_n}\,W_{\underline a_1\cdots\underline a_n}$.

\end{document}